\pdfoutput=1
\documentclass[acmsmall,fleqn]{acmart}

\acmJournal{PACMPL}
\acmVolume{1}
\acmNumber{ICFP} % CONF = POPL or ICFP or OOPSLA
\acmArticle{55}
\acmYear{2018}
\acmMonth{1}
\acmDOI{} % \acmDOI{10.1145/nnnnnnn.nnnnnnn}
\startPage{1}

\setcopyright{none}
\newcommand{\ponders}[3]{\ignorespaces}
\newcommand{\TODO}[1]{}
\newcommand{\ck}[1]{\ponders{CK}{blue}{#1}}
\newcommand{\nw}[1]{\ponders{NW}{orange}{#1}}
\newcommand{\mpi}[1]{\ponders{MP}{green}{#1}} % \mp already defined

\usepackage{fancyvrb}

\DefineVerbatimEnvironment%
  {core}{Verbatim}
  {xleftmargin=\mathindent}

\usepackage{color}
\usepackage{pgfplots}
\pgfplotsset{compat=1.14}
\usepackage{pgfplotstable}
\usepgfplotslibrary{colorbrewer}
\pgfplotsset{colormap/Dark2-8}
\pgfplotsset{cycle list/Dark2}

\usepackage{subcaption}
\usepackage{caption}
\usepackage{tikz}
\usepackage{tikz-cd}
\usepackage{csvsimple,booktabs, siunitx, array}
\usepackage{sansmath}

\makeatletter
\@ifundefined{lhs2tex.lhs2tex.sty.read}%
  {\@namedef{lhs2tex.lhs2tex.sty.read}{}%
   \newcommand\SkipToFmtEnd{}%
   \newcommand\EndFmtInput{}%
   \long\def\SkipToFmtEnd#1\EndFmtInput{}%
  }\SkipToFmtEnd

\newcommand\ReadOnlyOnce[1]{\@ifundefined{#1}{\@namedef{#1}{}}\SkipToFmtEnd}
\usepackage{amstext}
\usepackage{amssymb}
\usepackage{stmaryrd}
\DeclareFontFamily{OT1}{cmtex}{}
\DeclareFontShape{OT1}{cmtex}{m}{n}
  {<5><6><7><8>cmtex8
   <9>cmtex9
   <10><10.95><12><14.4><17.28><20.74><24.88>cmtex10}{}
\DeclareFontShape{OT1}{cmtex}{m}{it}
  {<-> ssub * cmtt/m/it}{}

\DeclareFontShape{OT1}{cmtt}{bx}{n}
  {<5><6><7><8>cmtt8
   <9>cmbtt9
   <10><10.95><12><14.4><17.28><20.74><24.88>cmbtt10}{}
\DeclareFontShape{OT1}{cmtex}{bx}{n}
  {<-> ssub * cmtt/bx/n}{}
\newcommand{\Conid}[1]{\mathit{#1}}
\newcommand{\Varid}[1]{\mathit{#1}}
\newcommand{\anonymous}{\kern0.06em \vbox{\hrule\@width.5em}}

\usepackage{polytable}

\@ifundefined{mathindent}%
  {\newdimen\mathindent\mathindent\leftmargini}%
  {}%

\def\resethooks{%
  \global\let\SaveRestoreHook\empty
  \global\let\ColumnHook\empty}
\newcommand*{\savecolumns}[1][default]%
  {\g@addto@macro\SaveRestoreHook{\savecolumns[#1]}}
\newcommand*{\restorecolumns}[1][default]%
  {\g@addto@macro\SaveRestoreHook{\restorecolumns[#1]}}
\newcommand*{\aligncolumn}[2]%
  {\g@addto@macro\ColumnHook{\column{#1}{#2}}}

\resethooks

\newcommand{\onelinecommentchars}{\quad-{}- }
\newcommand{\commentbeginchars}{\enskip\{-}
\newcommand{\commentendchars}{-\}\enskip}

\newcommand{\visiblecomments}{%
  \let\onelinecomment=\onelinecommentchars
  \let\commentbegin=\commentbeginchars
  \let\commentend=\commentendchars}

\newcommand{\invisiblecomments}{%
  \let\onelinecomment=\empty
  \let\commentbegin=\empty
  \let\commentend=\empty}

\visiblecomments

\newlength{\blanklineskip}
\newcommand{\hsindent}[1]{\quad}% default is fixed indentation
\let\hspre\empty
\let\hspost\empty

\EndFmtInput
\makeatother
\ReadOnlyOnce{polycode.fmt}%
\makeatletter

\newcommand{\hsnewpar}[1]%
  {{\parskip=0pt\parindent=0pt\par\vskip #1\noindent}}

\newcommand{\hscodestyle}{}

\newcommand{\sethscode}[1]%
  {\expandafter\let\expandafter\hscode\csname #1\endcsname
   \expandafter\let\expandafter\endhscode\csname end#1\endcsname}

  {\par\noindent
   \advance\leftskip\mathindent
   \hscodestyle
   \let\\=\@normalcr
   \let\hspre\(\let\hspost\)%
   \pboxed}%
  {\endpboxed\)%
   \par\noindent
   \ignorespacesafterend}

  {\hsnewpar\abovedisplayskip
   \advance\leftskip\mathindent
   \hscodestyle
   \let\hspre\(\let\hspost\)%
   \pboxed}%
  {\endpboxed%
   \hsnewpar\belowdisplayskip
   \ignorespacesafterend}

  {\hsnewpar\abovedisplayskip
   \advance\leftskip\mathindent
   \hscodestyle
   \let\\=\@normalcr
   \(\pboxed}%
  {\endpboxed\)%
   \hsnewpar\belowdisplayskip
   \ignorespacesafterend}

\newcommand{\plainhs}{\sethscode{plainhscode}}

\plainhs

  {\hsnewpar\abovedisplayskip
   \advance\leftskip\mathindent
   \hscodestyle
   \let\\=\@normalcr
   \(\parray}%
  {\endparray\)%
   \hsnewpar\belowdisplayskip
   \ignorespacesafterend}

  {\parray}{\endparray}

  {\(\parray}{\endparray\)}

\def\codeframewidth{\arrayrulewidth}
\RequirePackage{calc}

  {\parskip=\abovedisplayskip\par\noindent
   \hscodestyle
   \arrayrulewidth=\codeframewidth
   \tabular{@{}|p{\linewidth-2\arraycolsep-2\arrayrulewidth-2pt}|@{}}%
   \hline\framedhslinecorrect\\{-1.5ex}%
   \let\endoflinesave=\\
   \let\\=\@normalcr
   \(\pboxed}%
  {\endpboxed\)%
   \framedhslinecorrect\endoflinesave{.5ex}\hline
   \endtabular
   \parskip=\belowdisplayskip\par\noindent
   \ignorespacesafterend}

\newcommand{\framedhslinecorrect}[2]%
  {#1[#2]}

  {\(\def\column##1##2{}%
   \let\>\undefined\let\<\undefined\let\\\undefined
   \newcommand\>[1][]{}\newcommand\<[1][]{}\newcommand\\[1][]{}%
   \def\fromto##1##2##3{##3}%
   }{\) }%

  {\let\orighscode=\hscode
   \let\origendhscode=\endhscode
   \def\endhscode{\def\hscode{\endgroup\def\@currenvir{hscode}\\}\begingroup}
   \orighscode\def\hscode{\endgroup\def\@currenvir{hscode}}}%
  {\origendhscode
   \global\let\hscode=\orighscode
   \global\let\endhscode=\origendhscode}%

\makeatother
\EndFmtInput
\ReadOnlyOnce{forall.fmt}%
\makeatletter

\let\HaskellResetHook\empty
\newcommand*{\AtHaskellReset}[1]{%
  \g@addto@macro\HaskellResetHook{#1}}
\newcommand*{\HaskellReset}{\HaskellResetHook}

\newcommand\hsforall{\global\let\hsdot=\hsperiodonce}
\newcommand*\hsperiodonce[2]{#2\global\let\hsdot=\hscompose}
\newcommand*\hscompose[2]{#1}

\AtHaskellReset{\global\let\hsdot=\hscompose}

\HaskellReset

\makeatother
\EndFmtInput

\newcommand{\keyword}[1]{\mathsfbf{#1}}
\renewcommand{\Varid}[1]{\mathsfsl{#1}}
\renewcommand{\Conid}[1]{\mathsfsl{#1}}

\begin{document}

\title{Generic Deriving of Generic Traversals}
%% Author with single affiliation.
\author{Csongor Kiss}
%\authornote{with author1 note}          %% \authornote is optional;
                                        %% can be repeated if necessary
%\orcid{nnnn-nnnn-nnnn-nnnn}             %% \orcid is optional
\affiliation{
%  \position{Position1}
  \department{Department of Computing}              %% \department is recommended
  \institution{Imperial College London}            %% \institution is required
%  \streetaddress{Street1 Address1}
%  \city{City1}
%  \state{State1}
%  \postcode{Post-Code1}
  \country{United Kingdom}                    %% \country is recommended
}
\email{csongor.kiss14@imperial.ac.uk}          %% \email is recommended
\author{Matthew Pickering}
%\authornote{with author1 note}          %% \authornote is optional;
                                        %% can be repeated if necessary
%\orcid{nnnn-nnnn-nnnn-nnnn}             %% \orcid is optional
\affiliation{
%  \position{Position1}
  \department{Department of Computer Science}              %% \department is recommended
  \institution{University of Bristol}            %% \institution is required
%  \streetaddress{Street1 Address1}
%  \city{City1}
%  \state{State1}
%  \postcode{Post-Code1}
  \country{United Kingdom}                    %% \country is recommended
}
\email{matthew.pickering@bristol.ac.uk}          %% \email is recommended
\author{Nicolas Wu}
%\authornote{with author1 note}          %% \authornote is optional;
                                        %% can be repeated if necessary
%\orcid{nnnn-nnnn-nnnn-nnnn}             %% \orcid is optional
\affiliation{
%  \position{Position1}
  \department{Department of Computer Science}              %% \department is recommended
  \institution{University of Bristol}            %% \institution is required
%  \streetaddress{Street1 Address1}
%  \city{City1}
%  \state{State1}
%  \postcode{Post-Code1}
  \country{United Kingdom}                    %% \country is recommended
}
\email{nicolas.wu@bristol.ac.uk}          %% \email is recommended

\begin{abstract}
Functional programmers have an established tradition of using
traversals as a design pattern to work with recursive data structures.
The technique is so prolific that a whole host of libraries have been
designed to help in the task of automatically providing traversals by
analysing the generic structure of data types.
More recently, lenses have entered the functional scene and have
proved themselves to be a simple and versatile mechanism for working
with product types. They make it easy to focus on the salient parts of
a data structure in a composable and reusable manner.

In this paper, we use the combination of lenses and traversals to give
rise to an expressive and flexible library for querying and modifying
complex data structures.
Furthermore, since our lenses and traversals are based on the
generic shape of data, we are able to use this information to produce code
that is as efficient as hand-written versions.
The technique leverages the structure of data to produce generic
abstractions that are then eliminated by the standard workhorses of
modern functional compilers: inlining and specialisation.

  %and finally
  %reflect generally on the benefits and drawbacks of the interface we have constructed.
\end{abstract}

%% End of generated code

%% Keywords
%% comma separated list
  %% \keywords are mandatory in final camera-ready submission

%% \maketitle
%% Note: \maketitle command must come after title commands, author
%% commands, abstract environment, Computing Classification System
%% environment and commands, and keywords command.
\maketitle

\section{Introduction}

% format <..> = "\mathbin{{<}{..}{>}}"
% format <.>  = "\mathbin{<\mkern-8mu\circ\mkern-8mu>}"
% format <.> = "\mathbin{{<}{.}{>}}"

%% generic

%% Dicts

Traversals are a ubiquitous way of querying and manipulating data.
They provide a reliable interface for working with data types in a
structured and predictable manner.
An appropriate suite of traversals is a valuable tool that eases the
task of constructing programs that interact with diverse data.
Unfortunately, writing traversals quickly becomes tedious work that
requires continuous curation as code evolves over time.
Naturally, our desire is to have our traversals provided for
us.

Our goal is to identify a declarative family of useful traversals that
is expressive enough for a wide range of practical programming tasks.
Furthermore, we want to completely remove the burden of writing these
traversals by automatically deriving them whenever possible.
Not only that, but we also want to generate code that performs as well
as hand-written code.

The most famous existing solution, Scrap Your Boilerplate
(SYB)~\cite{Lammel:2003:SYB}, treats this problem by performing
run-time type tests to decide which part of the tree to traverse. This
leads to a flexible interface at the expense of performance: the
approach is famously slow.
In this paper we present \text{\ttfamily generic\char45{}lens}, a library that provides a
suite of traversals that is both faster and richer than SYB.
In essence, we believe that it is time to scrap your SYB.

We are not the first to optimise SYB, and like our
predecessors~\cite{Adams:2015:optimizing,Yallop:2017:staged}, we make
use of the fact that much of the information required for traversals
is \emph{statically} known, thereby avoiding \emph{dynamic} checks at
run-time. Our innovation is to work with types and use these to infer
generated code generically in a suitable form for an automatic
evaluator to optimise effectively.
By leveraging the static information that is given to us by the generic
structure of data, we are able to produce much better generated code.
The generic abstraction is eliminated.

To have a taste of the \text{\ttfamily generic\char45{}lens} library, consider a data type of
weighted trees. There are two type parameters, which correspond to the
type of elements and weights in the tree:
\begin{hscode}\SaveRestoreHook
\column{B}{@{}>{\hspre}l<{\hspost}@{}}%
\column{16}{@{}>{\hspre}l<{\hspost}@{}}%
\column{E}{@{}>{\hspre}l<{\hspost}@{}}%
\>[B]{}\keyword{data}\;\Conid{WTree}\;\Varid{a}\;\Varid{w}\mathrel{=}\Conid{Leaf}\;\Varid{a}{}\<[E]%
\\
\>[B]{}\hsindent{16}{}\<[16]%
\>[16]{}\mid \Conid{Fork}\;(\Conid{WTree}\;\Varid{a}\;\Varid{w})\;(\Conid{WTree}\;\Varid{a}\;\Varid{w}){}\<[E]%
\\
\>[B]{}\hsindent{16}{}\<[16]%
\>[16]{}\mid \Conid{WithWeight}\;(\Conid{WTree}\;\Varid{a}\;\Varid{w})\;\Varid{w}{}\<[E]%
\ColumnHook
\end{hscode}\resethooks
Suppose that we want to gather all the weights in the tree.
Our library provides a traversal for this data type called \ensuremath{\Varid{param}}
which takes a type-level integer as an argument. This allows us to
specify which parameter we want the traversal to focus on.
Counting from the right we specify \ensuremath{\ \texttt{@}\mathrm{0}} to say that we want to focus
on the \ensuremath{\mathrm{0}}th parameter, which is \ensuremath{\Varid{w}}, the weights in the tree.
\begin{hscode}\SaveRestoreHook
\column{B}{@{}>{\hspre}l<{\hspost}@{}}%
\column{E}{@{}>{\hspre}l<{\hspost}@{}}%
\>[B]{}\Varid{weights}\mathbin{::}\Conid{WTree}\;\Conid{Int}\;\Conid{Int}\to [\mskip1.5mu \Conid{Int}\mskip1.5mu]{}\<[E]%
\\
\>[B]{}\Varid{weights}\mathrel{=}\Varid{toListOf}\;(\Varid{param}\ \texttt{@}\mathrm{0}){}\<[E]%
\ColumnHook
\end{hscode}\resethooks
The \ensuremath{\Varid{toListOf}} combinator takes a traversal and turns it into a fold
which summarises what it is focusing on. Applying \ensuremath{\Varid{weights}} to a tree
will then correctly return a list of all the weights, even though the
node values are also integers. Thankfully, if we accidentally use an
incorrect index, the \text{\ttfamily generic\char45{}lens} library produces a bespoke
compile-time error to help us correct our mistake.

\paragraph{Contributions}
The primary contribution of this paper is a demonstration of how we
can achieve guilt-free generic programming using existing language
features. More specifically, our contributions are the following:
\begin{enumerate}
\item We specify a high-level interface for describing a family of
      useful lenses, prisms, and traversals in a type-directed manner.
\item We introduce a technique that allow generic traversals over
      multiple type parameters.
\item We outline the implementation of \text{\ttfamily generic\char45{}lens}, a library that
      implements this interface using generics.
\item We provide benchmarks which demonstrate that \text{\ttfamily generic\char45{}lens} is
      as fast as hand-written code. We also discuss the optimisations which
      we require a compiler to perform.
\end{enumerate}

The remainder of this paper is structured as follows.
Section~\ref{sec:motivation} motivates the use of the generic
lenses and traversals.
Section~\ref{sec:interface} describes the interface that
our library \text{\ttfamily generic\char45{}lens} supplies.
Section~\ref{sec:background} gives the background necessary for the implementation of our library.
We then move onto implementing generic traversals that are directed by types in
Section~\ref{sec:types}, by parameters in Section~\ref{sec:params},
and by constraints in Section~\ref{sec:constraints}.
We consider the performance of our library in
Section~\ref{sec:performance} and evaluate our library with benchmarks
in Section~\ref{sec:benchmarks}. Finally, we discuss related work in
Section~\ref{sec:related}
.

% format <..> = "\mathbin{{<}{..}{>}}"
% format <.>  = "\mathbin{<\mkern-8mu\circ\mkern-8mu>}"
% format <.> = "\mathbin{{<}{.}{>}}"

%% generic

%% Dicts

\section{Type-directed Queries}
\label{sec:motivation}

\begin{figure}
\begin{hscode}\SaveRestoreHook
\column{B}{@{}>{\hspre}l<{\hspost}@{}}%
\column{16}{@{}>{\hspre}l<{\hspost}@{}}%
\column{40}{@{}>{\hspre}c<{\hspost}@{}}%
\column{40E}{@{}l@{}}%
\column{43}{@{}>{\hspre}l<{\hspost}@{}}%
\column{45}{@{}>{\hspre}l<{\hspost}@{}}%
\column{68}{@{}>{\hspre}l<{\hspost}@{}}%
\column{70}{@{}>{\hspre}l<{\hspost}@{}}%
\column{E}{@{}>{\hspre}l<{\hspost}@{}}%
\>[B]{}\textbf{Lenses and Traversals}{}\<[E]%
\\
\>[B]{}\Varid{view\char95 }{}\<[16]%
\>[16]{}\mathbin{::}\Conid{Lens}\;\Varid{s}\;\Varid{t}\;\Varid{a}\;\Varid{b}\to \Varid{s}\to \Varid{a}{}\<[E]%
\\
\>[B]{}\Varid{update\char95 }{}\<[16]%
\>[16]{}\mathbin{::}\Conid{Lens}\;\Varid{s}\;\Varid{t}\;\Varid{a}\;\Varid{b}\to \Varid{b}\to \Varid{s}\to \Varid{t}{}\<[E]%
\\
\>[B]{}\Varid{modify\char95 }{}\<[16]%
\>[16]{}\mathbin{::}\Conid{Lens}\;\Varid{s}\;\Varid{t}\;\Varid{a}\;\Varid{b}\to (\Varid{a}\to \Varid{b})\to \Varid{s}\to \Varid{t}{}\<[E]%
\\[\blanklineskip]%
\>[B]{}\Varid{toListOf\char95 }{}\<[16]%
\>[16]{}\mathbin{::}\Conid{Traversal}\;\Varid{s}\;\Varid{s}\;\Varid{a}\;\Varid{a}\to \Varid{s}\to [\mskip1.5mu \Varid{a}\mskip1.5mu]{}\<[E]%
\\
\>[B]{}\Varid{over\char95 }{}\<[16]%
\>[16]{}\mathbin{::}\Conid{Traversal}\;\Varid{s}\;\Varid{t}\;\Varid{a}\;\Varid{b}\to (\Varid{a}\to \Varid{b})\to \Varid{s}\to \Varid{t}{}\<[E]%
\\
\>[B]{}\Varid{traverseOf\char95 }{}\<[16]%
\>[16]{}\mathbin{::}\Conid{Traversal}\;\Varid{s}\;\Varid{t}\;\Varid{a}\;\Varid{b}\to (\forall \Varid{g}\hsforall \hsdot{\mathbin{\cdot}}{.}\Conid{Applicative}\;\Varid{g}\Rightarrow (\Varid{a}\to \Varid{g}\;\Varid{b})\to \Varid{s}\to \Varid{g}\;\Varid{t}){}\<[E]%
\\[\blanklineskip]%
\>[B]{}(\mathbin{\circ}){}\<[16]%
\>[16]{}\mathbin{::}o_1, o_2 \in \{ Lens, Traversal \} \Rightarrow \;o_1\;\Varid{s}\;\Varid{t}\;\Varid{c}\;\Varid{d}\;\to \;o_2\;\Varid{c}\;\Varid{d}\;\Varid{a}\;\Varid{b}\;\to \;(o_1\mathop{\vee}o_2)\;\Varid{s}\;\Varid{t}\;\Varid{a}\;\Varid{b}{}\<[E]%
\\[\blanklineskip]%
\>[B]{}\textbf{Generic Lenses}{}\<[E]%
\\
\>[B]{}\Varid{field\char95 }{}\<[16]%
\>[16]{}\mathbin{::}\forall \Varid{name}\hsforall \;\Varid{s}\;\Varid{t}\;\Varid{a}\;\Varid{b}\hsdot{\mathbin{\cdot}}{.}\;{}\<[45]%
\>[45]{}\Conid{HasField}\;\Varid{name}\;\Varid{s}\;\Varid{t}\;\Varid{a}\;\Varid{b}{}\<[70]%
\>[70]{}\Rightarrow \Conid{Lens}\;\Varid{s}\;\Varid{t}\;\Varid{a}\;\Varid{b}{}\<[E]%
\\
\>[B]{}\Varid{typed\char95 }{}\<[16]%
\>[16]{}\mathbin{::}\forall \Varid{a}\hsforall \;\Varid{s}\hsdot{\mathbin{\cdot}}{.}{}\<[40]%
\>[40]{}\;{}\<[40E]%
\>[45]{}\Conid{HasType}\;\Varid{a}\;\Varid{s}{}\<[70]%
\>[70]{}\Rightarrow \Conid{Lens}\;\Varid{s}\;\Varid{s}\;\Varid{a}\;\Varid{a}{}\<[E]%
\\
\>[B]{}\Varid{position\char95 }{}\<[16]%
\>[16]{}\mathbin{::}\forall \Varid{pos}\hsforall \;\Varid{s}\;\Varid{t}\;\Varid{a}\;\Varid{b}\hsdot{\mathbin{\cdot}}{.}{}\<[40]%
\>[40]{}\;{}\<[40E]%
\>[45]{}\Conid{HasPosition}\;\Varid{pos}\;\Varid{s}\;\Varid{t}\;\Varid{a}\;\Varid{b}{}\<[70]%
\>[70]{}\Rightarrow \Conid{Lens}\;\Varid{s}\;\Varid{t}\;\Varid{a}\;\Varid{b}{}\<[E]%
\\
\>[B]{}\Varid{super\char95 }{}\<[16]%
\>[16]{}\mathbin{::}\forall \Varid{sup}\hsforall \;\Varid{sub}\hsdot{\mathbin{\cdot}}{.}{}\<[40]%
\>[40]{}\;{}\<[40E]%
\>[45]{}\Conid{Subtype}\;\Varid{sup}\;\Varid{sub}{}\<[70]%
\>[70]{}\Rightarrow \Conid{Lens}\;\Varid{sub}\;\Varid{sub}\;\Varid{sup}\;\Varid{sup}{}\<[E]%
\\[\blanklineskip]%
\>[B]{}\textbf{Generic Traversals}{}\<[E]%
\\
\>[B]{}\Varid{types}\_{}\<[16]%
\>[16]{}\mathbin{::}\forall \Varid{a}\hsforall \;\Varid{s}\hsdot{\mathbin{\cdot}}{.}{}\<[40]%
\>[40]{}\;{}\<[40E]%
\>[45]{}\Conid{HasTypes}\;\Varid{s}\;\Varid{a}{}\<[68]%
\>[68]{}\Rightarrow \Conid{Traversal}\;\Varid{s}\;\Varid{s}\;\Varid{a}\;\Varid{a}{}\<[E]%
\\
\>[B]{}\Varid{param\char95 }{}\<[16]%
\>[16]{}\mathbin{::}\forall \Varid{pos}\hsforall \;\Varid{s}\;\Varid{t}\;\Varid{a}\;\Varid{b}\hsdot{\mathbin{\cdot}}{.}{}\<[40]%
\>[40]{}\;{}\<[40E]%
\>[45]{}\Conid{HasParam}\;\Varid{pos}\;\Varid{s}\;\Varid{t}\;\Varid{a}\;\Varid{b}{}\<[68]%
\>[68]{}\Rightarrow \Conid{Traversal}\;\Varid{s}\;\Varid{t}\;\Varid{a}\;\Varid{b}{}\<[E]%
\\
\>[B]{}\Varid{constraints\char95 }{}\<[16]%
\>[16]{}\mathbin{::}\forall \Varid{c}\hsforall \;\Varid{s}\;\Varid{t}\hsdot{\mathbin{\cdot}}{.}{}\<[40]%
\>[40]{}\;{}\<[40E]%
\>[45]{}\Conid{HasConstraints}\;\Varid{c}\;\Varid{s}\;\Varid{t}{}\<[68]%
\>[68]{}\Rightarrow {}\<[E]%
\\
\>[40]{}\hsindent{3}{}\<[43]%
\>[43]{}\;(\forall \Varid{g}\hsforall \hsdot{\mathbin{\cdot}}{.}\Conid{Applicative}\;\Varid{g}\Rightarrow (\forall \Varid{a}\hsforall \;\Varid{b}\hsdot{\mathbin{\cdot}}{.}\Varid{c}\;\Varid{a}\;\Varid{b}\Rightarrow \Varid{a}\to \Varid{g}\;\Varid{b})\to \Varid{s}\to \Varid{g}\;\Varid{t}){}\<[E]%
\ColumnHook
\end{hscode}\resethooks
\caption{The generic lens and traversal interface to \text{\ttfamily generic\char45{}lens}}
\label{fig:interface}
\end{figure}

Suppose you are running a biscuit distribution company. You have customers who
place orders for biscuits which you need to keep track of and process. In addition,
you allow customers to prioritise their biscuit orders, which are
then distributed from an entirely separate distribution facility.

To this end, we implement \ensuremath{\Conid{Item}}, a data type to represent a
type of biscuit, and \ensuremath{\Conid{Invoice}}, a data type to represent a single order. It is
parameterised by the type of priority we assign to the orders. Finally,
\ensuremath{\Conid{Orders}}, a top-level data structure which contains the normal and
priority queues. The priority queue has an augmented priority field
that keeps track of the priority level.
\begin{hscode}\SaveRestoreHook
\column{B}{@{}>{\hspre}l<{\hspost}@{}}%
\column{18}{@{}>{\hspre}l<{\hspost}@{}}%
\column{E}{@{}>{\hspre}l<{\hspost}@{}}%
\>[B]{}\keyword{data}\;\Conid{Item}\mathrel{=}\Conid{Item}\;\{\mskip1.5mu \Varid{name}\mathbin{::}\Conid{String},\Varid{cost}\mathbin{::}\Conid{Cost}\mskip1.5mu\}{}\<[E]%
\\
\>[B]{}\keyword{newtype}\;\Conid{Cost}\mathrel{=}\Conid{Cost}\;\Conid{Double}{}\<[E]%
\\[\blanklineskip]%
\>[B]{}\keyword{data}\;\Conid{Invoice}\;\Varid{p}\mathrel{=}\Conid{Invoice}\;\{\mskip1.5mu \Varid{item}\mathbin{::}\Conid{Item},\Varid{name}\mathbin{::}\Conid{String},\Varid{number}\mathbin{::}\Conid{Int},\Varid{priority}\mathbin{::}\Varid{p}\mskip1.5mu\}{}\<[E]%
\\[\blanklineskip]%
\>[B]{}\keyword{data}\;\Conid{Orders}\mathrel{=}\Conid{Orders}\;[\mskip1.5mu \Conid{Invoice}\;\Conid{Int}\mskip1.5mu]\;[\mskip1.5mu \Conid{Invoice}\;(\Conid{Int},\Conid{Double})\mskip1.5mu]{}\<[E]%
\\[\blanklineskip]%
\>[B]{}\Varid{bourbon}\mathbin{::}\Conid{Item}{}\<[E]%
\\
\>[B]{}\Varid{bourbon}\mathrel{=}\Conid{Item}\;\text{\ttfamily \char34 Bourbon\char34}\;(\Conid{Cost}\;\mathrm{100}){}\<[E]%
\\[\blanklineskip]%
\>[B]{}\Varid{orders}\mathrel{=}\Conid{Orders}\;{}\<[18]%
\>[18]{}[\mskip1.5mu \Conid{Invoice}\;\Varid{bourbon}\;\text{\ttfamily \char34 Earl\char34}\;\mathrm{1}\;\mathrm{0},\Conid{Invoice}\;\Varid{bourbon}\;\text{\ttfamily \char34 Johnny\char34}\;\mathrm{2}\;\mathrm{2}\mskip1.5mu]\;{}\<[E]%
\\
\>[18]{}[\mskip1.5mu \Conid{Invoice}\;\Varid{bourbon}\;\text{\ttfamily \char34 George\char34}\;\mathrm{2}\;(\mathrm{0},\mathrm{3})\mskip1.5mu]{}\<[E]%
\ColumnHook
\end{hscode}\resethooks
We will now use our generic queries to interrogate specific aspects of this
data structure.
We will first give the specification by example before an in-depth
 explanation in the next section. The interface of our library is
 summarised in Figure~\ref{fig:interface}.

Starting from the simplest example, \ensuremath{\Varid{field}} derives a \ensuremath{\Conid{Lens}} which focuses on
a named field in a data type. We can use \ensuremath{\Varid{field}\ \texttt{@}\text{\ttfamily \char34 name\char34}} to focus on the
\ensuremath{\text{\ttfamily \char34 name\char34}} field of an \ensuremath{\Conid{Item}}. Here, we use visible type
application~\cite{Eisenberg:2016:visible} to supply the static argument
\ensuremath{\text{\ttfamily \char34 name\char34}} to \ensuremath{\Varid{field}}. Once we have focused, we can update or view this field.

\begin{hscode}\SaveRestoreHook
\column{B}{@{}>{\hspre}l<{\hspost}@{}}%
\column{E}{@{}>{\hspre}l<{\hspost}@{}}%
\>[B]{}\texttt{> }\texttt{ghci> }\;\Varid{view}\;(\Varid{field}\ \texttt{@}\text{\ttfamily \char34 name\char34})\;\Varid{bourbon}{}\<[E]%
\\
\>[B]{}\texttt{> }\text{\ttfamily \char34 Bourbon\char34}{}\<[E]%
\\
\>[B]{}\texttt{> }\texttt{ghci> }\;\Varid{update}\;(\Varid{field}\ \texttt{@}\text{\ttfamily \char34 cost\char34})\;(\Conid{Cost}\;\mathrm{110})\;\Varid{bourbon}{}\<[E]%
\\
\>[B]{}\texttt{> }\Conid{Item}\;\text{\ttfamily \char34 Bourbon\char34}\;(\Conid{Cost}\;\mathrm{110}){}\<[E]%
\ColumnHook
\end{hscode}\resethooks
Why do we not use the in-built record selector? For it is not \emph{compositional}.
We can compose lenses together using the composition operator \ensuremath{\mathbin{\circ}} in order to
inspect nested fields. For example, if we want to find the name of an item we
would compose the two \ensuremath{\Varid{field}} lenses like so:
\begin{hscode}\SaveRestoreHook
\column{B}{@{}>{\hspre}l<{\hspost}@{}}%
\column{E}{@{}>{\hspre}l<{\hspost}@{}}%
\>[B]{}\Varid{nameOfItem}\mathbin{::}\Conid{Invoice}\;\Varid{p}\to \Conid{String}{}\<[E]%
\\
\>[B]{}\Varid{nameOfItem}\mathrel{=}\Varid{view}\;(\Varid{field}\ \texttt{@}\text{\ttfamily \char34 item\char34}\mathbin{\circ}\Varid{field}\ \texttt{@}\text{\ttfamily \char34 name\char34}){}\<[E]%
\ColumnHook
\end{hscode}\resethooks
These lenses are read left-to-right: first we apply a lens that finds
the field called \ensuremath{\text{\ttfamily \char34 item\char34}} and then we apply a lens that finds the
field \ensuremath{\text{\ttfamily \char34 name\char34}}.

This is all well and good if we just have nested products but no good at all
for modifying many parts of a data structure at once. As a special thank you to
our customers, we wish to decrease the cost of all invoices.
To do this
we use the \ensuremath{\Varid{types}} traversal which creates a \ensuremath{\Conid{Traversal}} that focuses on
every part of a data structure with a specific type.
\begin{hscode}\SaveRestoreHook
\column{B}{@{}>{\hspre}l<{\hspost}@{}}%
\column{E}{@{}>{\hspre}l<{\hspost}@{}}%
\>[B]{}\Varid{thankYou}\mathbin{::}\Conid{Orders}\to \Conid{Orders}{}\<[E]%
\\
\>[B]{}\Varid{thankYou}\mathrel{=}\Varid{over}\;(\Varid{types}\ \texttt{@}\Conid{Cost})\;(\lambda (\Conid{Cost}\;\Varid{c})\to \Conid{Cost}\;(\Varid{c}\times\mathrm{0.85})){}\<[E]%
\ColumnHook
\end{hscode}\resethooks

Later we realise that we only really want to thank our priority customers.
In order to do this we first need to restrict our focus to
the priority queue but then can reuse the previous incantation.
The \ensuremath{\Varid{position}} lens selects the $k$th field of a data type by its position
in the data declaration.

\begin{hscode}\SaveRestoreHook
\column{B}{@{}>{\hspre}l<{\hspost}@{}}%
\column{E}{@{}>{\hspre}l<{\hspost}@{}}%
\>[B]{}\Varid{thankYouPriority}\mathbin{::}\Conid{Orders}\to \Conid{Orders}{}\<[E]%
\\
\>[B]{}\Varid{thankYouPriority}\mathrel{=}\Varid{over}\;(\Varid{position}\ \texttt{@}\mathrm{2}\mathbin{\circ}\Varid{types}\ \texttt{@}\Conid{Cost})\;(\lambda (\Conid{Cost}\;\Varid{c})\to \Conid{Cost}\;(\Varid{c}\times\mathrm{0.85})){}\<[E]%
\ColumnHook
\end{hscode}\resethooks
We have composed together a lens and a traversal to get a traversal. A lens is a special
case of a traversal that also allows us to extract a value by
focusing on one item.

Finally, like any good business, we give our customers the choice and opportunity
to upgrade their standard orders to premium orders. In order to do so, we have
to modify an \ensuremath{\Conid{Invoice}\;\Conid{Int}} into an \ensuremath{\Conid{Invoice}\;(\Conid{Int},\Conid{Double})}. We use the \ensuremath{\Varid{param}}
traversal in order to modify the $0$th type parameter from the right of \ensuremath{\Conid{Invoice}} from an \ensuremath{\Conid{Int}} to a \ensuremath{(\Conid{Int},\Conid{Double})}.
\begin{hscode}\SaveRestoreHook
\column{B}{@{}>{\hspre}l<{\hspost}@{}}%
\column{E}{@{}>{\hspre}l<{\hspost}@{}}%
\>[B]{}\Varid{upgrade}\mathbin{::}\Conid{Double}\to \Conid{Invoice}\;\Conid{Int}\to \Conid{Invoice}\;(\Conid{Int},\Conid{Double}){}\<[E]%
\\
\>[B]{}\Varid{upgrade}\;\Varid{bribe}\;\Varid{invoice}\mathrel{=}\Varid{over}\;(\Varid{param}\ \texttt{@}\mathrm{0})\;(\lambda \Varid{i}\to (\Varid{i},\Varid{bribe}))\;\Varid{invoice}{}\<[E]%
\ColumnHook
\end{hscode}\resethooks
The above example highlights how our traversal can change the type of its argument.
Traditionally type changing is difficult to implement in generic traversal frameworks
such as SYB.

At the end of the year, our auditors want to see a summary of all the items we have sold this year.
They don't care whether they were priority orders or not. We just need to extract
all the \ensuremath{\Conid{Item}}s we have sold. We can use the \ensuremath{\Varid{types}} traversal to focus on all
\ensuremath{\Conid{Item}}s in the tree and extract them.
\begin{hscode}\SaveRestoreHook
\column{B}{@{}>{\hspre}l<{\hspost}@{}}%
\column{E}{@{}>{\hspre}l<{\hspost}@{}}%
\>[B]{}\Varid{audit}\mathbin{::}\Conid{Orders}\to [\mskip1.5mu \Conid{Item}\mskip1.5mu]{}\<[E]%
\\
\>[B]{}\Varid{audit}\mathrel{=}\Varid{toListOf}\;(\Varid{types}\ \texttt{@}\Conid{Item}){}\<[E]%
\ColumnHook
\end{hscode}\resethooks
The \ensuremath{\Varid{toListOf}} combinator
summarises a \ensuremath{\Conid{Traversal}} by returning all the parts it focuses on.

We have seen examples of how we can concisely traverse, modify, inspect, and
analyse our biscuit pipeline.
This was made possible by the use of lenses and traversals
that are generically derived from the data types involved.
In the next section we describe the generic combinators that we have used
in these examples.

\section{Interface}
\label{sec:interface}

In this section we discuss ways of identifying certain parts of algebraic data types
using a type-directed approach. These can be classified into the following
three categories, based on the underlying structure of a data type.
\begin{description}
\item[Lenses]: Patterns applicable to data types made from \emph{products}
\item[Prisms]: Patterns applicable to data types made from \emph{sums}
\item[Traversals]: Patterns applicable to data types made from \emph{sums of products}
\end{description}
These abstractions are known together as \emph{optics}.
We will concentrate on lenses and traversals in this paper. %
Prisms follow the same principles so we discuss them only briefly.

\subsection{Lenses}

\nw{Not clear how much of this is original to us, and how much is
established}
A lens focuses on one part of a product. The focus can then be viewed
and updated whilst the rest of the structure remains unchanged.
A lens \ensuremath{\Varid{l}\mathbin{::}\Conid{Lens}\;\Varid{s}\;\Varid{t}\;\Varid{a}\;\Varid{b}} can be read as saying that \ensuremath{\Varid{l}} is a lens
whose source is of type \ensuremath{\Varid{s}}, its focus is on a value of type \ensuremath{\Varid{a}} which
when changed to a value of type \ensuremath{\Varid{b}} replaces the value of type \ensuremath{\Varid{a}}
in \ensuremath{\Varid{s}} and produces a product of type \ensuremath{\Varid{t}}.

There are three three primitive ways that we can use a lens.
\begin{hscode}\SaveRestoreHook
\column{B}{@{}>{\hspre}l<{\hspost}@{}}%
\column{9}{@{}>{\hspre}l<{\hspost}@{}}%
\column{E}{@{}>{\hspre}l<{\hspost}@{}}%
\>[B]{}\Varid{view}{}\<[9]%
\>[9]{}\mathbin{::}\Conid{Lens}\;\Varid{s}\;\Varid{t}\;\Varid{a}\;\Varid{b}\to \Varid{s}\to \Varid{a}{}\<[E]%
\\
\>[B]{}\Varid{update}{}\<[9]%
\>[9]{}\mathbin{::}\Conid{Lens}\;\Varid{s}\;\Varid{t}\;\Varid{a}\;\Varid{b}\to \Varid{b}\to \Varid{s}\to \Varid{t}{}\<[E]%
\\
\>[B]{}\Varid{modify}{}\<[9]%
\>[9]{}\mathbin{::}\Conid{Lens}\;\Varid{s}\;\Varid{t}\;\Varid{a}\;\Varid{b}\to (\Varid{a}\to \Varid{b})\to \Varid{s}\to \Varid{t}{}\<[E]%
\ColumnHook
\end{hscode}\resethooks
Lenses with this interface are already well established~\cite{Pickering:2017:optics}.
The \ensuremath{\Varid{view}} operation extracts a component from its context.
The \ensuremath{\Varid{update}} operation updates a structure. We include \ensuremath{\Varid{modify}} as a
means of efficiently viewing and updating a structure in a single
step.

Our contribution is to derive a number of generic lenses. We first
consider the different ways we can specify how to access different
parts of a product data type. As such, we derive various \ensuremath{\Conid{Lens}}es
that focus on precisely one part of a product.

\subsubsection{By name}

For a data type with named fields we can specify the lens that focuses
on a field with a certain name. As each field must have a unique name,
this provides a way of specifying a unique field in a larger product.
We define a combinator, named \ensuremath{\Varid{field}} that provides this lens for all
suitable types.
\begin{hscode}\SaveRestoreHook
\column{B}{@{}>{\hspre}l<{\hspost}@{}}%
\column{E}{@{}>{\hspre}l<{\hspost}@{}}%
\>[B]{}\Varid{field}\mathbin{::}\Conid{HasField}\;\Varid{name}\;\Varid{s}\;\Varid{t}\;\Varid{a}\;\Varid{b}\Rightarrow \Conid{Lens}\;\Varid{s}\;\Varid{t}\;\Varid{a}\;\Varid{b}{}\<[E]%
\ColumnHook
\end{hscode}\resethooks
\ensuremath{\Conid{HasField}\;\Varid{name}\;\Varid{s}\;\Varid{t}\;\Varid{a}\;\Varid{b}} instances are derived generically. The constraint means
that the type \ensuremath{\Varid{s}} has a field called \ensuremath{\Varid{name}} of type \ensuremath{\Varid{a}}, and if we change the
field from \ensuremath{\Varid{a}} to \ensuremath{\Varid{b}}, we obtain a structure of type \ensuremath{\Varid{t}}. To illustrate this,
let us change the \ensuremath{\Varid{cost}} field of an \ensuremath{\Conid{Item}}:
\begin{hscode}\SaveRestoreHook
\column{B}{@{}>{\hspre}l<{\hspost}@{}}%
\column{E}{@{}>{\hspre}l<{\hspost}@{}}%
\>[B]{}\texttt{> }\texttt{ghci> }\;\Varid{modify}\;(\Varid{field}\ \texttt{@}\text{\ttfamily \char34 cost\char34})\;(\lambda (\Conid{Cost}\;\Varid{c})\to (\Conid{Cost}\;(\Varid{c}\mathbin{+}\mathrm{5})))\;\Varid{bourbon}{}\<[E]%
\\
\>[B]{}\texttt{> }\Conid{Item}\;\text{\ttfamily \char34 Bourbon\char34}\;(\Conid{Cost}\;\mathrm{105}){}\<[E]%
\ColumnHook
\end{hscode}\resethooks
In this case, the type of the derived instance is the following:
\begin{hscode}\SaveRestoreHook
\column{B}{@{}>{\hspre}l<{\hspost}@{}}%
\column{E}{@{}>{\hspre}l<{\hspost}@{}}%
\>[B]{}\Varid{field}\ \texttt{@}\text{\ttfamily \char34 cost\char34}\ \texttt{@}\Conid{Item}\mathbin{::}\Conid{Lens}\;\Conid{Item}\;\Conid{Item}\;\Conid{Cost}\;\Conid{Cost}{}\<[E]%
\ColumnHook
\end{hscode}\resethooks
That is, the type of the \ensuremath{\Conid{Cost}} field can not be changed within \ensuremath{\Conid{Item}} -- of
course, as it is fixed to be \ensuremath{\Conid{Cost}} in \ensuremath{\Conid{Item}}'s definition. Contrast this with \ensuremath{\Conid{Invoice}\;\Varid{p}}:
it is parameterised by a type variable. \ensuremath{\Conid{Invoice}\;\Varid{a}} can be changed to \ensuremath{\Conid{Invoice}\;\Varid{b}}, as long as we can
change the inner \ensuremath{\Varid{a}} into \ensuremath{\Varid{b}}.
\begin{hscode}\SaveRestoreHook
\column{B}{@{}>{\hspre}l<{\hspost}@{}}%
\column{E}{@{}>{\hspre}l<{\hspost}@{}}%
\>[B]{}\Varid{field}\ \texttt{@}\text{\ttfamily \char34 priority\char34}\ \texttt{@}(\Conid{Invoice}\;\anonymous )\mathbin{::}\Conid{Lens}\;(\Conid{Invoice}\;\Varid{a})\;(\Conid{Invoice}\;\Varid{b})\;\Varid{a}\;\Varid{b}{}\<[E]%
\ColumnHook
\end{hscode}\resethooks
Accordingly, this version of the lens readily allows us to carry out
type-changing manipulations:
\begin{hscode}\SaveRestoreHook
\column{B}{@{}>{\hspre}l<{\hspost}@{}}%
\column{E}{@{}>{\hspre}l<{\hspost}@{}}%
\>[B]{}\texttt{> }\texttt{ghci> }\;\Varid{modify}\;(\Varid{field}\ \texttt{@}\text{\ttfamily \char34 priority\char34})\;(\lambda \Varid{i}\to (\Varid{i},\mathrm{0}))\;(\Conid{Invoice}\;\Varid{bourbon}\;\text{\ttfamily \char34 Johnny\char34}\;\mathrm{2}\;\mathrm{2}){}\<[E]%
\\
\>[B]{}\texttt{> }\Conid{Invoice}\;\Varid{bourbon}\;\text{\ttfamily \char34 Johnny\char34}\;\mathrm{2}\;(\mathrm{2},\mathrm{0}){}\<[E]%
\ColumnHook
\end{hscode}\resethooks
We stated that a lens focuses on exactly one part of a structure, meaning in
this case that it must contain the field we are interested in. But what happens
when we ask for a field that does not exist? We throw a type error!
\begin{hscode}\SaveRestoreHook
\column{B}{@{}>{\hspre}l<{\hspost}@{}}%
\column{E}{@{}>{\hspre}l<{\hspost}@{}}%
\>[B]{}\texttt{> }\texttt{ghci> }\;\Varid{view}\;(\Varid{field}\ \texttt{@}\text{\ttfamily \char34 weight\char34})\;\Varid{bourbon}{}\<[E]%
\\
\>[B]{}\texttt{> }\texttt{error: * The type Item does not contain a field named weight.}{}\<[E]%
\ColumnHook
\end{hscode}\resethooks
That is, we can \textit{statically} determine whether a field exists, or not.
Notice that the error message generated by our library is informative,
and hides away the underlying complexities.
Our approach not only provides a pleasant user experience, but also obviates the
need for any dynamic checks.

\subsubsection{By type}

Often it is burdensome to access the fields by name, as it can change over time.
In many cases we do not care how exactly the subpart can be located, as long as
it is uniquely identified by its type. The \ensuremath{\Varid{typed}} lens focuses on the unique
type in a product.
\begin{hscode}\SaveRestoreHook
\column{B}{@{}>{\hspre}l<{\hspost}@{}}%
\column{E}{@{}>{\hspre}l<{\hspost}@{}}%
\>[B]{}\Varid{typed}\mathbin{::}\Conid{HasType}\;\Varid{a}\;\Varid{s}\Rightarrow \Conid{Lens}\;\Varid{s}\;\Varid{s}\;\Varid{a}\;\Varid{a}{}\<[E]%
\ColumnHook
\end{hscode}\resethooks
For example, \ensuremath{\Conid{Item}} has one \ensuremath{\Conid{Cost}} field so we can use the \ensuremath{\Varid{typed}} lens to
update and modify it.
\begin{hscode}\SaveRestoreHook
\column{B}{@{}>{\hspre}l<{\hspost}@{}}%
\column{E}{@{}>{\hspre}l<{\hspost}@{}}%
\>[B]{}\texttt{> }\texttt{ghci> }\;\Varid{update}\;(\Varid{typed}\ \texttt{@}\Conid{Cost})\;(\Conid{Cost}\;\mathrm{200})\;\Varid{bourbon}{}\<[E]%
\\
\>[B]{}\texttt{> }\Conid{Item}\;\text{\ttfamily \char34 Bourbon\char34}\;(\Conid{Cost}\;\mathrm{200}){}\<[E]%
\ColumnHook
\end{hscode}\resethooks
Often in practice, the type of interest can even be inferred from the
context, and we do not need to specify explicitly. For example:
\begin{hscode}\SaveRestoreHook
\column{B}{@{}>{\hspre}l<{\hspost}@{}}%
\column{E}{@{}>{\hspre}l<{\hspost}@{}}%
\>[B]{}\texttt{> }\texttt{ghci> }\;\Varid{modify}\;\Varid{typed}\;(\text{\ttfamily \char34 Chocolate~\char34}\mathbin{++})\;\Varid{bourbon}{}\<[E]%
\\
\>[B]{}\texttt{> }\Conid{Item}\;\text{\ttfamily \char34 Chocolate~Bourbon\char34}\;(\Conid{Cost}\;\mathrm{100}){}\<[E]%
\ColumnHook
\end{hscode}\resethooks
It is clear that appending text requires a \ensuremath{\Conid{String}}, and therefore \ensuremath{\Varid{typed}}
knows which field to select.
As expected, requesting a lens for a type not contained in the
product yields a type error.

The \ensuremath{\Varid{typed}} lens is monomorphic as it is complicated to
specify precisely when it is safe to change the type. For instance, we could
``lose'' target if we changed the field's type to something already present in
the structure, as this type would no longer be uniquely identifiable.
These complications do not arise for the \ensuremath{\Varid{field}} and \ensuremath{\Varid{position}} lenses.

\subsubsection{By position}

Not all product types have named fields. For example, consider \ensuremath{\Conid{Orders}}:
it contains no named fields but we might still want to restrict our attention to
either the first or second order queue.
In this case, we can refer to the fields positionally.
\begin{hscode}\SaveRestoreHook
\column{B}{@{}>{\hspre}l<{\hspost}@{}}%
\column{E}{@{}>{\hspre}l<{\hspost}@{}}%
\>[B]{}\Varid{position}\mathbin{::}\Conid{HasPosition}\;\Varid{pos}\;\Varid{s}\;\Varid{t}\;\Varid{a}\;\Varid{b}\Rightarrow \Conid{Lens}\;\Varid{s}\;\Varid{t}\;\Varid{a}\;\Varid{b}{}\<[E]%
\ColumnHook
\end{hscode}\resethooks
Indexed from 1, the \ensuremath{\Varid{position}} lens focuses on the $k$th field in a product.
\ensuremath{\Conid{HasPosition}} instances are derived generically for all product types.
\begin{hscode}\SaveRestoreHook
\column{B}{@{}>{\hspre}l<{\hspost}@{}}%
\column{E}{@{}>{\hspre}l<{\hspost}@{}}%
\>[B]{}\texttt{> }\texttt{ghci> }\;\Varid{view}\;(\Varid{position}\ \texttt{@}\mathrm{2})\;\Varid{orders}{}\<[E]%
\\
\>[B]{}\texttt{> }[\mskip1.5mu \Conid{Invoice}\;\Varid{bourbon}\;\text{\ttfamily \char34 George\char34}\;\mathrm{2}\;(\mathrm{0},\mathrm{3})\mskip1.5mu]{}\<[E]%
\ColumnHook
\end{hscode}\resethooks
Trying to access an ``out of bounds'' element results in a type error:
\begin{hscode}\SaveRestoreHook
\column{B}{@{}>{\hspre}l<{\hspost}@{}}%
\column{E}{@{}>{\hspre}l<{\hspost}@{}}%
\>[B]{}\texttt{> }\texttt{ghci> }\;\Varid{view}\;(\Varid{position}\ \texttt{@}\mathrm{3})\;\Varid{orders}{}\<[E]%
\\
\>[B]{}\texttt{> }\texttt{error: * The type Orders does not contain a field at position 3}{}\<[E]%
\ColumnHook
\end{hscode}\resethooks
This lens works equally well for any product data type including in-built types
such as tuples.

\subsubsection{By structure}
Finally, the \ensuremath{\Varid{super}} lens generalises the \ensuremath{\Varid{field}} lens to focus on a collection
of fields rather than just one. The \ensuremath{\Conid{Subtype}\;\Varid{sup}\;\Varid{sub}} constraint holds
if the data type \ensuremath{\Varid{sub}} contains all the fields labels (with the same types)
as \ensuremath{\Varid{sup}} contains.
\begin{hscode}\SaveRestoreHook
\column{B}{@{}>{\hspre}l<{\hspost}@{}}%
\column{E}{@{}>{\hspre}l<{\hspost}@{}}%
\>[B]{}\Varid{super}\mathbin{::}\Conid{Subtype}\;\Varid{sup}\;\Varid{sub}\Rightarrow \Conid{Lens}\;\Varid{sub}\;\Varid{sub}\;\Varid{sup}\;\Varid{sup}{}\<[E]%
\ColumnHook
\end{hscode}\resethooks
Consider a new data type \ensuremath{\Conid{WeighedItem}} which adds a
new weight field to \ensuremath{\Conid{Item}} so we can compute the postage for our orders.
The \ensuremath{\Varid{super}} lens will be used to extract a value of type
\ensuremath{\Conid{Item}} from \ensuremath{\Conid{WeighedItem}}.
As such, \ensuremath{\Conid{WeighedItem}} is a subtype of \ensuremath{\Conid{Item}} as it contains all the fields which \ensuremath{\Conid{Item}}
contains. Thus \ensuremath{\Varid{super}\ \texttt{@}\Conid{Item}\ \texttt{@}\Conid{WeighedItem}\mathbin{::}\Conid{Lens}\;\Conid{WeighedItem}\;\Conid{WeighedItem}\;\Conid{Item}\;\Conid{Item}}.

\begin{hscode}\SaveRestoreHook
\column{B}{@{}>{\hspre}l<{\hspost}@{}}%
\column{E}{@{}>{\hspre}l<{\hspost}@{}}%
\>[B]{}\keyword{newtype}\;\Conid{Weight}\mathrel{=}\Conid{Weight}\;\Conid{Double}{}\<[E]%
\\
\>[B]{}\keyword{data}\;\Conid{WeighedItem}\mathrel{=}\Conid{WItem}\;\{\mskip1.5mu \Varid{name}\mathbin{::}\Conid{String},\Varid{cost}\mathbin{::}\Conid{Cost},\Varid{weight}\mathbin{::}\Conid{Weight}\mskip1.5mu\}{}\<[E]%
\ColumnHook
\end{hscode}\resethooks
We can then use \ensuremath{\Varid{super}} to modify several fields at once.
\begin{hscode}\SaveRestoreHook
\column{B}{@{}>{\hspre}l<{\hspost}@{}}%
\column{E}{@{}>{\hspre}l<{\hspost}@{}}%
\>[B]{}\texttt{> }\texttt{ghci> }\;\Varid{view}\;(\Varid{super}\ \texttt{@}\Conid{Item})\;(\Conid{WItem}\;\text{\ttfamily \char34 Bourbon\char34}\;(\Conid{Cost}\;\mathrm{2000})\;(\Conid{Weight}\;\mathrm{0.03})){}\<[E]%
\\
\>[B]{}\texttt{> }\Conid{Item}\;\text{\ttfamily \char34 Bourbon\char34}\;(\Conid{Cost}\;\mathrm{2000}){}\<[E]%
\\
\>[B]{}\texttt{> }\texttt{ghci> }\;\Varid{update}\;(\Varid{super}\ \texttt{@}\Conid{Item})\;\Varid{bourbon}\;(\Conid{WItem}\;\text{\ttfamily \char34 Bourbon+\char34}\;(\Conid{Cost}\;\mathrm{500})\;(\Conid{Weight}\;\mathrm{0.03})){}\<[E]%
\\
\>[B]{}\texttt{> }\Conid{WItem}\;\text{\ttfamily \char34 Bourbon\char34}\;(\Conid{Cost}\;\mathrm{100})\;(\Conid{Weight}\;\mathrm{0.03}){}\<[E]%
\ColumnHook
\end{hscode}\resethooks
This kind of lens is particularly useful in data processing pipelines where
additional steps add computed fields to a data type.

\subsection{Prisms}

\emph{Prisms} are the dual to lenses: while a lens focuses on one part of a
product, a prism focuses on one part of a sum. As such, the focused value might not
be present. Prisms can be used in the other way; they can construct the sum by
injecting in the focused part.
We derive prisms for data types which are made from sums.
A prism \ensuremath{\Varid{p}\mathbin{::}\Conid{Prism}\;\Varid{s}\;\Varid{t}\;\Varid{a}\;\Varid{b}} consumes values of type \ensuremath{\Varid{s}} and supposing
we can turn an \ensuremath{\Varid{a}} into a \ensuremath{\Varid{b}} produces values of type \ensuremath{\Varid{t}}.
\begin{hscode}\SaveRestoreHook
\column{B}{@{}>{\hspre}l<{\hspost}@{}}%
\column{E}{@{}>{\hspre}l<{\hspost}@{}}%
\>[B]{}\Varid{match}\mathbin{::}\Conid{Prism}\;\Varid{s}\;\Varid{t}\;\Varid{a}\;\Varid{b}\to \Varid{s}\to \Conid{Either}\;\Varid{t}\;\Varid{a}{}\<[E]%
\\
\>[B]{}\Varid{build}\mathbin{::}\Conid{Prism}\;\Varid{s}\;\Varid{t}\;\Varid{a}\;\Varid{b}\to \Varid{b}\to \Varid{t}{}\<[E]%
\ColumnHook
\end{hscode}\resethooks
Since prisms behave similarly to lenses, we describe them only briefly
 in this section to give an intuition of their use and focus on lenses
 and traversals in the remainder of the paper.

Consider a simple sum type \ensuremath{\Conid{D}} that makes use of the sum of
constructors:
\begin{hscode}\SaveRestoreHook
\column{B}{@{}>{\hspre}l<{\hspost}@{}}%
\column{E}{@{}>{\hspre}l<{\hspost}@{}}%
\>[B]{}\keyword{data}\;\Conid{D}\mathrel{=}\Conid{DInt}\;\Conid{Int}\mid \Conid{DPair}\;\Conid{Bool}\;\Conid{String}{}\<[E]%
\ColumnHook
\end{hscode}\resethooks
As with lenses we provide three different ways of deriving prisms for
sum types. 
\subsubsection{By name}
The \ensuremath{\_\Conid{Ctor}} prism selects a constructor by its name.
\begin{hscode}\SaveRestoreHook
\column{B}{@{}>{\hspre}l<{\hspost}@{}}%
\column{E}{@{}>{\hspre}l<{\hspost}@{}}%
\>[B]{}\_\Conid{Ctor}\mathbin{::}\Conid{AsConstructor}\;\Varid{name}\;\Varid{s}\;\Varid{t}\;\Varid{a}\;\Varid{b}\Rightarrow \Conid{Prism}\;\Varid{s}\;\Varid{t}\;\Varid{a}\;\Varid{b}{}\<[E]%
\ColumnHook
\end{hscode}\resethooks
\begin{hscode}\SaveRestoreHook
\column{B}{@{}>{\hspre}l<{\hspost}@{}}%
\column{E}{@{}>{\hspre}l<{\hspost}@{}}%
\>[B]{}\texttt{> }\texttt{ghci> }\;\Varid{match}\;(\_\Conid{Ctor}\ \texttt{@}\text{\ttfamily \char34 DInt\char34})\;(\Conid{DInt}\;\mathrm{1}){}\<[E]%
\\
\>[B]{}\texttt{> }\Conid{Right}\;\mathrm{1}{}\<[E]%
\ColumnHook
\end{hscode}\resethooks
\subsubsection{By type}
The \ensuremath{\_\Conid{Typed}} prism selects a constructor by the type inside the constructor.
Constructors that contain multiple values are viewed as a tuple of those
values.
\begin{hscode}\SaveRestoreHook
\column{B}{@{}>{\hspre}l<{\hspost}@{}}%
\column{E}{@{}>{\hspre}l<{\hspost}@{}}%
\>[B]{}\_\Conid{Typed}\mathbin{::}\Conid{AsType}\;\Varid{s}\;\Varid{a}\Rightarrow \Conid{Prism}\;\Varid{s}\;\Varid{s}\;\Varid{a}\;\Varid{a}{}\<[E]%
\ColumnHook
\end{hscode}\resethooks
\begin{hscode}\SaveRestoreHook
\column{B}{@{}>{\hspre}l<{\hspost}@{}}%
\column{E}{@{}>{\hspre}l<{\hspost}@{}}%
\>[B]{}\texttt{> }\texttt{ghci> }\;\Varid{build}\;\_\Conid{Typed}\;(\Conid{False},\text{\ttfamily \char34 wurble\char34})\mathbin{::}\Conid{D}{}\<[E]%
\\
\>[B]{}\texttt{> }\Conid{DPair}\;\Conid{False}\;\text{\ttfamily \char34 wurble\char34}{}\<[E]%
\ColumnHook
\end{hscode}\resethooks
\subsubsection{By structure}
The \ensuremath{\_\Conid{Sub}} prism allows a substructure to be injected into a superstructure.
\begin{hscode}\SaveRestoreHook
\column{B}{@{}>{\hspre}l<{\hspost}@{}}%
\column{E}{@{}>{\hspre}l<{\hspost}@{}}%
\>[B]{}\_\Conid{Sub}\mathbin{::}\Conid{AsSubtype}\;\Varid{sub}\;\Varid{sup}\Rightarrow \Conid{Prism}\;\Varid{sup}\;\Varid{sup}\;\Varid{sub}\;\Varid{sub}{}\<[E]%
\ColumnHook
\end{hscode}\resethooks
A sum \ensuremath{\Conid{Sub}} is a subtype of another sum \ensuremath{\Conid{Sup}} if a value of \ensuremath{\Conid{Sub}} can be given
(modulo naming of constructors) whenever a value of Sup is expected.
Consider the data type \ensuremath{\Conid{E}}, a supertype of \ensuremath{\Conid{D}}:
\begin{hscode}\SaveRestoreHook
\column{B}{@{}>{\hspre}l<{\hspost}@{}}%
\column{E}{@{}>{\hspre}l<{\hspost}@{}}%
\>[B]{}\keyword{data}\;\Conid{E}\mathrel{=}\Conid{EInt}\;\Conid{Int}\mid \Conid{EPair}\;\Conid{Bool}\;\Conid{String}\mid \Conid{EChar}\;\Conid{Char}{}\<[E]%
\ColumnHook
\end{hscode}\resethooks
We can then use \ensuremath{\_\Conid{Sub}\ \texttt{@}\Conid{D}} to pattern match on values of \ensuremath{\Conid{E}} as if they were
\ensuremath{\Conid{D}} (in this case a failure as \ensuremath{\Conid{D}} has no corresponding \ensuremath{\Conid{Char}} constructor):
\begin{hscode}\SaveRestoreHook
\column{B}{@{}>{\hspre}l<{\hspost}@{}}%
\column{E}{@{}>{\hspre}l<{\hspost}@{}}%
\>[B]{}\texttt{> }\texttt{ghci> }\;\Varid{match}\;(\_\Conid{Sub}\ \texttt{@}\Conid{D})\;(\Conid{EChar}\;\text{\ttfamily 'a'}){}\<[E]%
\\
\>[B]{}\texttt{> }\Conid{Left}\;(\Conid{EChar}\;\text{\ttfamily 'a'}){}\<[E]%
\ColumnHook
\end{hscode}\resethooks
Or in the other direction, build values of \ensuremath{\Conid{E}} from \ensuremath{\Conid{D}}:
\begin{hscode}\SaveRestoreHook
\column{B}{@{}>{\hspre}l<{\hspost}@{}}%
\column{E}{@{}>{\hspre}l<{\hspost}@{}}%
\>[B]{}\texttt{> }\texttt{ghci> }\;\Varid{build}\;\_\Conid{Sub}\;(\Conid{DInt}\;\mathrm{10})\mathbin{::}\Conid{E}{}\<[E]%
\\
\>[B]{}\texttt{> }\Conid{EInt}\;\mathrm{10}{}\<[E]%
\ColumnHook
\end{hscode}\resethooks
The combination of prisms and lenses make for an extremely powerful
and versatile querying language when combined with traversals, which
we discuss next.

\subsection{Traversals}
For algebraic data types (i.e. those constructed using a combination of sums and products), we derive \emph{traversals}. A traversal written \ensuremath{\Conid{Traversal}\;\Varid{s}\;\Varid{t}\;\Varid{a}\;\Varid{b}} walks over
a value of type \ensuremath{\Varid{s}}, modifying all \ensuremath{\Varid{a}}s into \ensuremath{\Varid{b}}s, resulting in a value
of type \ensuremath{\Varid{t}}. For example, we could imagine a traversal \ensuremath{\Varid{tree}\mathbin{::}\Conid{Traversal}\;(\Conid{Tree}\;\Varid{a})\;(\Conid{Tree}\;\Varid{b})\;\Varid{a}\;\Varid{b}}
that focuses on all the elements in a tree. The most general combinator is
\ensuremath{\Varid{traverseOf}} but we will mostly use the specialisations \ensuremath{\Varid{over}} and \ensuremath{\Varid{toListOf}}
which modify and summarise respectively.

\begin{hscode}\SaveRestoreHook
\column{B}{@{}>{\hspre}l<{\hspost}@{}}%
\column{11}{@{}>{\hspre}l<{\hspost}@{}}%
\column{E}{@{}>{\hspre}l<{\hspost}@{}}%
\>[B]{}\Varid{over}{}\<[11]%
\>[11]{}\mathbin{::}\Conid{Traversal}\;\Varid{s}\;\Varid{t}\;\Varid{a}\;\Varid{b}\to (\Varid{a}\to \Varid{b})\to \Varid{s}\to \Varid{t}{}\<[E]%
\\
\>[B]{}\Varid{toListOf}\mathbin{::}\Conid{Traversal}\;\Varid{s}\;\Varid{s}\;\Varid{a}\;\Varid{a}\to \Varid{s}\to [\mskip1.5mu \Varid{a}\mskip1.5mu]{}\<[E]%
\\
\>[B]{}\Varid{traverseOf}\mathbin{::}\Conid{Traversal}\;\Varid{s}\;\Varid{t}\;\Varid{a}\;\Varid{b}\to (\forall \Varid{g}\hsforall \hsdot{\mathbin{\cdot}}{.}\Conid{Applicative}\;\Varid{g}\Rightarrow (\Varid{a}\to \Varid{g}\;\Varid{b})\to \Varid{s}\to \Varid{g}\;\Varid{t}){}\<[E]%
\ColumnHook
\end{hscode}\resethooks

\noindent
We now describe the different traversals that can be generically derived.
\subsubsection{By type}
The \ensuremath{\Varid{types}} function allows us to traverse all values of a given type in a
data type.
\begin{hscode}\SaveRestoreHook
\column{B}{@{}>{\hspre}l<{\hspost}@{}}%
\column{E}{@{}>{\hspre}l<{\hspost}@{}}%
\>[B]{}\Varid{types}\mathbin{::}\Conid{HasTypes}\;\Varid{s}\;\Varid{a}\Rightarrow \Conid{Traversal}\;\Varid{s}\;\Varid{s}\;\Varid{a}\;\Varid{a}{}\<[E]%
\ColumnHook
\end{hscode}\resethooks
Recalling an example we saw in the previous section, \ensuremath{\Varid{types}\ \texttt{@}\Conid{Cost}}
generates a traversal that considers all values of type \ensuremath{\Conid{Cost}} wherever they
are located in a structure. We can use this to uniformly modify all the costs
in a data structure.
\begin{hscode}\SaveRestoreHook
\column{B}{@{}>{\hspre}l<{\hspost}@{}}%
\column{E}{@{}>{\hspre}l<{\hspost}@{}}%
\>[B]{}\Varid{costInc}\mathbin{::}\Conid{HasTypes}\;\Varid{t}\;\Conid{Cost}\Rightarrow \Varid{t}\to \Varid{t}{}\<[E]%
\\
\>[B]{}\Varid{costInc}\mathrel{=}\Varid{over}\;(\Varid{types}\ \texttt{@}\Conid{Cost})\;(\lambda (\Conid{Cost}\;\Varid{c})\to \Conid{Cost}\;(\Varid{c}\mathbin{+}\mathrm{5})){}\<[E]%
\ColumnHook
\end{hscode}\resethooks

By using the \ensuremath{\Varid{types}} combinator, we did not need to spell out the recursion
over \ensuremath{\Conid{Orders}}. Furthermore, the function \ensuremath{\Varid{costInc}} is polymorphic
and will work for any data structure containing costs. For similar reasons
to the \ensuremath{\Varid{typed}} lens, the \ensuremath{\Varid{types}} traversal is monomorphic and can't change
types.

However, there is a danger lurking in the shadows: when using \ensuremath{\Varid{types}}, we
must be careful to not specify a type that is too general. Consider
the running example again, if we want to modify the priorities of a
normal invoice, our first attempt might be:

\begin{hscode}\SaveRestoreHook
\column{B}{@{}>{\hspre}l<{\hspost}@{}}%
\column{E}{@{}>{\hspre}l<{\hspost}@{}}%
\>[B]{}\Varid{modifyPriority}\mathbin{::}(\Conid{Int}\to \Conid{Int})\to \Conid{Invoice}\;\Conid{Int}\to \Conid{Invoice}\;\Conid{Int}{}\<[E]%
\\
\>[B]{}\Varid{modifyPriority}\mathrel{=}\Varid{over}\;(\Varid{types}\ \texttt{@}\Conid{Int}){}\<[E]%
\ColumnHook
\end{hscode}\resethooks

This will have have unexpected consequences as there are other values of type
\ensuremath{\Conid{Int}} in our \ensuremath{\Conid{Invoice}}s, namely the order number.
The modification function will also update all
of those against our will.
Our tree contains many \ensuremath{\Conid{Int}}s used in different
ways. For this reason, the \ensuremath{\Varid{types}} combinator should be used with care. The programmer
must maintain good type discipline to avoid semantically different types being
traversed together.
This is because the way we specified \ensuremath{\Varid{modifyPriority}} did not quite reflect what we actually
\textit{meant}. Our intention is to update only the \ensuremath{\Conid{Int}}s that are
in the priority positions. A type-based query is insufficient here because it cannot
distinguish between uses of \ensuremath{\Conid{Int}}.
This problem did not exist for the lens version, because that requires the type
to appear exactly once, avoiding such clashes.

\subsubsection{By parameter}
In the previous example, our real intention was to select the values in
the priority fields only, or in other words, those that correspond to
the \ensuremath{\Varid{p}} type parameter.
Thus, we provide traversals that are defined over a specific type parameter.
We use positional indexing to refer to the type parameter of interest.
\begin{hscode}\SaveRestoreHook
\column{B}{@{}>{\hspre}l<{\hspost}@{}}%
\column{E}{@{}>{\hspre}l<{\hspost}@{}}%
\>[B]{}\Varid{param}\mathbin{::}\Conid{HasParam}\;\Varid{pos}\;\Varid{s}\;\Varid{t}\;\Varid{a}\;\Varid{b}\Rightarrow \Conid{Traversal}\;\Varid{s}\;\Varid{t}\;\Varid{a}\;\Varid{b}{}\<[E]%
\ColumnHook
\end{hscode}\resethooks

Numbering starts from the outside, meaning that the last parameter has the
index 0. Trying to access an out of bounds type parameter results in a type
error.
Using \ensuremath{\Varid{param}}, we can revise the devision of the \ensuremath{\Varid{modifyPriority}} function:
\begin{hscode}\SaveRestoreHook
\column{B}{@{}>{\hspre}l<{\hspost}@{}}%
\column{E}{@{}>{\hspre}l<{\hspost}@{}}%
\>[B]{}\Varid{treeIncParam}\mathbin{::}\Conid{HasParam}\;\mathrm{0}\;\Varid{s}\;\Varid{s}\;\Conid{Int}\;\Conid{Int}\Rightarrow \Varid{s}\to \Varid{s}{}\<[E]%
\\
\>[B]{}\Varid{treeIncParam}\mathrel{=}\Varid{over}\;(\Varid{param}\ \texttt{@}\mathrm{0})\;(\mathbin{+}\mathrm{1}){}\<[E]%
\ColumnHook
\end{hscode}\resethooks
This revised definition now properly distinguishes between the different \ensuremath{\Conid{Int}}s
in our \ensuremath{\Conid{Invoice}}s.

\subsubsection{By constraint}

The most general type of traversal is the \emph{constrained} traversal.
A constrained traversal focuses on all positions in a data type.
It does this by requiring that that the types in all positions
satisfy a constraint, and then uniformly applies
a function in terms of this constraint to all fields.

A constrained traversal thus has the following type:
\begin{hscode}\SaveRestoreHook
\column{B}{@{}>{\hspre}l<{\hspost}@{}}%
\column{14}{@{}>{\hspre}l<{\hspost}@{}}%
\column{E}{@{}>{\hspre}l<{\hspost}@{}}%
\>[B]{}\Varid{constraints}{}\<[14]%
\>[14]{}\mathbin{::}\Conid{HasConstraints}\;\Varid{c}\;\Varid{s}\;\Varid{t}\Rightarrow \Conid{Applicative}\;\Varid{g}\Rightarrow (\forall \Varid{a}\hsforall \;\Varid{b}\hsdot{\mathbin{\cdot}}{.}\Varid{c}\;\Varid{a}\;\Varid{b}\Rightarrow \Varid{a}\to \Varid{g}\;\Varid{b})\to \Varid{s}\to \Varid{g}\;\Varid{t}{}\<[E]%
\ColumnHook
\end{hscode}\resethooks

The user can instantiate the traversal to any type class of their choosing,
thereby specifying the traversal strategy. The traversing function has to be
one that only has knowledge of what information is available in the class \ensuremath{\Varid{c}}.
Via the ad-hoc overloading mechanism of type classes, the function is
instantiated to the version specified for each field in \ensuremath{\Varid{s}}.

There are many choices to which we could instantiate \ensuremath{\Varid{c}}, in fact, it is
the most general traversal and subsumes the two other traversals we have discussed.
Users might also decide to instantiate \ensuremath{\Varid{c}} to a constraint based on \ensuremath{\Conid{Data}} or
\ensuremath{\Conid{Generic}} in order to specify dynamically how each field is processed. The
\ensuremath{\Varid{constraints}} traversal just provides a framework whilst the constraint determines
how to deal precisely with each subpart.

\subsection{Composition}

The final ingredient is an overloaded composition operator \ensuremath{\mathbin{\circ}} which can be
used to compose together any combination of lenses, prisms and traversals.

The type of this operator can be thought of abstractly as
\begin{hscode}\SaveRestoreHook
\column{B}{@{}>{\hspre}l<{\hspost}@{}}%
\column{E}{@{}>{\hspre}l<{\hspost}@{}}%
\>[B]{}(\mathbin{\circ})\mathbin{::}o_1, o_2 \in \{ Lens, Traversal \} \Rightarrow \;o_1\;\Varid{s}\;\Varid{t}\;\Varid{c}\;\Varid{d}\;\to \;o_2\;\Varid{c}\;\Varid{d}\;\Varid{a}\;\Varid{b}\;\to \;(o_1\mathop{\vee}o_2)\;\Varid{s}\;\Varid{t}\;\Varid{a}\;\Varid{b}{}\<[E]%
\ColumnHook
\end{hscode}\resethooks
\mpi{RIP HASSE DIAGRAM, THIS IS WHERE YOU ONCE LAY}
The join operation is specified by
defining a \ensuremath{\Conid{Traversal}} to be above a \ensuremath{\Conid{Lens}}.
We present the composition operator in this way as in the full generality
there are more components (such as prisms) to the hierarchy~\cite{Pickering:2017:optics}.
The intuition is that a lens is a special case of a traversal where
there is exactly one focused element. Being more restrictive allows lenses
to support the additional operations of viewing that traversals do not support.

\paragraph{Summary}
In this section we have described the various widgets that allow
values to be traversed, modified, and inspected using generic lenses
and traversals. These operations form an interface for our library,
which is summarised in Figure~\ref{fig:interface}.
% include ImplementationBackground.lhs
\section{Background: Lenses, Traversals, and Generics}
\label{sec:background}

% format <..> = "\mathbin{{<}{..}{>}}"
% format <.>  = "\mathbin{<\mkern-8mu\circ\mkern-8mu>}"
% format <.> = "\mathbin{{<}{.}{>}}"

%% generic

%% Dicts

In this section we will begin to describe an efficient implementation
of the interface found in Figure~\ref{fig:interface}, while
introducing the necessary background that is the foundation for our
generic traversals.

\subsection{Lenses and Traversals}

We start by briefly describing the concrete representation of \ensuremath{\Conid{Lens}}
and \ensuremath{\Conid{Traversal}} and the associated operators before explaining the
implementation of the different derived lenses and traversals.

\paragraph{Lenses}
The representation of lenses that we use is the
van Laarhoven representation~\cite{2009:vanLaarhoven:cps}. A van Laarhoven lens
is a function of the following type:
\begin{hscode}\SaveRestoreHook
\column{B}{@{}>{\hspre}l<{\hspost}@{}}%
\column{E}{@{}>{\hspre}l<{\hspost}@{}}%
\>[B]{}\keyword{type}\;\Conid{Lens}\;\Varid{s}\;\Varid{t}\;\Varid{a}\;\Varid{b}\mathrel{=}\forall \Varid{f}\hsforall \hsdot{\mathbin{\cdot}}{.}\Conid{Functor}\;\Varid{f}\Rightarrow (\Varid{a}\to \Varid{f}\;\Varid{b})\to (\Varid{s}\to \Varid{f}\;\Varid{t}){}\<[E]%
\ColumnHook
\end{hscode}\resethooks
We can implement the \ensuremath{\Varid{view}} and \ensuremath{\Varid{update}} functions as required by our interface
by suitably instantiating \ensuremath{\Varid{f}} to the \ensuremath{\Conid{Const}} and \ensuremath{\Conid{Identity}} functor respectively~\cite{Mcbride:2008:Applicative}.

\begin{minipage}[t]{0.4\textwidth}
\begin{hscode}\SaveRestoreHook
\column{B}{@{}>{\hspre}l<{\hspost}@{}}%
\column{E}{@{}>{\hspre}l<{\hspost}@{}}%
\>[B]{}\Varid{view}\mathbin{::}\Conid{Lens}\;\Varid{s}\;\Varid{t}\;\Varid{a}\;\Varid{b}\to \Varid{s}\to \Varid{a}{}\<[E]%
\\
\>[B]{}\Varid{view}\;\Varid{l}\mathrel{=}\Varid{getConst}\hsdot{\mathbin{\cdot}}{.}\Varid{l}\;\Conid{Const}{}\<[E]%
\ColumnHook
\end{hscode}\resethooks
\end{minipage}
\begin{minipage}[t]{0.55\textwidth}
\begin{hscode}\SaveRestoreHook
\column{B}{@{}>{\hspre}l<{\hspost}@{}}%
\column{E}{@{}>{\hspre}l<{\hspost}@{}}%
\>[B]{}\Varid{update}\mathbin{::}\Conid{Lens}\;\Varid{s}\;\Varid{t}\;\Varid{a}\;\Varid{b}\to \Varid{b}\to \Varid{s}\to \Varid{t}{}\<[E]%
\\
\>[B]{}\Varid{update}\;\Varid{l}\;\Varid{b}\mathrel{=}\Varid{runIdentity}\hsdot{\mathbin{\cdot}}{.}\Varid{l}\;(\Varid{const}\;(\Conid{Identity}\;\Varid{b})){}\<[E]%
\ColumnHook
\end{hscode}\resethooks
\end{minipage}

\paragraph{Traversals}
The van Laarhoven representation is also convenient as the type is similar to
that of traversals.
We implement \ensuremath{\Conid{Traversal}\;\Varid{s}\;\Varid{t}\;\Varid{a}\;\Varid{b}} with functions of the following type:
\begin{hscode}\SaveRestoreHook
\column{B}{@{}>{\hspre}l<{\hspost}@{}}%
\column{E}{@{}>{\hspre}l<{\hspost}@{}}%
\>[B]{}\keyword{type}\;\Conid{Traversal}\;\Varid{s}\;\Varid{t}\;\Varid{a}\;\Varid{b}\mathrel{=}\forall \Varid{f}\hsforall \hsdot{\mathbin{\cdot}}{.}\Conid{Applicative}\;\Varid{f}\Rightarrow (\Varid{a}\to \Varid{f}\;\Varid{b})\to (\Varid{s}\to \Varid{f}\;\Varid{t}){}\<[E]%
\ColumnHook
\end{hscode}\resethooks
Again, we implement our interface by instantiating the applicative
\ensuremath{\Varid{f}} to \ensuremath{\Conid{Identity}} and \ensuremath{\Conid{Const}}, which provides the correct
specialisation to implement the functions.
\ck{I don't know why, but if I add the type signature for \ensuremath{\Varid{over}}, then the
constrained example stack overflows... not if \ensuremath{\Varid{over}} is imported though. wtf}
\nw{For now I will cheat}
\begin{hscode}\SaveRestoreHook
\column{B}{@{}>{\hspre}l<{\hspost}@{}}%
\column{E}{@{}>{\hspre}l<{\hspost}@{}}%
\>[B]{}\Varid{over}\mathbin{::}\Conid{Traversal}\;\Varid{s}\;\Varid{t}\;\Varid{a}\;\Varid{b}\to (\Varid{a}\to \Varid{b})\to \Varid{s}\to \Varid{t}{}\<[E]%
\ColumnHook
\end{hscode}\resethooks
\vskip -\belowdisplayskip
\vskip -\abovedisplayskip
\vskip -\blanklineskip
\begin{hscode}\SaveRestoreHook
\column{B}{@{}>{\hspre}l<{\hspost}@{}}%
\column{E}{@{}>{\hspre}l<{\hspost}@{}}%
\>[B]{}\Varid{over}\;\Varid{t}\;\Varid{f}\mathrel{=}\Varid{runIdentity}\hsdot{\mathbin{\cdot}}{.}\Varid{t}\;(\Conid{Identity}\hsdot{\mathbin{\cdot}}{.}\Varid{f}){}\<[E]%
\\
\>[B]{}\Varid{toListOf}\mathbin{::}\Conid{Traversal}\;\Varid{s}\;\Varid{s}\;\Varid{a}\;\Varid{a}\to \Varid{s}\to [\mskip1.5mu \Varid{a}\mskip1.5mu]{}\<[E]%
\\
\>[B]{}\Varid{toListOf}\;\Varid{t}\mathrel{=}\Varid{getConst}\hsdot{\mathbin{\cdot}}{.}\Varid{t}\;(\Conid{Const}\hsdot{\mathbin{\cdot}}{.}\Varid{singleton}){}\<[E]%
\\[\blanklineskip]%
\>[B]{}\Varid{traverseOf}\mathbin{::}\Conid{Traversal}\;\Varid{s}\;\Varid{t}\;\Varid{a}\;\Varid{b}\to (\forall \Varid{g}\hsforall \hsdot{\mathbin{\cdot}}{.}\Conid{Applicative}\;\Varid{g}\Rightarrow (\Varid{a}\to \Varid{g}\;\Varid{b})\to \Varid{s}\to \Varid{g}\;\Varid{t}){}\<[E]%
\\
\>[B]{}\Varid{traverseOf}\mathrel{=}\Varid{id}{}\<[E]%
\\[\blanklineskip]%
\>[B]{}\Varid{singleton}\mathbin{::}\Varid{a}\to [\mskip1.5mu \Varid{a}\mskip1.5mu]{}\<[E]%
\\
\>[B]{}\Varid{singleton}\;\Varid{x}\mathrel{=}[\mskip1.5mu \Varid{x}\mskip1.5mu]{}\<[E]%
\ColumnHook
\end{hscode}\resethooks
We do not provide any additional justification for this representation
as it has been extensively studied \cite{Jaskelioff:Representation:2014,Oconnor:Functor:2011,Bird:2013:Traversals}.
In any case, the choice is not crucial to our work. We could instead
apply the same techniques to the
profunctor~\cite{Pickering:2017:optics} and other encodings.

\paragraph{Composition}
With this representation, \ensuremath{\mathbin{\circ}} simply becomes composition:
\begin{hscode}\SaveRestoreHook
\column{B}{@{}>{\hspre}l<{\hspost}@{}}%
\column{E}{@{}>{\hspre}l<{\hspost}@{}}%
\>[B]{}(\mathbin{\circ})\mathbin{::}\Conid{Lens}\;\Varid{s}\;\Varid{t}\;\Varid{c}\;\Varid{d}\to \Conid{Lens}\;\Varid{c}\;\Varid{d}\;\Varid{a}\;\Varid{b}\to \Conid{Lens}\;\Varid{s}\;\Varid{t}\;\Varid{a}\;\Varid{b}{}\<[E]%
\\
\>[B]{}(\mathbin{\circ})\mathrel{=}(\hsdot{\mathbin{\cdot}}{.}){}\<[E]%
\ColumnHook
\end{hscode}\resethooks
In order to make lenses and traversals compose, a lens composed with a
traversal must result in a traversal. This is easy to observe as the
type of a \ensuremath{\Conid{Traversal}} is more constrained than that of a \ensuremath{\Conid{Lens}}
because \ensuremath{\Conid{Functor}} is a superclass of \ensuremath{\Conid{Applicative}}.

\subsection{Generic Programming}

Datatype-generic programming allows data types to be decomposed into
their constituent parts, which are shown
below:

\begin{minipage}[t]{0.3\textwidth}
\begin{hscode}\SaveRestoreHook
\column{B}{@{}>{\hspre}l<{\hspost}@{}}%
\column{27}{@{}>{\hspre}c<{\hspost}@{}}%
\column{27E}{@{}l@{}}%
\column{33}{@{}>{\hspre}l<{\hspost}@{}}%
\column{E}{@{}>{\hspre}l<{\hspost}@{}}%
\>[B]{}\keyword{data}\;\Varid{f}\mathbin{{:}{+}{:}}\Varid{g}{}\<[27]%
\>[27]{}\mathrel{=}{}\<[27E]%
\>[33]{}\Conid{L}\;\Varid{f}\mid \Conid{R}\;\Varid{g}{}\<[E]%
\\
\>[B]{}\keyword{data}\;\Varid{f}\mathbin{{:}{\times}{:}}\Varid{g}{}\<[27]%
\>[27]{}\mathrel{=}{}\<[27E]%
\>[33]{}\Varid{f}\mathbin{{:}{\times}{:}}\Varid{g}{}\<[E]%
\\
\>[B]{}\keyword{newtype}\;\Conid{K}\;\Varid{a}{}\<[27]%
\>[27]{}\mathrel{=}{}\<[27E]%
\>[33]{}\Conid{K}\;\Varid{a}{}\<[E]%
\ColumnHook
\end{hscode}\resethooks
\end{minipage}
\begin{minipage}[t]{0.15\textwidth}
\begin{hscode}\SaveRestoreHook
\column{B}{@{}>{\hspre}l<{\hspost}@{}}%
\column{15}{@{}>{\hspre}l<{\hspost}@{}}%
\column{E}{@{}>{\hspre}l<{\hspost}@{}}%
\>[B]{}\keyword{data}\;\Conid{V}{}\<[E]%
\\
\>[B]{}\keyword{data}\;\Conid{U}\mathrel{=}{}\<[15]%
\>[15]{}\Conid{U}{}\<[E]%
\ColumnHook
\end{hscode}\resethooks
\end{minipage}
\begin{minipage}[t]{0.5\textwidth}
\noindent
\begin{hscode}\SaveRestoreHook
\column{B}{@{}>{\hspre}l<{\hspost}@{}}%
\column{12}{@{}>{\hspre}c<{\hspost}@{}}%
\column{12E}{@{}l@{}}%
\column{15}{@{}>{\hspre}l<{\hspost}@{}}%
\column{27}{@{}>{\hspre}c<{\hspost}@{}}%
\column{27E}{@{}l@{}}%
\column{33}{@{}>{\hspre}l<{\hspost}@{}}%
\column{E}{@{}>{\hspre}l<{\hspost}@{}}%
\>[B]{}\keyword{newtype}\;\Conid{M}\;(\Varid{m}\mathbin{::}\Conid{Meta})\;\Varid{a}{}\<[27]%
\>[27]{}\mathrel{=}{}\<[27E]%
\>[33]{}\Conid{M}\;\Varid{a}{}\<[E]%
\\
\>[B]{}\keyword{data}\;\Conid{Meta}{}\<[12]%
\>[12]{}\mathrel{=}{}\<[12E]%
\>[15]{}\Conid{MetaData}\;\Conid{Symbol}{}\<[E]%
\\
\>[12]{}\mid {}\<[12E]%
\>[15]{}\Conid{MetaCons}\;\Conid{Symbol}{}\<[E]%
\\
\>[12]{}\mid {}\<[12E]%
\>[15]{}\Conid{MetaSel}\;(\Conid{Maybe}\;\Conid{Symbol}){}\<[E]%
\ColumnHook
\end{hscode}\resethooks
\end{minipage}

\noindent
This is a sum-of-products representation similar to that proposed
by~\citet{Hinze:2006:generics}.
Algebraic data types can be uniformly viewed in
this way: choice between constructor variants is encoded as (potentially
nested) binary sums (\ensuremath{\mathbin{{:}{+}{:}}}). A single field of type \ensuremath{\Varid{a}} inside a constructor is
stored as \ensuremath{\Conid{K}\;\Varid{a}}; multiple fields are collected in (potentially nested)
binary products (\ensuremath{\mathbin{{:}{\times}{:}}}). Datatypes with no constructors
are represented by \ensuremath{\Conid{V}}, and constructors with no fields by \ensuremath{\Conid{U}}.

Additional metadata (name of the datatype, names of constructors, and
(optional) names of fields) can be attached to the nodes via \ensuremath{\Conid{M}}.
The meta constructor \ensuremath{\Conid{M}} makes use of \emph{datatype
promotion}~\cite{Yorgey:2012:promotion}, which allows \ensuremath{\Conid{Meta}}'s constructors to be
used in a type context. In general, promotion allows data types like \ensuremath{\Conid{Meta}} and
\ensuremath{\Conid{Bool}} to be used as kinds.

The isomorphism between concrete types and their sum-of-products view
is witnessed by an instance of the \ensuremath{\Conid{Generic}} type class:
\begin{hscode}\SaveRestoreHook
\column{B}{@{}>{\hspre}l<{\hspost}@{}}%
\column{3}{@{}>{\hspre}l<{\hspost}@{}}%
\column{E}{@{}>{\hspre}l<{\hspost}@{}}%
\>[B]{}\keyword{class}\;\Conid{Generic}\;\Varid{a}\;\keyword{where}{}\<[E]%
\\
\>[B]{}\hsindent{3}{}\<[3]%
\>[3]{}\keyword{type}\;\keyword{family}\;\Conid{Rep}\;\Varid{a}\mathbin{::}\star{}\<[E]%
\\
\>[B]{}\hsindent{3}{}\<[3]%
\>[3]{}\Varid{from}\mathbin{::}\Varid{a}\to \Conid{Rep}\;\Varid{a}{}\<[E]%
\\
\>[B]{}\hsindent{3}{}\<[3]%
\>[3]{}\Varid{to}\mathbin{::}\Conid{Rep}\;\Varid{a}\to \Varid{a}{}\<[E]%
\ColumnHook
\end{hscode}\resethooks
As the type of the generic view is different for each type,
\ensuremath{\Conid{Generic}} associates the concrete type and their representation type
via the \ensuremath{\Conid{Rep}} type family~\cite{assoc05}.
Writing these instances is laborious, but straightforward. The Glasgow Haskell
Compiler (GHC) provides built-in support for deriving these
instances~\cite{Magalhaes:2010:generic}.
In practice, this requires us to append a \ensuremath{\keyword{deriving}\;\Conid{Generic}} clause to
our data definitions: we omit this in our presentation.

% This representation is shallow because it is not recursive.
Now consider the following definition of linked lists:
\begin{hscode}\SaveRestoreHook
\column{B}{@{}>{\hspre}l<{\hspost}@{}}%
\column{E}{@{}>{\hspre}l<{\hspost}@{}}%
\>[B]{}\keyword{data}\;\Conid{List}\;\Varid{a}\mathrel{=}\Conid{Empty}\mid \Conid{Cons}\;\Varid{a}\;(\Conid{List}\;\Varid{a}){}\<[E]%
\ColumnHook
\end{hscode}\resethooks
The generic view of \ensuremath{\Conid{List}\;\Conid{Int}} has the type
\begin{hscode}\SaveRestoreHook
\column{B}{@{}>{\hspre}l<{\hspost}@{}}%
\column{22}{@{}>{\hspre}l<{\hspost}@{}}%
\column{24}{@{}>{\hspre}l<{\hspost}@{}}%
\column{28}{@{}>{\hspre}l<{\hspost}@{}}%
\column{E}{@{}>{\hspre}l<{\hspost}@{}}%
\>[B]{}\Conid{Rep}\;(\Conid{List}\;\Conid{Int})\equiv{}\<[22]%
\>[22]{}\Conid{M}\;({}^\prime\!\Conid{MetaData}\;\text{\ttfamily \char34 List\char34})\;(\Conid{M}\;({}^\prime\!\Conid{MetaCons}\;\text{\ttfamily \char34 Empty\char34})\;\Conid{U}{}\<[E]%
\\
\>[22]{}\hsindent{2}{}\<[24]%
\>[24]{}\mathbin{{:}{+}{:}}\Conid{M}\;({}^\prime\!\Conid{MetaCons}\;\text{\ttfamily \char34 Cons\char34})\;(\Conid{M}\;({}^\prime\!\Conid{MetaSel}\;{}^\prime\!\Conid{Nothing})\;(\Conid{K}\;\Conid{Int}){}\<[E]%
\\
\>[24]{}\hsindent{4}{}\<[28]%
\>[28]{}\mathbin{{:}{\times}{:}}\Conid{M}\;({}^\prime\!\Conid{MetaSel}\;{}^\prime\!\Conid{Nothing})\;(\Conid{K}\;(\Conid{List}\;\Conid{Int})))){}\<[E]%
\ColumnHook
\end{hscode}\resethooks
Reflecting the algebraic structure of \ensuremath{\Conid{List}\;\Conid{Int}} to the type-level in this way
allows us to statically introspect the shape and metadata of the type. Using
this information we can derive safe and optimal transformations without having
to write boilerplate code.\mpi{Mention here about type changing posibilities as well?}

\section{Generic Traversals with Types}
\label{sec:types}

Now that we have implemented the basic parts of our interface, we turn
to deriving interesting traversals. We concentrate on
traversals since the principles for deriving lenses are the same.

The principle of the implementation is simple. In order to generate an
optic for a specific data type we convert that data type to its
generic representation using \ensuremath{\Varid{from}\mathbin{::}\Conid{Generic}\;\Varid{a}\Rightarrow \Varid{a}\to \Conid{Rep}\;\Varid{a}}. The
type family \ensuremath{\Conid{Rep}} turns a type into the type of its generic
representation. The type of its generic representation directly
corresponds to the structure of the generic representation.

In order to implement a function that consumes the generic
representation, we need to define a type class for the function we
want to implement and then a type class instance for each clause of
the function. So we implement a separate case to deal with empty
types, products, sums and so on.

There are two complexities that must be considered in the
implementation.
First, we must decide whether or not we want to access a specific value at
the leaves of a data structure.
The second complexity in the implementation is maintaining good type-inference
behaviour.
To give a sense of how we deal with these complexities, we discuss the
implementation of the \ensuremath{\Varid{types}} traversal, then we describe how type inference
works.

\paragraph{Implementing \ensuremath{\Varid{types}}}

We now turn to how we implement our traversals using this machinery.
As a reminder, the \ensuremath{\Varid{types}} traversal is indexed by a type, and it provides
access to all subparts of a structure that have the specified type.
In this section we implement a naive first attempt at \ensuremath{\Varid{types}}
with deficiencies that we will address later on.

First, we create a type class, \ensuremath{\Conid{HasTypes}\_\;\Varid{s}\;\Varid{t}\;\Varid{a}\;\Varid{b}}, which represents that \ensuremath{\Varid{s}}
contains some (zero or more) values of type \ensuremath{\Varid{a}}, and changing these \ensuremath{\Varid{a}}s into
\ensuremath{\Varid{b}}s results in a structure of type \ensuremath{\Varid{t}}.
The sole member of this type class is the \ensuremath{\Varid{types}\_} combinator.
\begin{hscode}\SaveRestoreHook
\column{B}{@{}>{\hspre}l<{\hspost}@{}}%
\column{3}{@{}>{\hspre}l<{\hspost}@{}}%
\column{E}{@{}>{\hspre}l<{\hspost}@{}}%
\>[B]{}\keyword{class}\;\Conid{HasTypes}\_\;\Varid{s}\;\Varid{t}\;\Varid{a}\;\Varid{b}\;\keyword{where}{}\<[E]%
\\
\>[B]{}\hsindent{3}{}\<[3]%
\>[3]{}\Varid{types}\_\mathbin{::}\Conid{Traversal}\;\Varid{s}\;\Varid{t}\;\Varid{a}\;\Varid{b}{}\<[E]%
\ColumnHook
\end{hscode}\resethooks
\ensuremath{\Conid{HasTypes}\_} is the abstract interface that we are going to instantiate by
induction over the generic view. Matching on individual cases is done using
the auxiliary type class \ensuremath{\Conid{GHasTypes}\;\Varid{s}\;\Varid{t}\;\Varid{a}\;\Varid{b}}.
Then each type whose generic representation admits a \ensuremath{\Conid{GHasTypes}} instance has a
\ensuremath{\Conid{HasTypes}\_} instance itself derivable via the isomorphism.
% This is in a spec, because the good instance is defined in the next section
% nw: it's duplicated and hidden above to make the linebreak nice.
\begin{hscode}\SaveRestoreHook
\column{B}{@{}>{\hspre}l<{\hspost}@{}}%
\column{3}{@{}>{\hspre}l<{\hspost}@{}}%
\column{E}{@{}>{\hspre}l<{\hspost}@{}}%
\>[B]{}\keyword{class}\;\Conid{GHasTypes}\;\Varid{s}\;\Varid{t}\;\Varid{a}\;\Varid{b}\;\keyword{where}{}\<[E]%
\\
\>[B]{}\hsindent{3}{}\<[3]%
\>[3]{}\Varid{gtypes}\mathbin{::}\Conid{Traversal}\;\Varid{s}\;\Varid{t}\;\Varid{a}\;\Varid{b}{}\<[E]%
\\[\blanklineskip]%
\>[B]{}\keyword{instance}\;(\Conid{GHasTypes}\;(\Conid{Rep}\;\Varid{s})\;(\Conid{Rep}\;\Varid{t})\;\Varid{a}\;\Varid{b},\Conid{Generic}\;\Varid{s},\Conid{Generic}\;\Varid{t}){}\<[E]%
\\
\>[B]{}\hsindent{3}{}\<[3]%
\>[3]{}\Rightarrow \Conid{HasTypes}\_\;\Varid{s}\;\Varid{t}\;\Varid{a}\;\Varid{b}\;\keyword{where}{}\<[E]%
\\
\>[B]{}\hsindent{3}{}\<[3]%
\>[3]{}\Varid{types}\_\mathrel{=}\Varid{iso}_{\Conid{Rep}}\hsdot{\mathbin{\cdot}}{.}\Varid{gtypes}{}\<[E]%
\ColumnHook
\end{hscode}\resethooks
Where \ensuremath{\Varid{iso}_{\Conid{Rep}}} is a lens that views a value as its (isomorphic) generic
representation by using the methods from the \ensuremath{\Conid{Generic}} type class. Changes made
on the generic representations are reflected on the original type, including
changes of types.
\begin{hscode}\SaveRestoreHook
\column{B}{@{}>{\hspre}l<{\hspost}@{}}%
\column{E}{@{}>{\hspre}l<{\hspost}@{}}%
\>[B]{}\Varid{iso}_{\Conid{Rep}}\mathbin{::}(\Conid{Generic}\;\Varid{s},\Conid{Generic}\;\Varid{t})\Rightarrow \Conid{Lens}\;\Varid{s}\;\Varid{t}\;(\Conid{Rep}\;\Varid{s})\;(\Conid{Rep}\;\Varid{t}){}\<[E]%
\\
\>[B]{}\Varid{iso}_{\Conid{Rep}}\;\Varid{f}\mathrel{=}\Varid{fmap}\;\Varid{to}\hsdot{\mathbin{\cdot}}{.}\Varid{f}\hsdot{\mathbin{\cdot}}{.}\Varid{from}{}\<[E]%
\ColumnHook
\end{hscode}\resethooks
The above instance is defined \textit{for all} types, meaning that all
\ensuremath{\Varid{types}} queries require \ensuremath{\Varid{s}} and \ensuremath{\Varid{t}} to have a \ensuremath{\Conid{Generic}} instance, as per the
constraints. However, certain types do not admit a \ensuremath{\Conid{Generic}} instance, namely the
non-algebraic primitive types. For these, we define overlapping instances
that are picked instead of the general one above.
\begin{hscode}\SaveRestoreHook
\column{B}{@{}>{\hspre}l<{\hspost}@{}}%
\column{3}{@{}>{\hspre}l<{\hspost}@{}}%
\column{E}{@{}>{\hspre}l<{\hspost}@{}}%
\>[B]{}\keyword{instance}\;\Conid{HasTypes}\_\;\Conid{Char}\;\Conid{Char}\;\Varid{a}\;\Varid{b}\;\keyword{where}{}\<[E]%
\\
\>[B]{}\hsindent{3}{}\<[3]%
\>[3]{}\Varid{types}\_\;\anonymous \mathrel{=}\Varid{pure}{}\<[E]%
\ColumnHook
\end{hscode}\resethooks
The code above shows the instance for \ensuremath{\Conid{Char}}, and is similar for other
primitive types such as \ensuremath{\Conid{Double}}, \ensuremath{\Conid{Float}}, \ensuremath{\Conid{Int}}, and \ensuremath{\Conid{Integer}}.
Given that these types are not actually containers, they can not
possibly contain any interesting values, thus their traversal is
defined as the no-op \ensuremath{\Varid{pure}}.

Additionally, we define the \ensuremath{\Varid{iso}_\Conid{K}} and \ensuremath{\Varid{iso}_\Conid{M}} lenses, which focus on the value
inside the \ensuremath{\Conid{K}} node and the generic structure wrapped by metadata nodes
respectively.

\begin{minipage}{0.5\textwidth}
\begin{hscode}\SaveRestoreHook
\column{B}{@{}>{\hspre}l<{\hspost}@{}}%
\column{E}{@{}>{\hspre}l<{\hspost}@{}}%
\>[B]{}\Varid{iso}_\Conid{K}\mathbin{::}\Conid{Lens}\;(\Conid{K}\;\Varid{a})\;(\Conid{K}\;\Varid{b})\;\Varid{a}\;\Varid{b}{}\<[E]%
\\
\>[B]{}\Varid{iso}_\Conid{K}\;\Varid{f}\;\Varid{s}\mathrel{=}\Conid{K}\mathbin{{\langle}{\$}{\rangle}}\Varid{f}\;(\Varid{unK}\;\Varid{s}){}\<[E]%
\ColumnHook
\end{hscode}\resethooks
\end{minipage}~
\begin{minipage}{0.5\textwidth}
\begin{hscode}\SaveRestoreHook
\column{B}{@{}>{\hspre}l<{\hspost}@{}}%
\column{E}{@{}>{\hspre}l<{\hspost}@{}}%
\>[B]{}\Varid{iso}_\Conid{M}\mathbin{::}\Conid{Lens}\;(\Conid{M}\;\Varid{m}\;\Varid{s})\;(\Conid{M}\;\Varid{m}\;\Varid{t})\;\Varid{s}\;\Varid{t}{}\<[E]%
\\
\>[B]{}\Varid{iso}_\Conid{M}\;\Varid{f}\;\Varid{s}\mathrel{=}\Conid{M}\mathbin{{\langle}{\$}{\rangle}}\Varid{f}\;(\Varid{unM}\;\Varid{s}){}\<[E]%
\ColumnHook
\end{hscode}\resethooks
\end{minipage}
We now deal with the generic cases one-by-one. As the \ensuremath{\Varid{types}} traversal is
oblivious to metadata such as constructor names, \ensuremath{\Varid{gtypes}} simply skips
over these.
\begin{hscode}\SaveRestoreHook
\column{B}{@{}>{\hspre}l<{\hspost}@{}}%
\column{3}{@{}>{\hspre}l<{\hspost}@{}}%
\column{E}{@{}>{\hspre}l<{\hspost}@{}}%
\>[B]{}\keyword{instance}\;\Conid{GHasTypes}\;\Varid{s}\;\Varid{t}\;\Varid{a}\;\Varid{b}\Rightarrow \Conid{GHasTypes}\;(\Conid{M}\;\Varid{m}\;\Varid{s})\;(\Conid{M}\;\Varid{m}\;\Varid{t})\;\Varid{a}\;\Varid{b}\;\keyword{where}{}\<[E]%
\\
\>[B]{}\hsindent{3}{}\<[3]%
\>[3]{}\Varid{gtypes}\mathrel{=}\Varid{iso}_\Conid{M}\hsdot{\mathbin{\cdot}}{.}\Varid{gtypes}{}\<[E]%
\ColumnHook
\end{hscode}\resethooks
Next we handle sums, i.e. the constructors of a datatype. Since the aim is to
discover every node in the structure, we recursively call \ensuremath{\Varid{gtypes}} on whichever
case alternative is present. The constraints \ensuremath{\Conid{GHasTypes}\;\Varid{l}_{\mathrm{1}}\;\Varid{l}_{\mathrm{2}}\;\Varid{a}\;\Varid{b}} and
\ensuremath{\Conid{GHasTypes}\;\Varid{r}_{\mathrm{1}}\;\Varid{r}_{\mathrm{2}}\;\Varid{a}\;\Varid{b}} ensure that we can indeed traverse both cases.
\begin{hscode}\SaveRestoreHook
\column{B}{@{}>{\hspre}l<{\hspost}@{}}%
\column{3}{@{}>{\hspre}l<{\hspost}@{}}%
\column{5}{@{}>{\hspre}l<{\hspost}@{}}%
\column{E}{@{}>{\hspre}l<{\hspost}@{}}%
\>[B]{}\keyword{instance}\;(\Conid{GHasTypes}\;\Varid{l}_{\mathrm{1}}\;\Varid{l}_{\mathrm{2}}\;\Varid{a}\;\Varid{b},\Conid{GHasTypes}\;\Varid{r}_{\mathrm{1}}\;\Varid{r}_{\mathrm{2}}\;\Varid{a}\;\Varid{b}){}\<[E]%
\\
\>[B]{}\hsindent{5}{}\<[5]%
\>[5]{}\Rightarrow \Conid{GHasTypes}\;(\Varid{l}_{\mathrm{1}}\mathbin{{:}{+}{:}}\Varid{r}_{\mathrm{1}})\;(\Varid{l}_{\mathrm{2}}\mathbin{{:}{+}{:}}\Varid{r}_{\mathrm{2}})\;\Varid{a}\;\Varid{b}\;\keyword{where}{}\<[E]%
\\
\>[B]{}\hsindent{3}{}\<[3]%
\>[3]{}\Varid{gtypes}\;\Varid{f}\;(\Conid{L}\;\Varid{l})\mathrel{=}\Conid{L}\mathbin{{\langle}{\$}{\rangle}}\Varid{gtypes}\;\Varid{f}\;\Varid{l}{}\<[E]%
\\
\>[B]{}\hsindent{3}{}\<[3]%
\>[3]{}\Varid{gtypes}\;\Varid{f}\;(\Conid{R}\;\Varid{r})\mathrel{=}\Conid{R}\mathbin{{\langle}{\$}{\rangle}}\Varid{gtypes}\;\Varid{f}\;\Varid{r}{}\<[E]%
\ColumnHook
\end{hscode}\resethooks
Products are treated similarly: we traverse both left and right trees, looking
for \ensuremath{\Varid{a}}s.
\begin{hscode}\SaveRestoreHook
\column{B}{@{}>{\hspre}l<{\hspost}@{}}%
\column{3}{@{}>{\hspre}l<{\hspost}@{}}%
\column{5}{@{}>{\hspre}l<{\hspost}@{}}%
\column{E}{@{}>{\hspre}l<{\hspost}@{}}%
\>[B]{}\keyword{instance}\;(\Conid{GHasTypes}\;\Varid{l}_{\mathrm{1}}\;\Varid{l}_{\mathrm{2}}\;\Varid{a}\;\Varid{b},\Conid{GHasTypes}\;\Varid{r}_{\mathrm{1}}\;\Varid{r}_{\mathrm{2}}\;\Varid{a}\;\Varid{b}){}\<[E]%
\\
\>[B]{}\hsindent{5}{}\<[5]%
\>[5]{}\Rightarrow \Conid{GHasTypes}\;(\Varid{l}_{\mathrm{1}}\mathbin{{:}{\times}{:}}\Varid{r}_{\mathrm{1}})\;(\Varid{l}_{\mathrm{2}}\mathbin{{:}{\times}{:}}\Varid{r}_{\mathrm{2}})\;\Varid{a}\;\Varid{b}\;\keyword{where}{}\<[E]%
\\
\>[B]{}\hsindent{3}{}\<[3]%
\>[3]{}\Varid{gtypes}\;\Varid{f}\;(\Varid{l}\mathbin{{:}{\times}{:}}\Varid{r})\mathrel{=}(\mathbin{{:}{\times}{:}})\mathbin{{\langle}{\$}{\rangle}}\Varid{gtypes}\;\Varid{f}\;\Varid{l}\mathbin{{\langle}{*}{\rangle}}\Varid{gtypes}\;\Varid{f}\;\Varid{r}{}\<[E]%
\ColumnHook
\end{hscode}\resethooks
Now the interesting case -- that is, when we encounter a field of type \ensuremath{\Varid{a}}. We
can now stop the search, and focus on this leaf node.
\begin{hscode}\SaveRestoreHook
\column{B}{@{}>{\hspre}l<{\hspost}@{}}%
\column{3}{@{}>{\hspre}l<{\hspost}@{}}%
\column{E}{@{}>{\hspre}l<{\hspost}@{}}%
\>[B]{}\keyword{instance}\;\Conid{GHasTypes}\;(\Conid{K}\;\Varid{a})\;(\Conid{K}\;\Varid{b})\;\Varid{a}\;\Varid{b}\;\keyword{where}{}\<[E]%
\\
\>[B]{}\hsindent{3}{}\<[3]%
\>[3]{}\Varid{gtypes}\mathrel{=}\Varid{iso}_\Conid{K}{}\<[E]%
\ColumnHook
\end{hscode}\resethooks
What if the leaf is not actually the type we were looking for, but something
else? If we were doing only a \textit{shallow} traversal, this is where we
would stop. Given that our traversal is \textit{deep}, we look further to see
if this leaf contains any more values of type \ensuremath{\Varid{a}}, by recursively invoking a
\ensuremath{\Conid{HasTypes}\_} constraint, and the corresponding \ensuremath{\Varid{types}\_} traversal, now for the leaf.
Note that this instance is \emph{overlapped} by the previous one, as it is
strictly more general than the previous case, and it is picked when the above
does not match.
\begin{hscode}\SaveRestoreHook
\column{B}{@{}>{\hspre}l<{\hspost}@{}}%
\column{3}{@{}>{\hspre}l<{\hspost}@{}}%
\column{E}{@{}>{\hspre}l<{\hspost}@{}}%
\>[B]{}\keyword{instance}\;\Conid{HasTypes}\_\;\Varid{s}\;\Varid{t}\;\Varid{a}\;\Varid{b}\Rightarrow \Conid{GHasTypes}\;(\Conid{K}\;\Varid{s})\;(\Conid{K}\;\Varid{t})\;\Varid{a}\;\Varid{b}\;\keyword{where}{}\<[E]%
\\
\>[B]{}\hsindent{3}{}\<[3]%
\>[3]{}\Varid{gtypes}\mathrel{=}\Varid{iso}_\Conid{K}\hsdot{\mathbin{\cdot}}{.}\Varid{types}\_{}\<[E]%
\ColumnHook
\end{hscode}\resethooks
When the leaves are primitives this process stops, since their
\ensuremath{\Conid{HasTypes}\_} instance is \ensuremath{\Varid{pure}}.

Two cases remain: \ensuremath{\Conid{U}}, when the field contains no value at all
(isomorphic to unit), and \ensuremath{\Conid{V}}, which corresponds to types with no
constructors. Both of these are just skipped.

\begin{minipage}{0.5\textwidth}
\begin{hscode}\SaveRestoreHook
\column{B}{@{}>{\hspre}l<{\hspost}@{}}%
\column{3}{@{}>{\hspre}l<{\hspost}@{}}%
\column{E}{@{}>{\hspre}l<{\hspost}@{}}%
\>[B]{}\keyword{instance}\;\Conid{GHasTypes}\;\Conid{U}\;\Conid{U}\;\Varid{a}\;\Varid{b}\;\keyword{where}{}\<[E]%
\\
\>[B]{}\hsindent{3}{}\<[3]%
\>[3]{}\Varid{gtypes}\;\anonymous \mathrel{=}\Varid{pure}{}\<[E]%
\ColumnHook
\end{hscode}\resethooks
\end{minipage}
\begin{minipage}{0.5\textwidth}
\begin{hscode}\SaveRestoreHook
\column{B}{@{}>{\hspre}l<{\hspost}@{}}%
\column{3}{@{}>{\hspre}l<{\hspost}@{}}%
\column{E}{@{}>{\hspre}l<{\hspost}@{}}%
\>[B]{}\keyword{instance}\;\Conid{GHasTypes}\;\Conid{V}\;\Conid{V}\;\Varid{a}\;\Varid{b}\;\keyword{where}{}\<[E]%
\\
\>[B]{}\hsindent{3}{}\<[3]%
\>[3]{}\Varid{gtypes}\;\anonymous \mathrel{=}\Varid{pure}{}\<[E]%
\ColumnHook
\end{hscode}\resethooks
\end{minipage}
We have now covered all cases, but we ought to be careful; the two
overlapping instances for the \ensuremath{\Conid{K}} cases can lead to surprising results when
the query is changing types. Consider the type \ensuremath{\Conid{IntPair}}:
\begin{hscode}\SaveRestoreHook
\column{B}{@{}>{\hspre}l<{\hspost}@{}}%
\column{E}{@{}>{\hspre}l<{\hspost}@{}}%
\>[B]{}\keyword{data}\;\Conid{IntPair}\;\Varid{a}\mathrel{=}\Conid{IntPair}\;\Conid{Int}\;\Varid{a}{}\<[E]%
\ColumnHook
\end{hscode}\resethooks
As expected, updating the \ensuremath{\Conid{Int}}s in an \ensuremath{\Conid{IntPair}\;\Conid{Int}} updates both the
monomorphic value, and the field corresponding to the type variable:
\begin{hscode}\SaveRestoreHook
\column{B}{@{}>{\hspre}l<{\hspost}@{}}%
\column{E}{@{}>{\hspre}l<{\hspost}@{}}%
\>[B]{}\texttt{> }\texttt{ghci> }\;\Varid{over}\;\Varid{types}\_\;((\mathbin{+}\mathrm{10})\mathbin{::}\Conid{Int}\to \Conid{Int})\;(\Conid{IntPair}\;\mathrm{1}\;(\mathrm{2}\mathbin{::}\Conid{Int}))\mathbin{::}\Conid{IntPair}\;\Conid{Int}{}\<[E]%
\\
\>[B]{}\texttt{> }\Conid{IntPair}\;\mathrm{11}\;\mathrm{12}{}\<[E]%
\ColumnHook
\end{hscode}\resethooks
However, when we map a function that changes the types, the monomorphic \ensuremath{\Conid{Int}}
is left alone:
\begin{hscode}\SaveRestoreHook
\column{B}{@{}>{\hspre}l<{\hspost}@{}}%
\column{E}{@{}>{\hspre}l<{\hspost}@{}}%
\>[B]{}\texttt{> }\texttt{ghci> }\;\Varid{over}\;\Varid{types}\_\;(\Varid{show}\mathbin{::}\Conid{Int}\to \Conid{String})\;(\Conid{IntPair}\;\mathrm{1}\;(\mathrm{2}\mathbin{::}\Conid{Int}))\mathbin{::}\Conid{IntPair}\;\Conid{String}{}\<[E]%
\\
\>[B]{}\texttt{> }\Conid{IntPair}\;\mathrm{1}\;\text{\ttfamily \char34 2\char34}{}\<[E]%
\ColumnHook
\end{hscode}\resethooks
While technically the correct behaviour, it is rather confusing.
Therefore we restrict this combinator to only allow monomorphic updates.
\begin{hscode}\SaveRestoreHook
\column{B}{@{}>{\hspre}l<{\hspost}@{}}%
\column{E}{@{}>{\hspre}l<{\hspost}@{}}%
\>[B]{}\keyword{type}\;\Conid{HasTypes}\;\Varid{s}\;\Varid{a}\mathrel{=}\Conid{HasTypes}\_\;\Varid{s}\;\Varid{s}\;\Varid{a}\;\Varid{a}{}\<[E]%
\\
\>[B]{}\Varid{types}\mathbin{::}\forall \Varid{a}\hsforall \;\Varid{s}\hsdot{\mathbin{\cdot}}{.}\Conid{HasTypes}\;\Varid{s}\;\Varid{a}\Rightarrow \Conid{Traversal}\;\Varid{s}\;\Varid{s}\;\Varid{a}\;\Varid{a}{}\<[E]%
\\
\>[B]{}\Varid{types}\mathrel{=}\Varid{types}\_{}\<[E]%
\ColumnHook
\end{hscode}\resethooks
Another thing
to note is the abundance of explicit type annotations in the above examples.
The promise of type inference is that the type of functions can be inferred
from their use without providing type signatures.

However, here the type really is an \emph{input} to the function, as it
determines the nature of the traversal.
The type abstraction \ensuremath{\forall \Varid{a}\hsforall } can be instantiated using visible type
applications \cite{Eisenberg:2016:visible} to provide hints to the
compiler. This allows the much more concise form
of application:
\begin{hscode}\SaveRestoreHook
\column{B}{@{}>{\hspre}l<{\hspost}@{}}%
\column{E}{@{}>{\hspre}l<{\hspost}@{}}%
\>[B]{}\texttt{> }\texttt{ghci> }\;\Varid{over}\;(\Varid{types}\ \texttt{@}\Conid{Int})\;(\mathbin{+}\mathrm{10})\;(\Conid{IntPair}\;\mathrm{1}\;(\mathrm{2}\mathbin{::}\Conid{Int})){}\<[E]%
\\
\>[B]{}\texttt{> }\Conid{IntPair}\;\mathrm{11}\;\mathrm{12}{}\<[E]%
\ColumnHook
\end{hscode}\resethooks
Note that we carefully chose the order in which to quantify the type variables
for \ensuremath{\Varid{types}}: it is much more common to provide the targeted type than the
structure's type.

%%%%%%%%%%%%%%%%%%%%%%%%%%%%%%%%%%%%%%%%%%%%%%%%%%%%%%%%%%%%%%%%%%%%%%%%%%%%%%%

\subsection{More efficient traversals}

Let us now evaluate our traversal so far. Consider the
following simple datatype:
\begin{hscode}\SaveRestoreHook
\column{B}{@{}>{\hspre}l<{\hspost}@{}}%
\column{E}{@{}>{\hspre}l<{\hspost}@{}}%
\>[B]{}\keyword{data}\;\Conid{T}\mathrel{=}\Conid{MkT}\;\Conid{Int}\;\Conid{String}\;(\Conid{Maybe}\;\Conid{Bool}){}\<[E]%
\ColumnHook
\end{hscode}\resethooks
Suppose that we wanted to collect in a list all the \ensuremath{\Conid{Int}} values in a given
\ensuremath{\Conid{T}}, as follows:
\begin{hscode}\SaveRestoreHook
\column{B}{@{}>{\hspre}l<{\hspost}@{}}%
\column{E}{@{}>{\hspre}l<{\hspost}@{}}%
\>[B]{}\texttt{> }\texttt{ghci> }\;\Varid{toListOf}\;(\Varid{types}\ \texttt{@}\Conid{Int})\;(\Conid{MkT}\;\mathrm{10}\;\text{\ttfamily \char34 a~long~string\char34}\;(\Conid{Just}\;\Conid{True})){}\<[E]%
\\
\>[B]{}\texttt{> }[\mskip1.5mu \mathrm{10}\mskip1.5mu]{}\<[E]%
\ColumnHook
\end{hscode}\resethooks
Clearly, the only interesting field is the first one. At runtime, there is no
need to inspect either the \ensuremath{\Conid{String}} field or the \ensuremath{\Conid{Maybe}\;\Conid{Bool}} field.
However, our naive implementation does inspect both. Worse, it traverses
the whole string, character by character! In the \ensuremath{\Conid{Maybe}\;\Conid{Bool}} case, the
compiler's inliner comes to the rescue: by inlining the traversal's definition
sufficiently many times, it is able to tell that only the \ensuremath{\Varid{pure}}
function is ever called, thus the whole field can be skipped.

The bigger problem is the string as the \ensuremath{\Conid{String}} type in Haskell is defined as
a linked list of characters, the generated traversal is recursive. As such, there
is no hope that the inliner could ever work out that this traversal is fruitless.
For large types that contain many recursive subparts, the performance penalty
is significant.

\paragraph{Interesting types}
In order to avoid this penalty at runtime, we need to identify at compile-time
which subparts need to be traversed; we call these subparts ``interesting''.
A type is interesting if it immediately contains the queried
type or it contains other interesting types. Crucially, a mutually recursive
group that does not contain the queried type need not be traversed at runtime
-- our predicate aims to filter out precisely these cases.

We proceed by defining the ``interesting'' predicate inductively on the type of the
generic structure. To express this type-level computation, we turn to the
\ensuremath{\Conid{Interesting}} closed type family. Closed type
families~\cite{Eisenberg:2014:closed} comprise an ordered set of potentially overlapping type equations.

The first two arguments are the generic structure and the queried type. The
third argument keeps a list of already seen types. This is to break loops in
case of (mutually) recursive types.
\savecolumns
\begin{hscode}\SaveRestoreHook
\column{B}{@{}>{\hspre}l<{\hspost}@{}}%
\column{3}{@{}>{\hspre}l<{\hspost}@{}}%
\column{31}{@{}>{\hspre}l<{\hspost}@{}}%
\column{35}{@{}>{\hspre}l<{\hspost}@{}}%
\column{44}{@{}>{\hspre}l<{\hspost}@{}}%
\column{E}{@{}>{\hspre}l<{\hspost}@{}}%
\>[B]{}\keyword{type}\;\keyword{family}\;\Conid{Interesting}\;(\Varid{rep}\mathbin{::}\star)\;(\Varid{a}\mathbin{::}\star)\;(\Varid{seen}\mathbin{::}[\mskip1.5mu \star\mskip1.5mu])\mathbin{::}\Conid{Bool}\;\keyword{where}{}\<[E]%
\\
\>[B]{}\hsindent{3}{}\<[3]%
\>[3]{}\Conid{Interesting}\;(\Varid{l}\mathbin{{:}{+}{:}}\Varid{r})\;{}\<[31]%
\>[31]{}\Varid{t}\;{}\<[35]%
\>[35]{}\Varid{seen}{}\<[44]%
\>[44]{}\mathrel{=}\Conid{Interesting}\;\Varid{l}\;\Varid{t}\;\Varid{seen}\mathrel{\vee}\Conid{Interesting}\;\Varid{r}\;\Varid{t}\;\Varid{seen}{}\<[E]%
\\
\>[B]{}\hsindent{3}{}\<[3]%
\>[3]{}\Conid{Interesting}\;(\Varid{l}\mathbin{{:}{\times}{:}}\Varid{r})\;{}\<[31]%
\>[31]{}\Varid{t}\;{}\<[35]%
\>[35]{}\Varid{seen}{}\<[44]%
\>[44]{}\mathrel{=}\Conid{Interesting}\;\Varid{l}\;\Varid{t}\;\Varid{seen}\mathrel{\vee}\Conid{Interesting}\;\Varid{r}\;\Varid{t}\;\Varid{seen}{}\<[E]%
\\
\>[B]{}\hsindent{3}{}\<[3]%
\>[3]{}\Conid{Interesting}\;(\Conid{K}\;\Varid{t})\;{}\<[31]%
\>[31]{}\Varid{t}\;{}\<[35]%
\>[35]{}\Varid{seen}{}\<[44]%
\>[44]{}\mathrel{=}{}^\prime\!\Conid{True}{}\<[E]%
\\
\>[B]{}\hsindent{3}{}\<[3]%
\>[3]{}\Conid{Interesting}\;(\Conid{K}\;\Conid{Char})\;{}\<[31]%
\>[31]{}\anonymous \;{}\<[35]%
\>[35]{}\anonymous {}\<[44]%
\>[44]{}\mathrel{=}{}^\prime\!\Conid{False}{}\<[E]%
\ColumnHook
\end{hscode}\resethooks
\vskip -\belowdisplayskip
\vskip -\abovedisplayskip
\restorecolumns
\begin{hscode}\SaveRestoreHook
\column{B}{@{}>{\hspre}l<{\hspost}@{}}%
\column{3}{@{}>{\hspre}c<{\hspost}@{}}%
\column{3E}{@{}l@{}}%
\column{E}{@{}>{\hspre}l<{\hspost}@{}}%
\>[3]{}\mathbin{...}{}\<[3E]%
\ColumnHook
\end{hscode}\resethooks
\vskip -\belowdisplayskip
\vskip -\abovedisplayskip
\restorecolumns
\begin{hscode}\SaveRestoreHook
\column{B}{@{}>{\hspre}l<{\hspost}@{}}%
\column{3}{@{}>{\hspre}l<{\hspost}@{}}%
\column{31}{@{}>{\hspre}l<{\hspost}@{}}%
\column{33}{@{}>{\hspre}l<{\hspost}@{}}%
\column{35}{@{}>{\hspre}l<{\hspost}@{}}%
\column{36}{@{}>{\hspre}l<{\hspost}@{}}%
\column{39}{@{}>{\hspre}l<{\hspost}@{}}%
\column{44}{@{}>{\hspre}l<{\hspost}@{}}%
\column{45}{@{}>{\hspre}l<{\hspost}@{}}%
\column{E}{@{}>{\hspre}l<{\hspost}@{}}%
\>[3]{}\Conid{Interesting}\;(\Conid{K}\;\Conid{Word})\;{}\<[31]%
\>[31]{}\anonymous \;{}\<[35]%
\>[35]{}\anonymous {}\<[44]%
\>[44]{}\mathrel{=}{}^\prime\!\Conid{False}{}\<[E]%
\\
\>[3]{}\Conid{Interesting}\;(\Conid{K}\;\Varid{r})\;{}\<[31]%
\>[31]{}\Varid{t}\;{}\<[35]%
\>[35]{}\Varid{seen}{}\<[44]%
\>[44]{}\mathrel{=}\Conid{InterestingUnless}\;(\Conid{Elem}\;\Varid{r}\;\Varid{seen})\;(\Conid{Rep}\;\Varid{r})\;\Varid{t}\;(\Varid{r}\;{}^\prime\!\mathbin{:}\;\Varid{seen}){}\<[E]%
\\
\>[3]{}\Conid{Interesting}\;(\Conid{M}\;\anonymous \;\Varid{f})\;{}\<[31]%
\>[31]{}\Varid{t}\;{}\<[35]%
\>[35]{}\Varid{seen}{}\<[44]%
\>[44]{}\mathrel{=}\Conid{Interesting}\;\Varid{f}\;\Varid{t}\;\Varid{seen}{}\<[E]%
\\
\>[3]{}\Conid{Interesting}\;\anonymous \;{}\<[31]%
\>[31]{}\anonymous \;{}\<[35]%
\>[35]{}\anonymous {}\<[44]%
\>[44]{}\mathrel{=}{}^\prime\!\Conid{False}{}\<[E]%
\\[\blanklineskip]%
\>[B]{}\keyword{type}\;\keyword{family}\;\Conid{InterestingUnless}\;(\Varid{s}\mathbin{::}\Conid{Bool})\;\Varid{f}\;(\Varid{a}\mathbin{::}\star)\;(\Varid{seen}\mathbin{::}[\mskip1.5mu \star\mskip1.5mu])\mathbin{::}\Conid{Bool}\;\keyword{where}{}\<[E]%
\\
\>[B]{}\hsindent{3}{}\<[3]%
\>[3]{}\Conid{InterestingUnless}\;{}^\prime\!\Conid{True}\;{}\<[33]%
\>[33]{}\anonymous \;{}\<[36]%
\>[36]{}\anonymous \;{}\<[39]%
\>[39]{}\anonymous {}\<[45]%
\>[45]{}\mathrel{=}{}^\prime\!\Conid{False}{}\<[E]%
\\
\>[B]{}\hsindent{3}{}\<[3]%
\>[3]{}\Conid{InterestingUnless}\;{}^\prime\!\Conid{False}\;{}\<[33]%
\>[33]{}\Varid{f}\;{}\<[36]%
\>[36]{}\Varid{a}\;{}\<[39]%
\>[39]{}\Varid{seen}{}\<[45]%
\>[45]{}\mathrel{=}\Conid{Interesting}\;\Varid{f}\;\Varid{a}\;\Varid{seen}{}\<[E]%
\ColumnHook
\end{hscode}\resethooks
In addition to overlapping equations, we make use of two properties of closed
type families:
\begin{enumerate}
\item
Pattern matching on members of the open type universe \ensuremath{\star} (in the \ensuremath{\Varid{rep}}
argument). This is not essential for us since we could have defined an
inductive universe for the generic constructors.
\item
Non-linear patterns: the pattern \ensuremath{\Conid{Interesting}\;(\Conid{K}\;\Varid{t})\;\Varid{t}\;\Varid{seen}} matches
when the type of the field matches the query.
\end{enumerate}
Most of the cases are self-explanatory. For \ensuremath{\mathbin{{:}{+}{:}}} and \ensuremath{\mathbin{{:}{\times}{:}}} nodes, we require
that either branch is interesting. Fields are interesting when their type
matches the query. Otherwise, for primitive types such as \ensuremath{\Conid{Char}}, we stop
looking.
The rest of the fields are inspected further, as they might contain the query.
Their inspection is done by recursively invoking the \ensuremath{\Conid{Interesting}} predicate on
their representation type. \ensuremath{\Conid{Elem}\;\Varid{r}\;\Varid{seen}} is a predicate that returns true when
the field \ensuremath{\Varid{r}} is already in the \ensuremath{\Varid{seen}} set. This recursive branch could be
written as
\begin{hscode}\SaveRestoreHook
\column{B}{@{}>{\hspre}l<{\hspost}@{}}%
\column{3}{@{}>{\hspre}l<{\hspost}@{}}%
\column{E}{@{}>{\hspre}l<{\hspost}@{}}%
\>[3]{}\Conid{Interesting}\;(\Conid{K}\;\Varid{r})\;\Varid{t}\;\Varid{seen}\mathrel{=}\Conid{If}\;(\Conid{Elem}\;\Varid{r}\;\Varid{seen})\;{}^\prime\!\Conid{False}\;(\Conid{Interesting}\;(\Conid{Rep}\;\Varid{r})\;\Varid{t}\;(\Varid{r}\;{}^\prime\!\mathbin{:}\;\Varid{seen})){}\<[E]%
\ColumnHook
\end{hscode}\resethooks
However, type families are eagerly evaluated~\cite{Vytiniotis:2011:outsidein},
so both branches of \ensuremath{\Conid{If}} are evaluated. This would be disastrous, as without
the \ensuremath{\Varid{seen}} predicate, the recursive branch would diverge for mutually recursive
groups. Instead, we implement \ensuremath{\Conid{InterestingUnless}} to encode the conditional
and ensure that the recursive case is only evaluated for fields not visited already.

We will now use \ensuremath{\Conid{Interesting}} to refine our implementation of \ensuremath{\Varid{types}} to eliminate
unnecessary runtime traversals. We introduce an auxiliary class,
\ensuremath{\Conid{HasTypesOpt}}, which is indexed by a boolean flag: whether to inspect its
argument or not.
\begin{hscode}\SaveRestoreHook
\column{B}{@{}>{\hspre}l<{\hspost}@{}}%
\column{3}{@{}>{\hspre}l<{\hspost}@{}}%
\column{5}{@{}>{\hspre}l<{\hspost}@{}}%
\column{E}{@{}>{\hspre}l<{\hspost}@{}}%
\>[B]{}\keyword{class}\;\Conid{HasTypesOpt}\;(\Varid{i}\mathbin{::}\Conid{Bool})\;\Varid{s}\;\Varid{t}\;\Varid{a}\;\Varid{b}\;\keyword{where}{}\<[E]%
\\
\>[B]{}\hsindent{3}{}\<[3]%
\>[3]{}\Varid{typesOpt}\mathbin{::}\Conid{Traversal}\;\Varid{s}\;\Varid{t}\;\Varid{a}\;\Varid{b}{}\<[E]%
\\[\blanklineskip]%
\>[B]{}\keyword{instance}\;(\Conid{Generic}\;\Varid{s},\Conid{Generic}\;\Varid{t},\Conid{GHasTypes}\;(\Conid{Rep}\;\Varid{s})\;(\Conid{Rep}\;\Varid{t})\;\Varid{a}\;\Varid{b}){}\<[E]%
\\
\>[B]{}\hsindent{5}{}\<[5]%
\>[5]{}\Rightarrow \Conid{HasTypesOpt}\;{}^\prime\!\Conid{True}\;\Varid{s}\;\Varid{t}\;\Varid{a}\;\Varid{b}\;\keyword{where}{}\<[E]%
\\
\>[B]{}\hsindent{3}{}\<[3]%
\>[3]{}\Varid{typesOpt}\mathrel{=}\Varid{iso}_{\Conid{Rep}}\hsdot{\mathbin{\cdot}}{.}\Varid{gtypes}{}\<[E]%
\\[\blanklineskip]%
\>[B]{}\keyword{instance}\;\Conid{HasTypesOpt}\;{}^\prime\!\Conid{False}\;\Varid{s}\;\Varid{s}\;\Varid{a}\;\Varid{b}\;\keyword{where}{}\<[E]%
\\
\>[B]{}\hsindent{3}{}\<[3]%
\>[3]{}\Varid{typesOpt}\;\anonymous \mathrel{=}\Varid{pure}{}\<[E]%
\ColumnHook
\end{hscode}\resethooks
We now revise the default \ensuremath{\Conid{HasTypes}\_} instance by dispatching to the helper
\ensuremath{\Conid{HasTypesOpt}} with the flag set to the \ensuremath{\Conid{Interesting}} predicate applied to the
structure.
\begin{hscode}\SaveRestoreHook
\column{B}{@{}>{\hspre}l<{\hspost}@{}}%
\column{3}{@{}>{\hspre}l<{\hspost}@{}}%
\column{5}{@{}>{\hspre}l<{\hspost}@{}}%
\column{E}{@{}>{\hspre}l<{\hspost}@{}}%
\>[B]{}\keyword{instance}\;(\Conid{Generic}\;\Varid{s},\Conid{Generic}\;\Varid{t},\Conid{HasTypesOpt}\;(\Conid{Interesting}\;(\Conid{Rep}\;\Varid{s})\;\Varid{a}\;{}^\prime\![\mskip1.5mu \Varid{s}\mskip1.5mu])\;\Varid{s}\;\Varid{t}\;\Varid{a}\;\Varid{b}){}\<[E]%
\\
\>[B]{}\hsindent{5}{}\<[5]%
\>[5]{}\Rightarrow \Conid{HasTypes}\_\;\Varid{s}\;\Varid{t}\;\Varid{a}\;\Varid{b}\;\keyword{where}{}\<[E]%
\\
\>[B]{}\hsindent{3}{}\<[3]%
\>[3]{}\Varid{types}\_\mathrel{=}\Varid{typesOpt}\ \texttt{@}(\Conid{Interesting}\;(\Conid{Rep}\;\Varid{s})\;\Varid{a}\;{}^\prime\![\mskip1.5mu \Varid{s}\mskip1.5mu]){}\<[E]%
\ColumnHook
\end{hscode}\resethooks
Now depending on the result of computing \ensuremath{\Conid{Interesting}} for our type, we will
either stop traversing or carry on recursively if we need to traverse
more values. This saves us from performing unecessary work as was the case
in our example of traversing \ensuremath{\Conid{T}}.

The approach outlined above works for many polymorphic recursive data
types, but has its limitations. Consider the type of perfectly
balanced trees:
\begin{hscode}\SaveRestoreHook
\column{B}{@{}>{\hspre}l<{\hspost}@{}}%
\column{E}{@{}>{\hspre}l<{\hspost}@{}}%
\>[B]{}\keyword{data}\;\Conid{Perfect}\;\Varid{a}\mathrel{=}\Conid{Single}\;\Varid{a}\mid \Conid{Balanced}\;(\Conid{Perfect}\;(\Varid{a},\Varid{a})){}\<[E]%
\ColumnHook
\end{hscode}\resethooks
We can consider this to be divergent polymorphic recursion in the
sense that the type changes at every level, and that no finite fix
point can be found. To tackle such cases we might consider augmenting
our predicate with a number that acts as a depth bound.

% \begin{figure}[H]
% \centering
% \begin{subfigure}[t]{.49\textwidth}
%     \centering
%     \begin{core}
% updateFlatProdInt (FlatProd _ is s) =
%   FlatProd
%     0 (over (types @Int) is) (over (types @Int) s)
%     \end{core}
%     \caption{Naive}
%     \label{fig:naive}
% \end{subfigure}
% ~
% \begin{subfigure}[t]{.49\textwidth}
% \centering
%     \begin{core}
% updateFlatProdInt (FlatProd _ is s) =
%   FlatProd 0 (over (types @Int) is) s
%     \end{core}
%     \caption{Optimised}
%     \label{fig:optimised}
% \end{subfigure}
% \end{figure}

\section{Generic Traversals with Parameters}
\label{sec:params}

In this section we implement the \ensuremath{\Varid{param}} traversal which focuses on all
values corresponding to a type parameter. The motivation is to
be able to derive a traversal which is able to change the type of the elements
in a container. Recall that for \ensuremath{\Varid{types}} we disallowed this as the
behaviour of the traversal is not easy to specify. On the other hand, \ensuremath{\Varid{param}}
doesn't have this problem as we precisely change all the positions which
are necessary in order to change the type of a parameter.

Implementing \ensuremath{\Varid{param}} poses a number of new challenges:
\begin{description}
\item[Locating parameters]
Our traversal must distinguish between values that correspond to the queried
type parameter, and values that were monomorphically defined (or correspond to
other parameters). Even (and especially) when the query is not type-changing,
we must not confuse a monomorphic \ensuremath{\Conid{Int}} with a type parameter instantiated to
\ensuremath{\Conid{Int}}.
\item[Multiple parameters]
We want \ensuremath{\Varid{param}} to work for any number of type parameters, as opposed
to the special case where there is just one.
\item[Type inference]
When specifying a type-changing traversal, we need to infer how the type of the
structure will change.
\end{description}

Our solution to these problems uses only the \ensuremath{\Conid{Generic}} class and a set of type-level algorithms.
Previous approaches~\cite{Magalhaes:2010:generic} extended the generic representation
to allow working with a single type parameter, but our approach is more flexible
and uses existing machinery.

\subsection{Locating parameters}
We first tackle the problems of locating parameters and dealing with multiple
parameters simultaneously.
The type family \ensuremath{\Conid{Rep}} in the \ensuremath{\Conid{Generic}} class takes as argument a type with kind
\ensuremath{\star}. This means that it must be called on type constructors that are fully
saturated.

Recall the \ensuremath{\Conid{Invoice}\;\Varid{a}} type from our biscuit factory. When in the normal queue
case it is instantiated to \ensuremath{\Conid{Invoice}\;\Conid{Int}}, \ensuremath{\Conid{Rep}} is
unable to distinguish between the \ensuremath{\Varid{priority}} and \ensuremath{\Varid{number}} fields by their
types, as both are \ensuremath{\Conid{Int}}s, since \ensuremath{\Conid{Rep}\;(\Conid{Invoice}\;\Conid{Int})\equiv\Conid{K}\;\Conid{Item}\mathbin{{:}{\times}{:}}\Conid{K}\;\Conid{String}\mathbin{{:}{\times}{:}}\Conid{K}\;\Conid{Int}\mathbin{{:}{\times}{:}}\Conid{K}\;\Conid{Int}}.  In order to implement \ensuremath{\Varid{param}} we need to be
able to identify which is which.

To solve this, we tag each type parameter with
a unique index, corresponding to the parameter's position in the original type.
We can then track the position of each type parameter in the generic representation.
The \ensuremath{\Conid{Param}} newtype wraps a value of type \ensuremath{\Varid{a}}. It is indexed by a type-level
natural representing which parameter it corresponds to.
\begin{hscode}\SaveRestoreHook
\column{B}{@{}>{\hspre}l<{\hspost}@{}}%
\column{E}{@{}>{\hspre}l<{\hspost}@{}}%
\>[B]{}\keyword{newtype}\;\Conid{Param}\;(\Varid{i}\mathbin{::}\Conid{Nat})\;\Varid{a}\mathrel{=}\Conid{Param}\;\{\mskip1.5mu \Varid{unParam}\mathbin{::}\Varid{a}\mskip1.5mu\}{}\<[E]%
\ColumnHook
\end{hscode}\resethooks
Then, given any concrete instantiation of a type, we iterate through its type
parameters and wrap them in \ensuremath{\Conid{Param}} constructors with increasing indices.
The \ensuremath{\Conid{Index}} type family does that.
\begin{hscode}\SaveRestoreHook
\column{B}{@{}>{\hspre}l<{\hspost}@{}}%
\column{3}{@{}>{\hspre}l<{\hspost}@{}}%
\column{19}{@{}>{\hspre}l<{\hspost}@{}}%
\column{E}{@{}>{\hspre}l<{\hspost}@{}}%
\>[B]{}\keyword{type}\;\keyword{family}\;\Conid{Index}\;(\Varid{t}\mathbin{::}\Varid{k})\;(\Varid{i}\mathbin{::}\Conid{Nat})\mathbin{::}\Varid{k}\;\keyword{where}{}\<[E]%
\\
\>[B]{}\hsindent{3}{}\<[3]%
\>[3]{}\Conid{Index}\;(\Varid{t}\;\Varid{a})\;\Varid{i}{}\<[19]%
\>[19]{}\mathrel{=}\Conid{Index}\;\Varid{t}\;(\Varid{i}\mathbin{+}\mathrm{1})\;(\Conid{Param}\;\Varid{i}\;\Varid{a}){}\<[E]%
\\
\>[B]{}\hsindent{3}{}\<[3]%
\>[3]{}\Conid{Index}\;\Varid{t}\;\anonymous {}\<[19]%
\>[19]{}\mathrel{=}\Varid{t}{}\<[E]%
\ColumnHook
\end{hscode}\resethooks

This conversion allows us to track the parameters:
\begin{hscode}\SaveRestoreHook
\column{B}{@{}>{\hspre}l<{\hspost}@{}}%
\column{E}{@{}>{\hspre}l<{\hspost}@{}}%
\>[B]{}\Conid{Index}\;(\Conid{Invoice}\;\Conid{Int})\;\mathrm{0}\equiv\Conid{Invoice}\;(\Conid{Param}\;\mathrm{0}\;\Conid{Int}){}\<[E]%
\\
\>[B]{}\Conid{Index}\;(\Conid{Either}\;\Conid{Int}\;\Conid{String})\;\mathrm{0}\equiv\Conid{Either}\;(\Conid{Param}\;\mathrm{1}\;\Conid{Int})\;(\Conid{Param}\;\mathrm{0}\;\Conid{String}){}\<[E]%
\ColumnHook
\end{hscode}\resethooks
Note that numbering starts at the last parameter, as it is the outermost one.
With the new indexing in place, determining the origin of types in the generic
representation is no longer a problem. If the type is wrapped in a \ensuremath{\Conid{Param}}
constructor, it was a type parameter, otherwise it was an ordinary field.
\begin{hscode}\SaveRestoreHook
\column{B}{@{}>{\hspre}l<{\hspost}@{}}%
\column{E}{@{}>{\hspre}l<{\hspost}@{}}%
\>[B]{}\Conid{Rep}\;(\Conid{Index}\;(\Conid{Invoice}\;\Conid{Int})\;\mathrm{0})\equiv\Conid{K}\;\Conid{Item}\mathbin{{:}{\times}{:}}\Conid{K}\;\Conid{String}\mathbin{{:}{\times}{:}}\Conid{K}\;\Conid{Int}\mathbin{{:}{\times}{:}}\Conid{K}\;(\Conid{Param}\;\mathrm{0}\;\Conid{Int}){}\<[E]%
\ColumnHook
\end{hscode}\resethooks

Only one problem remains: the functions \ensuremath{\Varid{to}} and \ensuremath{\Varid{from}} operate on \ensuremath{\Conid{Rep}\;\Varid{a}}.
\ensuremath{\Varid{to}\mathbin{::}\Conid{Rep}\;\Varid{a}\to \Varid{a}} -- how do we turn this into \ensuremath{\Conid{Rep}\;(\Conid{Index}\;\Varid{a}\;\mathrm{0})\to \Varid{a}}?
\ensuremath{\Conid{Rep}\;(\Conid{Index}\;\Varid{a}\;\mathrm{0})} extends \ensuremath{\Conid{Rep}\;\Varid{a}} by wrapping certain fields in the
\ensuremath{\Conid{Param}} newtype. Newtype wrappers have no runtime representation, which means
that \ensuremath{\Conid{Rep}\;(\Conid{Index}\;\Varid{a}\;\mathrm{0})} and \ensuremath{\Conid{Rep}\;\Varid{a}} are representationally equal: they are
the same at runtime. This means that they can be safely
coerced~\cite{Breitner:2014:safe-coercions}.

The \ensuremath{\Conid{Generic}_\Conid{N}} class can be thought of as an
extension of the \ensuremath{\Conid{Generic}} class, whose \ensuremath{\Varid{to}_\Conid{N}} and \ensuremath{\Varid{from}_\Conid{N}} functions take care
of the coercions, by requiring that \ensuremath{\Conid{Rep}\;\Varid{a}} and \ensuremath{\Conid{Rep}_\Conid{N}\;\Varid{a}} are indeed coercible.
\vskip -\abovedisplayskip
\noindent
\begin{minipage}[t]{0.5\textwidth}%
\begin{hscode}\SaveRestoreHook
\column{B}{@{}>{\hspre}l<{\hspost}@{}}%
\column{3}{@{}>{\hspre}l<{\hspost}@{}}%
\column{4}{@{}>{\hspre}l<{\hspost}@{}}%
\column{11}{@{}>{\hspre}l<{\hspost}@{}}%
\column{23}{@{}>{\hspre}l<{\hspost}@{}}%
\column{E}{@{}>{\hspre}l<{\hspost}@{}}%
\>[B]{}\keyword{class}\;(\Conid{Coercible}\;(\Conid{Rep}\;\Varid{a})\;(\Conid{Rep}_\Conid{N}\;\Varid{a}),{}\<[E]%
\\
\>[B]{}\hsindent{4}{}\<[4]%
\>[4]{}\Conid{Generic}\;\Varid{a})\Rightarrow \Conid{Generic}_\Conid{N}\;(\Varid{a}\mathbin{::}\star)\;\keyword{where}{}\<[E]%
\\
\>[B]{}\hsindent{3}{}\<[3]%
\>[3]{}\keyword{type}\;\Conid{Rep}_\Conid{N}\;\Varid{a}\mathbin{::}\star{}\<[E]%
\\
\>[B]{}\hsindent{3}{}\<[3]%
\>[3]{}\Varid{to}_\Conid{N}{}\<[11]%
\>[11]{}\mathbin{::}\Conid{Rep}_\Conid{N}\;\Varid{a}{}\<[23]%
\>[23]{}\to \Varid{a}{}\<[E]%
\\
\>[B]{}\hsindent{3}{}\<[3]%
\>[3]{}\Varid{from}_\Conid{N}{}\<[11]%
\>[11]{}\mathbin{::}\Varid{a}{}\<[23]%
\>[23]{}\to \Conid{Rep}_\Conid{N}\;\Varid{a}{}\<[E]%
\ColumnHook
\end{hscode}\resethooks
\end{minipage}
\begin{minipage}[t]{0.5\textwidth}
\begin{hscode}\SaveRestoreHook
\column{B}{@{}>{\hspre}l<{\hspost}@{}}%
\column{3}{@{}>{\hspre}l<{\hspost}@{}}%
\column{4}{@{}>{\hspre}l<{\hspost}@{}}%
\column{12}{@{}>{\hspre}l<{\hspost}@{}}%
\column{E}{@{}>{\hspre}l<{\hspost}@{}}%
\>[B]{}\keyword{instance}\;(\Conid{Coercible}\;(\Conid{Rep}\;\Varid{a})\;(\Conid{Rep}_\Conid{N}\;\Varid{a}),{}\<[E]%
\\
\>[B]{}\hsindent{4}{}\<[4]%
\>[4]{}\Conid{Generic}\;\Varid{a})\Rightarrow \Conid{Generic}_\Conid{N}\;\Varid{a}\;\keyword{where}{}\<[E]%
\\
\>[B]{}\hsindent{3}{}\<[3]%
\>[3]{}\keyword{type}\;\Conid{Rep}_\Conid{N}\;\Varid{a}\mathrel{=}\Conid{Rep}\;(\Conid{Index}\;\Varid{a}\;\mathrm{0}){}\<[E]%
\\
\>[B]{}\hsindent{3}{}\<[3]%
\>[3]{}\Varid{to}_\Conid{N}{}\<[12]%
\>[12]{}\mathrel{=}\Varid{coerce}\;(\Varid{to}\ \texttt{@}\Varid{a}){}\<[E]%
\\
\>[B]{}\hsindent{3}{}\<[3]%
\>[3]{}\Varid{from}_\Conid{N}{}\<[12]%
\>[12]{}\mathrel{=}\Varid{coerce}\;(\Varid{from}\ \texttt{@}\Varid{a}){}\<[E]%
\ColumnHook
\end{hscode}\resethooks
\end{minipage}

\noindent
To reflect, we have taken care of the first two difficulties. By using the
\ensuremath{\Conid{Index}} type family to label each of the parameter positions, we can keep track
of which fields arise from parameters and which ones do not. This also works
for any number of parameters.

Once we have identified this information,
notice that the problem of traversing the \ensuremath{\Varid{i}}th parameter can be formulated as
a typed traversal of \ensuremath{\Conid{Param}\;\Varid{i}\;\Varid{a}} over the \ensuremath{\Conid{Index}}ed view of the structure.
We can now implement the \ensuremath{\Varid{param}} traversal:
\begin{hscode}\SaveRestoreHook
\column{B}{@{}>{\hspre}l<{\hspost}@{}}%
\column{3}{@{}>{\hspre}l<{\hspost}@{}}%
\column{E}{@{}>{\hspre}l<{\hspost}@{}}%
\>[B]{}\keyword{instance}\;(\Conid{Generic}_\Conid{N}\;\Varid{s},\Conid{Generic}_\Conid{N}\;\Varid{t},\Conid{GHasTypes}\;(\Conid{Rep}_\Conid{N}\;\Varid{s})\;(\Conid{Rep}_\Conid{N}\;\Varid{t})\;(\Conid{Param}\;\Varid{i}\;\Varid{a})\;(\Conid{Param}\;\Varid{i}\;\Varid{b})){}\<[E]%
\\
\>[B]{}\hsindent{3}{}\<[3]%
\>[3]{}\Rightarrow \Conid{HasParam}\;\Varid{i}\;\Varid{s}\;\Varid{t}\;\Varid{a}\;\Varid{b}\;\keyword{where}{}\<[E]%
\\
\>[B]{}\hsindent{3}{}\<[3]%
\>[3]{}\Varid{param}\mathrel{=}\Varid{iso}_{\Conid{Rep}_\Conid{N}}\hsdot{\mathbin{\cdot}}{.}\Varid{gtypes}\hsdot{\mathbin{\cdot}}{.}\Varid{paramIso}\ \texttt{@}\Varid{i}{}\<[E]%
\ColumnHook
\end{hscode}\resethooks
Here, \ensuremath{\Varid{paramIso}} is the lens that focuses on the values by forgetting the
\ensuremath{\Conid{Param}\;\Varid{i}} wrapper.
\begin{hscode}\SaveRestoreHook
\column{B}{@{}>{\hspre}l<{\hspost}@{}}%
\column{E}{@{}>{\hspre}l<{\hspost}@{}}%
\>[B]{}\Varid{paramIso}\mathbin{::}\forall \Varid{i}\hsforall \;\Varid{a}\;\Varid{b}\hsdot{\mathbin{\cdot}}{.}\Conid{Lens}\;(\Conid{Param}\;\Varid{i}\;\Varid{a})\;(\Conid{Param}\;\Varid{i}\;\Varid{b})\;\Varid{a}\;\Varid{b}{}\<[E]%
\\
\>[B]{}\Varid{paramIso}\;\Varid{f}\;\Varid{p}\mathrel{=}\Conid{Param}\mathbin{{\langle}{\$}{\rangle}}\Varid{f}\;(\Varid{unParam}\;\Varid{p}){}\<[E]%
\ColumnHook
\end{hscode}\resethooks
The function \ensuremath{\Varid{iso}_{\Conid{Rep}_\Conid{N}}} is analogous to \ensuremath{\Varid{iso}_{\Conid{Rep}}}, but for the new \ensuremath{\Conid{Rep}_\Conid{N}} representation.
\begin{hscode}\SaveRestoreHook
\column{B}{@{}>{\hspre}l<{\hspost}@{}}%
\column{E}{@{}>{\hspre}l<{\hspost}@{}}%
\>[B]{}\Varid{iso}_{\Conid{Rep}_\Conid{N}}\mathbin{::}(\Conid{Generic}_\Conid{N}\;\Varid{s},\Conid{Generic}_\Conid{N}\;\Varid{t})\Rightarrow \Conid{Lens}\;\Varid{s}\;\Varid{t}\;(\Conid{Rep}_\Conid{N}\;\Varid{s})\;(\Conid{Rep}_\Conid{N}\;\Varid{t}){}\<[E]%
\\
\>[B]{}\Varid{iso}_{\Conid{Rep}_\Conid{N}}\;\Varid{f}\;\Varid{s}\mathrel{=}\Varid{to}_\Conid{N}\mathbin{{\langle}{\$}{\rangle}}\Varid{f}\;(\Varid{from}_\Conid{N}\;\Varid{s}){}\<[E]%
\ColumnHook
\end{hscode}\resethooks

Now we turn to the problem of improving type inference for our type
changing traversals.

\subsection{Type inference}
The four parameters of \ensuremath{\Conid{Traversal}\;\Varid{s}\;\Varid{t}\;\Varid{a}\;\Varid{b}} have interesting
connections, and even from partial information we can infer the rest.
More formally, we define the \ensuremath{\Conid{HasParam}\;\Varid{i}\;\Varid{s}\;\Varid{t}\;\Varid{a}\;\Varid{b}} class with a single function,
\ensuremath{\Varid{param}\mathbin{::}\Conid{Traversal}\;\Varid{s}\;\Varid{t}\;\Varid{a}\;\Varid{b}}, which describes the traversal of the \ensuremath{\Varid{i}}th type
parameter, \ensuremath{\Varid{a}}, of \ensuremath{\Varid{s}}. \ensuremath{\Varid{t}} is the result of changing the \ensuremath{\Varid{i}}th parameter of
\ensuremath{\Varid{s}} to \ensuremath{\Varid{b}}.
\begin{hscode}\SaveRestoreHook
\column{B}{@{}>{\hspre}l<{\hspost}@{}}%
\column{3}{@{}>{\hspre}l<{\hspost}@{}}%
\column{E}{@{}>{\hspre}l<{\hspost}@{}}%
\>[B]{}\keyword{class}\;\Conid{HasParam}\;(\Varid{i}\mathbin{::}\Conid{Nat})\;\Varid{s}\;\Varid{t}\;\Varid{a}\;\Varid{b}\mid \Varid{i}\;\Varid{s}\;\Varid{b}\to \Varid{t},\Varid{i}\;\Varid{s}\to \Varid{a},\Varid{i}\;\Varid{t}\;\Varid{a}\to \Varid{s},\Varid{i}\;\Varid{t}\to \Varid{b}\;\keyword{where}{}\<[E]%
\\
\>[B]{}\hsindent{3}{}\<[3]%
\>[3]{}\Varid{param}\mathbin{::}\Conid{Traversal}\;\Varid{s}\;\Varid{t}\;\Varid{a}\;\Varid{b}{}\<[E]%
\ColumnHook
\end{hscode}\resethooks
In order to resolve which instance of \ensuremath{\Conid{HasParam}} to use, we must know the types
of all five type parameters. The user is expected to provide \ensuremath{\Varid{i}} by using type
applications but the other four can be inferred in different situations.

The four functional dependencies~\cite{Sulzmann:2007:fundeps-chr} each specify
what type information can be inferred if some of the types are known. They
act as a specification as to what relationship must hold between the type
variables in each instance.
\begin{description}
\item[\ensuremath{\Varid{i}\;\Varid{s}\;\Varid{b}\to \Varid{t}}]
The source type and modification function are known. Then we can
uniquely determine the target type \ensuremath{\Varid{t}}.
This ensures that if we provide \ensuremath{\Varid{i}} as a type argument and then fully
apply the traversal, we can infer the result type.
\item[\ensuremath{\Varid{i}\;\Varid{s}\to \Varid{a}}]
From a position and the source type only, we can uniquely determine the type of
the parameter at that position.
\item[\ensuremath{\Varid{i}\;\Varid{t}\;\Varid{a}\to \Varid{s}}]
The result type and modification function are known. Then
we can uniquely determine the source type \ensuremath{\Varid{s}}.
This dependency helps us when composing together traversals where we would
otherwise encounter ambiguous type variables in the middle of the composition.
\item[\ensuremath{\Varid{i}\;\Varid{t}\to \Varid{b}}]
From a position and the target type only, we can uniquely determine the type of
the parameter at that position.
\end{description}
Without these functional dependencies it would be very difficult to use
these optics without explicitly writing type signatures.

Now, we must modify the instance for \ensuremath{\Conid{HasParam}} which we defined above
in order to implement the stated functional dependencies. The way that
we prove to the compiler that each functional dependency holds is by defining
a type family which witnesses each assertion.

There are two kind of
dependencies: \ensuremath{\Varid{i}\;\Varid{s}\to \Varid{a}} and \ensuremath{\Varid{i}\;\Varid{t}\to \Varid{b}} both \textit{get} the parameter at
index \ensuremath{\Varid{i}}, while \ensuremath{\Varid{i}\;\Varid{s}\;\Varid{b}\to \Varid{t}} and \ensuremath{\Varid{i}\;\Varid{t}\;\Varid{a}\to \Varid{s}} both \textit{set} it.
In order to assign an operational meaning to these functional dependencies, we
define two type families that express the getting and the setting relations
respectively.
First, \ensuremath{\Conid{GetParam}} peels off the parameters of its argument one by one until it
reaches the \ensuremath{\Varid{i}}th.
\begin{hscode}\SaveRestoreHook
\column{B}{@{}>{\hspre}l<{\hspost}@{}}%
\column{3}{@{}>{\hspre}l<{\hspost}@{}}%
\column{E}{@{}>{\hspre}l<{\hspost}@{}}%
\>[B]{}\keyword{type}\;\keyword{family}\;\Conid{GetParam}\;(\Varid{t}\mathbin{::}\Varid{k})\;(\Varid{i}\mathbin{::}\Conid{Nat})\mathbin{::}\star\;\keyword{where}{}\<[E]%
\\
\>[B]{}\hsindent{3}{}\<[3]%
\>[3]{}\Conid{GetParam}\;(\Varid{t}\;\Varid{a})\;\mathrm{0}\mathrel{=}\Varid{a}{}\<[E]%
\\
\>[B]{}\hsindent{3}{}\<[3]%
\>[3]{}\Conid{GetParam}\;(\Varid{t}\;\anonymous )\;\Varid{i}\mathrel{=}\Conid{GetParam}\;\Varid{t}\;(\Varid{i}\mathbin{-}\mathrm{1}){}\<[E]%
\ColumnHook
\end{hscode}\resethooks
Similarly, \ensuremath{\Conid{PutParam}} digs into its argument to find and the \ensuremath{\Varid{i}}th parameter.
\begin{hscode}\SaveRestoreHook
\column{B}{@{}>{\hspre}l<{\hspost}@{}}%
\column{3}{@{}>{\hspre}l<{\hspost}@{}}%
\column{E}{@{}>{\hspre}l<{\hspost}@{}}%
\>[B]{}\keyword{type}\;\keyword{family}\;\Conid{PutParam}\;(\Varid{t}\mathbin{::}\Varid{k})\;(\Varid{i}\mathbin{::}\Conid{Nat})\;(\Varid{b}\mathbin{::}\star)\mathbin{::}\Varid{k}\;\keyword{where}{}\<[E]%
\\
\>[B]{}\hsindent{3}{}\<[3]%
\>[3]{}\Conid{PutParam}\;(\Varid{t}\;\anonymous )\;\mathrm{0}\;\Varid{b}\mathrel{=}\Varid{t}\;\Varid{b}{}\<[E]%
\\
\>[B]{}\hsindent{3}{}\<[3]%
\>[3]{}\Conid{PutParam}\;(\Varid{t}\;\Varid{a})\;\Varid{i}\;\Varid{b}\mathrel{=}(\Conid{PutParam}\;\Varid{t}\;(\Varid{i}\mathbin{-}\mathrm{1})\;\Varid{b})\;\Varid{a}{}\<[E]%
\ColumnHook
\end{hscode}\resethooks
\ck{could remove this paragraph}
Notice that both \ensuremath{\Conid{GetParam}} and \ensuremath{\Conid{PutParam}} operate on poly-kinded arguments,
even though we only intend to call them on types of kind \ensuremath{\star}. However, as
both functions peel off the arguments, intermediate recursive calls operate on
higher-kinded types.
\begin{hscode}\SaveRestoreHook
\column{B}{@{}>{\hspre}l<{\hspost}@{}}%
\column{5}{@{}>{\hspre}l<{\hspost}@{}}%
\column{E}{@{}>{\hspre}l<{\hspost}@{}}%
\>[5]{}\Conid{GetParam}\;(\Conid{Either}\;\Conid{Int}\;\Conid{String}\mathbin{::}\star)\;\mathrm{1}\equiv\Conid{GetParam}\;(\Conid{Either}\;\Conid{Int}\mathbin{::}\star\to \star)\;\mathrm{0}\equiv\Conid{Int}{}\<[E]%
\ColumnHook
\end{hscode}\resethooks
\ensuremath{\Conid{GetParam}} and \ensuremath{\Conid{PutParam}} highlight another important aspect of closed type
families: they are not parametric, as we can match on arguments that have
polymorphic kinds. They can also decompose application forms, as in the \ensuremath{\Varid{t}\;\Varid{a}}
pattern~\cite{Weirich:2011:GTA}.

Now we have a method of proving these dependencies. We supply the proofs
as instance constraints which allows the compiler to conclude the validity of
the functional dependencies. This leaves us with the final definition for
\ensuremath{\Varid{param}}.
\begin{hscode}\SaveRestoreHook
\column{B}{@{}>{\hspre}l<{\hspost}@{}}%
\column{3}{@{}>{\hspre}l<{\hspost}@{}}%
\column{E}{@{}>{\hspre}l<{\hspost}@{}}%
\>[B]{}\keyword{instance}\;(\Varid{a}\,\sim\,\Conid{GetParam}\;\Varid{s}\;\Varid{i},\Varid{b}\,\sim\,\Conid{GetParam}\;\Varid{t}\;\Varid{i},\Varid{t}\,\sim\,\Conid{PutParam}\;\Varid{s}\;\Varid{i}\;\Varid{b},\Varid{s}\,\sim\,\Conid{PutParam}\;\Varid{t}\;\Varid{i}\;\Varid{a},{}\<[E]%
\\
\>[B]{}\hsindent{3}{}\<[3]%
\>[3]{}\Conid{Generic}_\Conid{N}\;\Varid{s},\Conid{Generic}_\Conid{N}\;\Varid{t},\Conid{GHasTypes}\;(\Conid{Rep}_\Conid{N}\;\Varid{s})\;(\Conid{Rep}_\Conid{N}\;\Varid{t})\;(\Conid{Param}\;\Varid{i}\;\Varid{a})\;(\Conid{Param}\;\Varid{i}\;\Varid{b})){}\<[E]%
\\
\>[B]{}\hsindent{3}{}\<[3]%
\>[3]{}\Rightarrow \Conid{HasParam}\;\Varid{i}\;\Varid{s}\;\Varid{t}\;\Varid{a}\;\Varid{b}\;\keyword{where}{}\<[E]%
\\
\>[B]{}\hsindent{3}{}\<[3]%
\>[3]{}\Varid{param}\mathrel{=}\Varid{iso}_{\Conid{Rep}_\Conid{N}}\hsdot{\mathbin{\cdot}}{.}\Varid{gtypes}\hsdot{\mathbin{\cdot}}{.}\Varid{paramIso}\ \texttt{@}\Varid{i}{}\<[E]%
\ColumnHook
\end{hscode}\resethooks

As an example, consider the \ensuremath{\Conid{Poly}\;\Varid{a}\;\Varid{b}} type, which is a list that alternates between
elements of type \ensuremath{\Varid{a}} and type \ensuremath{\Varid{b}} (note the polymorphic recursion in the tail).
\begin{hscode}\SaveRestoreHook
\column{B}{@{}>{\hspre}l<{\hspost}@{}}%
\column{E}{@{}>{\hspre}l<{\hspost}@{}}%
\>[B]{}\keyword{data}\;\Conid{Poly}\;\Varid{a}\;\Varid{b}\mathrel{=}\Conid{PNil}\mid \Conid{PCons}\;\Varid{a}\;(\Conid{Poly}\;\Varid{b}\;\Varid{a}){}\<[E]%
\ColumnHook
\end{hscode}\resethooks
With \ensuremath{\Varid{param}}, we can specify a traversal that updates the \ensuremath{\Conid{String}}s
that correspond to the \ensuremath{\Varid{a}} parameter:
\begin{hscode}\SaveRestoreHook
\column{B}{@{}>{\hspre}l<{\hspost}@{}}%
\column{E}{@{}>{\hspre}l<{\hspost}@{}}%
\>[B]{}\texttt{> }\texttt{ghci> }\;\Varid{over}\;(\Varid{param}\ \texttt{@}\mathrm{1})\;\Varid{length}\;(\Conid{PCons}\;\text{\ttfamily \char34 wafer\char34}\;(\Conid{PCons}\;\text{\ttfamily \char34 oreo\char34}\;(\Conid{PCons}\;\text{\ttfamily \char34 nice\char34}\;\Conid{PNil}))){}\<[E]%
\\
\>[B]{}\texttt{> }\Conid{PCons}\;\mathrm{5}\;(\Conid{PCons}\;\text{\ttfamily \char34 oreo\char34}\;(\Conid{PCons}\;\mathrm{4}\;\Conid{PNil})){}\<[E]%
\ColumnHook
\end{hscode}\resethooks

\section{Generic Traversals with Class}
\label{sec:constraints}

Inspecting the inductive definition of \ensuremath{\Conid{GHasTypes}\;\Varid{s}\;\Varid{t}\;\Varid{a}\;\Varid{b}} in
Section~\ref{sec:types}, we see that all the inductive cases do is merely
``forward the focus''~\ck{this sounds a bit odd?} to their children. The first
time any decision is made is at \ensuremath{\Conid{K}}: whether to stop, or keep going via the
mutually recursive \ensuremath{\Conid{HasTypes}\_} class. Defining traversals that employ a
different operation on fields would require writing a very similar inductive
definition for each traversal, only differing at the last case: the fields.

Instead, we define an \textit{extensible} generic traversal that is
parameterised over a type class~\cite{Bolingbroke_CK} that provides the action applied to
the fields.
The requirement is that all fields have an instance of this class, giving the
name \emph{constrained traversal}.
\begin{hscode}\SaveRestoreHook
\column{B}{@{}>{\hspre}l<{\hspost}@{}}%
\column{3}{@{}>{\hspre}l<{\hspost}@{}}%
\column{E}{@{}>{\hspre}l<{\hspost}@{}}%
\>[B]{}\keyword{type}\;\Conid{Traversal}_\Conid{C}\;(\Varid{c}\mathbin{::}\star\to \star\to \Conid{Constraint})\;\Varid{s}\;\Varid{t}{}\<[E]%
\\
\>[B]{}\hsindent{3}{}\<[3]%
\>[3]{}\mathrel{=}\forall \Varid{f}\hsforall \hsdot{\mathbin{\cdot}}{.}\Conid{Applicative}\;\Varid{f}\Rightarrow (\forall \Varid{a}\hsforall \;\Varid{b}\hsdot{\mathbin{\cdot}}{.}\Varid{c}\;\Varid{a}\;\Varid{b}\Rightarrow \Varid{a}\to \Varid{f}\;\Varid{b})\to \Varid{s}\to \Varid{f}\;\Varid{t}{}\<[E]%
\ColumnHook
\end{hscode}\resethooks
Instead of specifying up-front the type of the focus \ensuremath{\Varid{a}}, we say that we target
every field of every type, as expressed by the \emph{rank-2}~\cite{PeytonJones:2007:PTI}
quantification of the variables \ensuremath{\Varid{a}} and \ensuremath{\Varid{b}} in
the first argument. Different instantiations of \ensuremath{\Varid{c}} can relate \ensuremath{\Varid{a}} and \ensuremath{\Varid{b}} in
different ways.
\ensuremath{\Conid{HasConstraints}} classifies types that can be traversed in this way, and
\ensuremath{\Conid{GHasConstraints}} provides a concrete definition by induction over the generic
structure.
\begin{hscode}\SaveRestoreHook
\column{B}{@{}>{\hspre}l<{\hspost}@{}}%
\column{3}{@{}>{\hspre}l<{\hspost}@{}}%
\column{E}{@{}>{\hspre}l<{\hspost}@{}}%
\>[B]{}\keyword{class}\;\Conid{HasConstraints}\;(\Varid{c}\mathbin{::}\star\to \star\to \Conid{Constraint})\;\Varid{s}\;\Varid{t}\;\keyword{where}{}\<[E]%
\\
\>[B]{}\hsindent{3}{}\<[3]%
\>[3]{}\Varid{constraints}\mathbin{::}\Conid{Traversal}_\Conid{C}\;\Varid{c}\;\Varid{s}\;\Varid{t}{}\<[E]%
\\[\blanklineskip]%
\>[B]{}\keyword{class}\;\Conid{GHasConstraints}\;(\Varid{c}\mathbin{::}\star\to \star\to \Conid{Constraint})\;\Varid{s}\;\Varid{t}\;\keyword{where}{}\<[E]%
\\
\>[B]{}\hsindent{3}{}\<[3]%
\>[3]{}\Varid{gconstraints}\mathbin{::}\Conid{Traversal}_\Conid{C}\;\Varid{c}\;\Varid{s}\;\Varid{t}{}\<[E]%
\ColumnHook
\end{hscode}\resethooks
The nodes \ensuremath{\mathbin{{:}{\times}{:}}}, \ensuremath{\mathbin{{:}{+}{:}}}, \ensuremath{\Conid{U}}, \ensuremath{\Conid{V}} and \ensuremath{\Conid{M}} are treated analogously to
\ensuremath{\Conid{HasTypes}\_}. We target our focus at the values, as specified by the action \ensuremath{\Varid{c}}.
\begin{hscode}\SaveRestoreHook
\column{B}{@{}>{\hspre}l<{\hspost}@{}}%
\column{3}{@{}>{\hspre}l<{\hspost}@{}}%
\column{E}{@{}>{\hspre}l<{\hspost}@{}}%
\>[B]{}\keyword{instance}\;\Varid{c}\;\Varid{a}\;\Varid{b}\Rightarrow \Conid{GHasConstraints}\;\Varid{c}\;(\Conid{K}\;\Varid{a})\;(\Conid{K}\;\Varid{b})\;\keyword{where}{}\<[E]%
\\
\>[B]{}\hsindent{3}{}\<[3]%
\>[3]{}\Varid{gconstraints}\mathrel{=}\Varid{iso}_\Conid{K}{}\<[E]%
\ColumnHook
\end{hscode}\resethooks
Here, \ensuremath{\Varid{iso}_\Conid{K}} is instantiated to the constrained traversal
\begin{hscode}\SaveRestoreHook
\column{B}{@{}>{\hspre}l<{\hspost}@{}}%
\column{E}{@{}>{\hspre}l<{\hspost}@{}}%
\>[B]{}\Varid{iso}_\Conid{K}\mathbin{::}\Varid{c}\;\Varid{a}\;\Varid{b}\Rightarrow \Conid{Traversal}_\Conid{C}\;\Varid{c}\;\Varid{a}\;\Varid{b}{}\<[E]%
\\
\>[B]{}\Varid{iso}_\Conid{K}\mathbin{::}\forall \Varid{f}\hsforall \hsdot{\mathbin{\cdot}}{.}(\Conid{Functor}\;\Varid{f},\Varid{c}\;\Varid{a}\;\Varid{b})\Rightarrow (\forall \Varid{a}_{\mathrm{1}}\hsforall \;\Varid{b}_{\mathrm{1}}\hsdot{\mathbin{\cdot}}{.}\Varid{c}\;\Varid{a}_{\mathrm{1}}\;\Varid{b}_{\mathrm{1}}\Rightarrow \Varid{a}_{\mathrm{1}}\to \Varid{f}\;\Varid{b}_{\mathrm{1}})\to \Conid{K}\;\Varid{a}\to \Varid{f}\;(\Conid{K}\;\Varid{b}){}\<[E]%
\ColumnHook
\end{hscode}\resethooks
Since its function argument can be applied to \emph{any} \ensuremath{\Varid{a}_{\mathrm{1}}} and \ensuremath{\Varid{b}_{\mathrm{1}}}, it is
certainly applicable to \ensuremath{\Varid{a}} and \ensuremath{\Varid{b}} (as the \ensuremath{\Varid{c}\;\Varid{a}\;\Varid{b}} instance is given).

To show that this traversal is indeed the most general, we allude briefly to an
implementation of \ensuremath{\Conid{HasTypes}\_} in terms of \ensuremath{\Conid{HasConstraints}}.
Note that compared to \ensuremath{\Conid{HasTypes}\_\;\Varid{s}\;\Varid{t}\;\Varid{a}\;\Varid{b}}, the type parameters of \ensuremath{\Conid{HasTypesC}\;\Varid{a}\;\Varid{b}\;\Varid{s}\;\Varid{t}}
are swapped. This is because the traversal will be constrained by \ensuremath{\Conid{HasTypesC}\;\Varid{a}\;\Varid{b}}
-- intuitively, we require that each field be traversable with an \ensuremath{\Varid{a}\to \Varid{b}} action.
\begin{hscode}\SaveRestoreHook
\column{B}{@{}>{\hspre}l<{\hspost}@{}}%
\column{3}{@{}>{\hspre}l<{\hspost}@{}}%
\column{E}{@{}>{\hspre}l<{\hspost}@{}}%
\>[B]{}\keyword{class}\;\Conid{HasTypesC}\;\Varid{a}\;\Varid{b}\;\Varid{s}\;\Varid{t}\;\keyword{where}{}\<[E]%
\\
\>[B]{}\hsindent{3}{}\<[3]%
\>[3]{}\Varid{typesC}\mathbin{::}\Conid{Traversal}\;\Varid{s}\;\Varid{t}\;\Varid{a}\;\Varid{b}{}\<[E]%
\ColumnHook
\end{hscode}\resethooks
The decision at the leaf nodes can be encoded via two corresponding instances.
The first instance describes what to do when the target of the focus is \ensuremath{\Varid{a}}.
\begin{hscode}\SaveRestoreHook
\column{B}{@{}>{\hspre}l<{\hspost}@{}}%
\column{3}{@{}>{\hspre}l<{\hspost}@{}}%
\column{E}{@{}>{\hspre}l<{\hspost}@{}}%
\>[B]{}\keyword{instance}\;\Conid{HasTypesC}\;\Varid{a}\;\Varid{b}\;\Varid{a}\;\Varid{b}\;\keyword{where}{}\<[E]%
\\
\>[B]{}\hsindent{3}{}\<[3]%
\>[3]{}\Varid{typesC}\;\Varid{f}\;\Varid{s}\mathrel{=}\Varid{f}\;\Varid{s}{}\<[E]%
\ColumnHook
\end{hscode}\resethooks
Here we note that this instance allows the field transformation to select the
queried types
\begin{hscode}\SaveRestoreHook
\column{B}{@{}>{\hspre}l<{\hspost}@{}}%
\column{E}{@{}>{\hspre}l<{\hspost}@{}}%
\>[B]{}\Varid{iso}_\Conid{K}\mathbin{::}(\forall \Varid{a}_{\mathrm{1}}\hsforall \;\Varid{b}_{\mathrm{1}}\hsdot{\mathbin{\cdot}}{.}\Conid{HasTypes}\_\;\Varid{a}\;\Varid{b}\;\Varid{a}_{\mathrm{1}}\;\Varid{b}_{\mathrm{1}}\Rightarrow \Varid{a}_{\mathrm{1}}\to \Varid{f}\;\Varid{b}_{\mathrm{1}})\to \Conid{K}\;\Varid{a}\to \Varid{f}\;(\Conid{K}\;\Varid{b}){}\<[E]%
\ColumnHook
\end{hscode}\resethooks
When \ensuremath{\Varid{a}\,\sim\,\Varid{a}_{\mathrm{1}}} and \ensuremath{\Varid{b}\,\sim\,\Varid{b}_{\mathrm{1}}}, instance resolution picks the above instance,
applying the transformation.
Otherwise, the more general instance is selected, which guides the recursion:
\begin{hscode}\SaveRestoreHook
\column{B}{@{}>{\hspre}l<{\hspost}@{}}%
\column{3}{@{}>{\hspre}l<{\hspost}@{}}%
\column{E}{@{}>{\hspre}l<{\hspost}@{}}%
\>[B]{}\keyword{instance}\;(\Conid{Generic}\;\Varid{s},\Conid{Generic}\;\Varid{t},\Conid{HasConstraints}\;(\Conid{HasTypesC}\;\Varid{a}\;\Varid{b})\;\Varid{s}\;\Varid{t}){}\<[E]%
\\
\>[B]{}\hsindent{3}{}\<[3]%
\>[3]{}\Rightarrow \Conid{HasTypesC}\;\Varid{a}\;\Varid{b}\;\Varid{s}\;\Varid{t}\;\keyword{where}{}\<[E]%
\\
\>[B]{}\hsindent{3}{}\<[3]%
\>[3]{}\Varid{typesC}\;\Varid{f}\mathrel{=}\Varid{constraints}\ \texttt{@}(\Conid{HasTypesC}\;\Varid{a}\;\Varid{b})\;(\Varid{typesC}\;\Varid{f}){}\<[E]%
\ColumnHook
\end{hscode}\resethooks
We omit here the definition for primitives, which can be defined analogously to
\ensuremath{\Conid{HasTypes}\_}.

If \ensuremath{\Conid{HasConstraints}} is indeed the most general traversal, then why not use it
to define \ensuremath{\Conid{HasTypes}\_}? The answer is of a
practical nature: the additional burden on the constraint solver slows down
compilation times, and the optimiser misses inlining and specialisation
opportunities more easily.\ck{Can we say more about this?, we have no space!}

\section{Performance}
\label{sec:performance}

When working generically we must always ask whether the abstraction comes
at the cost of performance. In this case, it is pleasing that our use of
generics is optimised away by the compiler.
There are four crucial reasons why we can be confident that GHC will produce
efficient code.

\begin{description}
  \item[Evidence generation.] By using a type-directed approach, we statically know the call
        hierarchy at compile time and can hence use this information to unroll
        our definitions. This is achieved during \emph{evidence generation}.
  \item[Specialisation.] Functions using our methods will have constrained types but we can
        eliminate this over heading via \emph{specialisation}.
  \item[Inlining.] We define our operations such that the composition operator is not
        recursive and can hence be readily \emph{inlined}.
  \item[Internal representation.] Finally, we choose an internal representation
        of our optics such that they expose the optimisation opportunities to
        the compiler.
\end{description}
In this section we describe the optimisations which we rely on to
produce efficient code.
We explain each of these techniques in turn.
Our running example in this section is the \ensuremath{\Varid{incList}} function which maps over a
list of trees and increments the \ensuremath{\Conid{Int}}s inside the tree.

\begin{hscode}\SaveRestoreHook
\column{B}{@{}>{\hspre}l<{\hspost}@{}}%
\column{E}{@{}>{\hspre}l<{\hspost}@{}}%
\>[B]{}\keyword{data}\;\Conid{Tree}\;\Varid{a}\mathrel{=}\Conid{Leaf}\;\Varid{a}\mid \Conid{Branch}\;(\Conid{Tree}\;\Varid{a})\;(\Conid{Tree}\;\Varid{a}){}\<[E]%
\\[\blanklineskip]%
\>[B]{}\Varid{incList}\mathbin{::}[\mskip1.5mu \Conid{Tree}\;\Conid{Int}\mskip1.5mu]\to [\mskip1.5mu \Conid{Tree}\;\Conid{Int}\mskip1.5mu]{}\<[E]%
\\
\>[B]{}\Varid{incList}\;[\mskip1.5mu \mskip1.5mu]\mathrel{=}[\mskip1.5mu \mskip1.5mu]{}\<[E]%
\\
\>[B]{}\Varid{incList}\;(\Varid{x}\mathbin{:}\Varid{xs})\mathrel{=}\Varid{over}\;(\Varid{types}\ \texttt{@}\Conid{Int})\;(\mathbin{+}\mathrm{1})\;\Varid{x}\mathbin{:}\Varid{incList}\;\Varid{xs}{}\<[E]%
\ColumnHook
\end{hscode}\resethooks

\subsection{Evidence Generation}

During compilation, type class constraints are desugared into
arguments to the function~\cite{Wadler:1989:MAP}. The argument is known as a
dictionary and contains a field for each method of a type class. Type class
methods are then desugared as lookup functions into this dictionary.

We use \ensuremath{\Varid{types}} in the definition of \ensuremath{\Varid{incList}} so the constraint solver must
generate evidence that \ensuremath{\Conid{HasTypes}\;(\Conid{Tree}\;\Conid{Int})\;\Conid{Int}}, it does so by
creating an appropriate dictionary.
% \nw{We don't use this until two pages time}
% It will create a dictionary |treeIntHasTypes| which contains this evidence.

The instance for \ensuremath{\Conid{HasTypes}\;\Varid{s}\;\Varid{a}} has constraints \ensuremath{\Conid{Generic}\;\Varid{s}}, and \ensuremath{\Conid{GHasTypes}\;(\Conid{Rep}\;\Varid{s})\;\Varid{a}}. We focus on \ensuremath{\Conid{HasTypes}} and \ensuremath{\Conid{GHasTypes}}, treating the dictionary for
\ensuremath{\Conid{Generic}\;(\Conid{Tree}\;\Conid{Int})} implicitly.
Thus the dictionaries that are produced for us are \ensuremath{\Conid{HasTypesDict}} and
\ensuremath{\Conid{GHasTypesDict}}, corresponding to the appropriate classes.
\begin{hscode}\SaveRestoreHook
\column{B}{@{}>{\hspre}l<{\hspost}@{}}%
\column{41}{@{}>{\hspre}l<{\hspost}@{}}%
\column{43}{@{}>{\hspre}l<{\hspost}@{}}%
\column{E}{@{}>{\hspre}l<{\hspost}@{}}%
\>[B]{}\keyword{data}\;\Conid{HasTypesDict}\;\Varid{s}\;\Varid{a}\mathrel{=}\Conid{HasTypesDict}\;\{\mskip1.5mu {}\<[41]%
\>[41]{}\Varid{types}\mathbin{::}\Conid{Traversal}\;\Varid{s}\;\Varid{s}\;\Varid{a}\;\Varid{a}\mskip1.5mu\}{}\<[E]%
\\
\>[B]{}\keyword{data}\;\Conid{GHasTypesDict}\;\Varid{s}\;\Varid{a}\mathrel{=}\Conid{GHasTypesDict}\;\{\mskip1.5mu {}\<[43]%
\>[43]{}\Varid{gtypes}\mathbin{::}\Conid{Traversal}\;\Varid{s}\;\Varid{s}\;\Varid{a}\;\Varid{a}\mskip1.5mu\}{}\<[E]%
\ColumnHook
\end{hscode}\resethooks
The necessary evidence generated for \ensuremath{\Conid{GHasTypes}\;(\Conid{Rep}\;(\Conid{Tree}\;\Conid{Int}))\;\Conid{Int}}
comes by providing the dictionaries for this type.
The simplified representation for \ensuremath{\Conid{Tree}\;\Conid{Int}} without metadata nodes is:
\begin{hscode}\SaveRestoreHook
\column{B}{@{}>{\hspre}l<{\hspost}@{}}%
\column{E}{@{}>{\hspre}l<{\hspost}@{}}%
\>[B]{}\Conid{Rep}\;(\Conid{Tree}\;\Conid{Int})\equiv\Conid{K}\;\Conid{Int}\mathbin{{:}{+}{:}}(\Conid{K}\;(\Conid{Tree}\;\Conid{Int})\mathbin{{:}{\times}{:}}\Conid{K}\;(\Conid{Tree}\;\Conid{Int})){}\<[E]%
\ColumnHook
\end{hscode}\resethooks
By working through this structure methodically, we arrive at the
following dictionary definitions:
\begin{hscode}\SaveRestoreHook
\column{B}{@{}>{\hspre}l<{\hspost}@{}}%
\column{3}{@{}>{\hspre}l<{\hspost}@{}}%
\column{29}{@{}>{\hspre}l<{\hspost}@{}}%
\column{36}{@{}>{\hspre}l<{\hspost}@{}}%
\column{E}{@{}>{\hspre}l<{\hspost}@{}}%
\>[B]{}\Varid{hasTypesDict}_{\Conid{TreeInt}}\mathbin{::}\Conid{HasTypesDict}\;(\Conid{Tree}\;\Conid{Int})\;\Conid{Int}{}\<[E]%
\\
\>[B]{}\Varid{hasTypesDict}_{\Conid{TreeInt}}\mathrel{=}\Conid{HasTypesDict}\;\{\mskip1.5mu \Varid{types}\mathrel{=}\Varid{iso}_{\Conid{Rep}}\hsdot{\mathbin{\cdot}}{.}\Varid{gtypes}\;\Varid{ghasTypesDict}_{\Conid{TreeInt}}\mskip1.5mu\}{}\<[E]%
\\[\blanklineskip]%
\>[B]{}\Varid{ghasTypesDict}_{\Conid{TreeInt}}\mathbin{::}\Conid{GHasTypesDict}\;(\Conid{Rep}\;(\Conid{Tree}\;\Conid{Int}))\;\Conid{Int}{}\<[E]%
\\
\>[B]{}\Varid{ghasTypesDict}_{\Conid{TreeInt}}\mathrel{=}\Conid{GHasTypesDict}\;\{\mskip1.5mu \Varid{gtypes}\mathrel{=}\lambda \Varid{f}\;\Varid{l1r1}\to \keyword{case}\;\Varid{l1r1}\;\keyword{of}{}\<[E]%
\\
\>[B]{}\hsindent{29}{}\<[29]%
\>[29]{}\Conid{L}\;\Varid{l}{}\<[36]%
\>[36]{}\to \Conid{L}\mathbin{{\langle}{\$}{\rangle}}\Varid{gtypes}\;\Varid{ghasTypesDict}_{\Conid{KInt}}\;\Varid{f}\;\Varid{l}{}\<[E]%
\\
\>[B]{}\hsindent{29}{}\<[29]%
\>[29]{}\Conid{R}\;\Varid{r}{}\<[36]%
\>[36]{}\to \Conid{R}\mathbin{{\langle}{\$}{\rangle}}\Varid{gtypes}\;\Varid{ghasTypesDict}_{\mathbin{{:}{\times}{:}}}\;\Varid{f}\;\Varid{r}\mskip1.5mu\}{}\<[E]%
\\[\blanklineskip]%
\>[B]{}\Varid{ghasTypesDict}_{\Conid{KInt}}\mathbin{::}\Conid{GHasTypesDict}\;(\Conid{K}\;\Conid{Int})\;\Conid{Int}{}\<[E]%
\\
\>[B]{}\Varid{ghasTypesDict}_{\Conid{KInt}}\mathrel{=}\Conid{GHasTypesDict}\;\{\mskip1.5mu \Varid{gtypes}\mathrel{=}\Varid{iso}_\Conid{K}\mskip1.5mu\}{}\<[E]%
\\[\blanklineskip]%
\>[B]{}\Varid{ghasTypesDict}_{\Conid{KTreeInt}}\mathbin{::}\Conid{GHasTypesDict}\;(\Conid{K}\;(\Conid{Tree}\;\Conid{Int}))\;\Conid{Int}{}\<[E]%
\\
\>[B]{}\Varid{ghasTypesDict}_{\Conid{KTreeInt}}\mathrel{=}\Conid{GHasTypesDict}\;\{\mskip1.5mu \Varid{gtypes}\mathrel{=}\Varid{iso}_\Conid{K}\hsdot{\mathbin{\cdot}}{.}\Varid{types}\;\Varid{hasTypesDict}_{\Conid{TreeInt}}\mskip1.5mu\}{}\<[E]%
\\[\blanklineskip]%
\>[B]{}\Varid{ghasTypesDict}_{\mathbin{{:}{\times}{:}}}\mathbin{::}\Conid{GHasTypesDict}\;(\Conid{K}\;(\Conid{Tree}\;\Conid{Int})\mathbin{{:}{\times}{:}}\Conid{K}\;(\Conid{Tree}\;\Conid{Int}))\;\Conid{Int}{}\<[E]%
\\
\>[B]{}\Varid{ghasTypesDict}_{\mathbin{{:}{\times}{:}}}\mathrel{=}\Conid{GHasTypesDict}\;\{\mskip1.5mu \Varid{gtypes}\mathrel{=}\lambda \Varid{f}\;(\Varid{l}\mathbin{{:}{\times}{:}}\Varid{r})\to (\mathbin{{:}{\times}{:}})\mathbin{{\langle}{\$}{\rangle}}{}\<[E]%
\\
\>[B]{}\hsindent{3}{}\<[3]%
\>[3]{}\Varid{gtypes}\;\Varid{ghasTypesDict}_{\Conid{KTreeInt}}\;\Varid{f}\;\Varid{l}\mathbin{{\langle}{*}{\rangle}}\Varid{gtypes}\;\Varid{ghasTypesDict}_{\Conid{KTreeInt}}\;\Varid{f}\;\Varid{r}\mskip1.5mu\}{}\<[E]%
\ColumnHook
\end{hscode}\resethooks
We first generate evidence by using the instance for \ensuremath{\mathbin{{:}{+}{:}}}, before
recursing into both branches and finding evidence for \ensuremath{\mathbin{{:}{\times}{:}}} and the
\ensuremath{\Conid{K}\;\Conid{Int}} nodes.
As such, we have a dictionary for each type constructor.
The constraint solver will terminate as it will observe that we can
use the \ensuremath{\Varid{ghasTypesDict}_{\Conid{TreeInt}}} dictionary when trying to solve the
recursive case. Thus, these dictionaries form a mutually recursive group.
The dictionaries generated are straightforward transcriptions of the instances,
with instance constraints solved and $\beta$-reduced. The definition of
\ensuremath{\Varid{ghasTypesDict}_{\Conid{TreeInt}}} is still not as efficient as it could be, and we discuss
how it can be further improved with inlining in
Section~\ref{sec:inline-dictionary}.

We see that the process of generating evidence also unrolls definitions.
If we had instead defined \ensuremath{\Varid{types}} as a function over a normal data type without
any type direction, it would be self-recursive and hence not able to be eliminated
in the same manner.
This process is safe as types are finite and statically
known at compile time. Without additional language pragmas, the restrictions
on instance contexts guarantee that the constaint solving process terminates.

\subsection{Inlining}

Once the structure is in place, there is still indirection present which can
be removed. The first step of doing this is inlining.
Inlining is the process of replacing a function's
name by its definition. It is the most crucial optimisation
in the compiler's pipeline as it enables
all other optimisations to occur. We already saw how the compiler generates naive
verbose code which is simplified when inlined, this is in general true for
all programs.

However, whilst always safe in a pure language like Haskell,
we must still be careful about when we inline. If we inline
too little then we miss optimisation opportunities. If we inline too much then
the size of our program becomes very large and takes a long time to compile.

The compiler contains a set of balanced heuristics to decide whether to inline
a definition~\cite{Jones:2002:Secrets}. These include factors such as: the syntactic size of a function,
as a measure to stop a lot of code duplication; whether a function is recursive,
recursive functions are never inlined; whether a function is applied to known arguments,
there is a good chance that the body will scrutinise the arguments and perform more
simplification and so on.

There are also manners in which the user can influence these automatic decisions.
One in particular is the use of \text{\ttfamily INLINE} pragmas which can be used to mark definitions
as very desirable to \text{\ttfamily INLINE}. In our use cases, marking some instance methods
as \text{\ttfamily INLINE} was necessary to unstick the optimiser and enable it to perform
much more simplification.

In addition, the optimiser will also evaluate programs by $\beta$-reducing,
evaluating case expressions with a known scrutinee and perform commuting conversions.
For a full account of the simple core transformations which the simplfiier performs
in order to generate simpler code, one should consult~\cite{Jones:1998:Transformation}.

\subsubsection{Optimising Dictionaries}
\label{sec:inline-dictionary}

We recall that our generated dictionaries are mutually recursive. This isn't
surprising, as we expect \ensuremath{\Varid{gtypes}} to be recursive in general if we are trying to
traverse a recursive data structure. Mutually recursive blocks of functions
must be treated with care, as repeatedly inlining them causes the inliner to diverge.
Each mutually recursive group is thus appointed a loop-breaker function, which
is never inlined, but we can freely inline other definitions into each other in
order to create a single self-recursive definition. After the dictionaries are
inlined into each other, we end up with the following evidence which has the correct
unrolled shape we were looking for.

\begin{hscode}\SaveRestoreHook
\column{B}{@{}>{\hspre}l<{\hspost}@{}}%
\column{27}{@{}>{\hspre}l<{\hspost}@{}}%
\column{34}{@{}>{\hspre}l<{\hspost}@{}}%
\column{51}{@{}>{\hspre}l<{\hspost}@{}}%
\column{109}{@{}>{\hspre}l<{\hspost}@{}}%
\column{E}{@{}>{\hspre}l<{\hspost}@{}}%
\>[B]{}\Varid{ghasTypesDict'}_{\Conid{TreeInt}}\mathbin{::}\Conid{GHasTypesDict}\;(\Conid{Rep}\;(\Conid{Tree}\;\Conid{Int}))\;\Conid{Int}{}\<[E]%
\\
\>[B]{}\Varid{ghasTypesDict'}_{\Conid{TreeInt}}\mathrel{=}\Conid{GHasTypesDict}\;\{\mskip1.5mu \Varid{gtypes}\mathrel{=}\lambda \Varid{f}\;\Varid{l1r1}\to \keyword{case}\;\Varid{l1r1}\;\keyword{of}{}\<[E]%
\\
\>[B]{}\hsindent{27}{}\<[27]%
\>[27]{}\Conid{L}\;\Varid{l}{}\<[34]%
\>[34]{}\to \Conid{L}\mathbin{{\langle}{\$}{\rangle}}\Varid{iso}_\Conid{K}\;\Varid{f}\;\Varid{l}{}\<[E]%
\\
\>[B]{}\hsindent{27}{}\<[27]%
\>[27]{}\Conid{R}\;\Varid{b}{}\<[34]%
\>[34]{}\to \Conid{R}\mathbin{{\langle}{\$}{\rangle}}(\lambda \Varid{f}\;(\Varid{l}\mathbin{{:}{\times}{:}}\Varid{r})\to (\mathbin{{:}{\times}{:}}){}\<[E]%
\\
\>[34]{}\hsindent{17}{}\<[51]%
\>[51]{}\mathbin{{\langle}{\$}{\rangle}}(\Varid{iso}_\Conid{K}\hsdot{\mathbin{\cdot}}{.}\Varid{iso}_{\Conid{Rep}}\hsdot{\mathbin{\cdot}}{.}\Varid{gtypes}\;\Varid{ghasTypesDict'}_{\Conid{TreeInt}})\;{}\<[109]%
\>[109]{}\Varid{f}\;\Varid{l}{}\<[E]%
\\
\>[34]{}\hsindent{17}{}\<[51]%
\>[51]{}\mathbin{{\langle}{*}{\rangle}}(\Varid{iso}_\Conid{K}\hsdot{\mathbin{\cdot}}{.}\Varid{iso}_{\Conid{Rep}}\hsdot{\mathbin{\cdot}}{.}\Varid{gtypes}\;\Varid{ghasTypesDict'}_{\Conid{TreeInt}})\;\Varid{f}\;\Varid{r})\;\Varid{f}\;\Varid{b}\mskip1.5mu\}{}\<[E]%
\ColumnHook
\end{hscode}\resethooks
In this case, \ensuremath{\Varid{ghasTypesDict'}_{\Conid{TreeInt}}} acts as the loop-breaker.

\subsection{Specialisation}

As we have seen, the evidence generation procedure and inlining are sufficient on
their own to eliminate much of the generic overhead of a statically known
parameter as long as we call the class method directly.
However, we use class methods inside bigger functions and when we do they give
rise to class constraints. When these larger functions are called, the
dictionary must be solved and the required evidenced passed to the function.

For instance,
we might want to write the more general type signature for \ensuremath{\Varid{incList}} to be
parametric over the choice of data structure contained in the list as long
as it contains integers. We will call this generalised version \ensuremath{\Varid{incListGen}}.
If we call \ensuremath{\Varid{incListGen}} and instantiate \ensuremath{\Varid{s}} to be \ensuremath{\Conid{Tree}\;\Conid{Int}} then we
should expect that the definition would be identical to \ensuremath{\Varid{incList}}.
\begin{hscode}\SaveRestoreHook
\column{B}{@{}>{\hspre}l<{\hspost}@{}}%
\column{E}{@{}>{\hspre}l<{\hspost}@{}}%
\>[B]{}\Varid{incListGen}\mathbin{::}\Conid{HasTypes}\;\Varid{s}\;\Conid{Int}\Rightarrow [\mskip1.5mu \Varid{s}\mskip1.5mu]\to [\mskip1.5mu \Varid{s}\mskip1.5mu]{}\<[E]%
\\
\>[B]{}\Varid{incListGen}\;[\mskip1.5mu \mskip1.5mu]\mathrel{=}[\mskip1.5mu \mskip1.5mu]{}\<[E]%
\\
\>[B]{}\Varid{incListGen}\;(\Varid{x}\mathbin{:}\Varid{xs})\mathrel{=}\Varid{over}\;(\Varid{types}\ \texttt{@}\Conid{Int})\;(\mathbin{+}\mathrm{1})\;\Varid{x}\mathbin{:}\Varid{incListGen}\;\Varid{xs}{}\<[E]%
\ColumnHook
\end{hscode}\resethooks
This problem is not trivial.
When \ensuremath{\Varid{incListGen}} is called, the evidence witnessing the constraint \ensuremath{\Conid{HasTypes}} will
be passed to it. In order to eliminate this dictionary, we need to push it inwards
to the call of \ensuremath{\Varid{types}}. Since \ensuremath{\Varid{incListGen}} is recursive it cannot be
inlined, so we rely on \emph{specialisation} instead.

The specialiser looks for calls to overloaded functions called at a known type.
It then creates a new type specialised definition which does not take a dictionary
argument and a rewrite rule which rewrites the old version to the new version.

Suppose that we know that the value of \ensuremath{\Varid{s}} is \ensuremath{\Conid{Tree}\;\Conid{Int}}, and that the evidence
dictionary for \ensuremath{\Conid{HasTypes}} is called \ensuremath{\Varid{treeIntHasTypes}}. The naive desugaring of
calling \ensuremath{\Varid{incListGen}\ \texttt{@}(\Conid{Tree}\;\Conid{Int})\;\Varid{xs}} is:

\begin{hscode}\SaveRestoreHook
\column{B}{@{}>{\hspre}l<{\hspost}@{}}%
\column{E}{@{}>{\hspre}l<{\hspost}@{}}%
\>[B]{}\Varid{incListGen}\;\Varid{treeIntHasTypes}\;\Varid{xs}{}\<[E]%
\ColumnHook
\end{hscode}\resethooks

The specialiser then observes this call to \ensuremath{\Varid{incListGen}} takes a dictionary argument
and creates a specialised version \ensuremath{\Varid{incListGen}_{\Conid{TreeInt}}} with the following definition:

\begin{hscode}\SaveRestoreHook
\column{B}{@{}>{\hspre}l<{\hspost}@{}}%
\column{3}{@{}>{\hspre}l<{\hspost}@{}}%
\column{5}{@{}>{\hspre}l<{\hspost}@{}}%
\column{E}{@{}>{\hspre}l<{\hspost}@{}}%
\>[B]{}\Varid{incListGen}_{\Conid{TreeInt}}\mathbin{::}[\mskip1.5mu \Conid{Tree}\;\Conid{Int}\mskip1.5mu]\to [\mskip1.5mu \Conid{Tree}\;\Conid{Int}\mskip1.5mu]{}\<[E]%
\\
\>[B]{}\Varid{incListGen}_{\Conid{TreeInt}}\;\Varid{xs}\mathrel{=}(\lambda \Varid{hasTypesDict}\;\Varid{xs}\to \keyword{case}\;\Varid{xs}\;\keyword{of}{}\<[E]%
\\
\>[B]{}\hsindent{3}{}\<[3]%
\>[3]{}[\mskip1.5mu \mskip1.5mu]\to [\mskip1.5mu \mskip1.5mu]{}\<[E]%
\\
\>[B]{}\hsindent{3}{}\<[3]%
\>[3]{}(\Varid{x}\mathbin{:}\Varid{xs})\to \Varid{over}\;(\Varid{types}\;\Varid{hasTypesDict})\;(\mathbin{+}\mathrm{1})\;\Varid{x}\mathbin{:}\Varid{incListGen}\;\Varid{hasTypesDict}\;\Varid{xs}){}\<[E]%
\\
\>[3]{}\hsindent{2}{}\<[5]%
\>[5]{}\Varid{treeIntHasTypes}\;\Varid{xs}{}\<[E]%
\ColumnHook
\end{hscode}\resethooks
The right-hand side of the definition is same as the right-hand side of \ensuremath{\Varid{incListGen}} applied
to \ensuremath{\Varid{treeIntHasTypes}}. Then, an additional rewrite rule is generated which replaces the overloaded
call with the specialised definition.
\begin{core}
{-# RULES "specincListGen" forall xs . incListGen treeIntHasTypes xs
                                  = incListGen_TreeInt #-}
\end{core}

That's the whole process.
After $\beta$-reduction, the dictionary
selector \ensuremath{\Varid{types}} is now adjacent to its dictionary and hence we can inline \text{\ttfamily types}
and select the correct method from \ensuremath{\Varid{treeIntHasTypes}}.
Notice that in the definition of \ensuremath{\Varid{incListGen}_{\Conid{TreeInt}}} we
still have an overloaded call to \ensuremath{\Varid{incListGen}}, this will be rewritten when the rewrite
rule is applied and then \ensuremath{\Varid{incListGen}_{\Conid{TreeInt}}} will become self-recursive.
After these two steps, we eliminate all
the occurences of \ensuremath{\Varid{treeIntHasTypes}} and the overloading overhead is eliminated.

Once again, specialisation is an \emph{enabling} transformation. Later optimisation
passes will perform more complicated rearranging with the express goal of
improving our code.

\subsection{Internal representation}

After we have created this unrolled pipeline of functions, the question remains
how this can become the same as hand-written definitions later in the compilation process.
How precisely do inlining and $\beta$-reduction lead to good code?
How and why depends on the internal representation of lenses and traversals we
choose in the library.

\subsubsection{Lenses}

In the case of lenses, the inliner does a sufficient job of combining the composition
of lenses into a single lens without further intervention. The lens composition
operator is not recursive and hence is readily inlined which leads to much
further simplification.
\begin{hscode}\SaveRestoreHook
\column{B}{@{}>{\hspre}l<{\hspost}@{}}%
\column{47}{@{}>{\hspre}l<{\hspost}@{}}%
\column{E}{@{}>{\hspre}l<{\hspost}@{}}%
\>[B]{}\keyword{data}\;\Varid{Lens}_\Varid{1}\;\Varid{s}\;\Varid{t}\;\Varid{a}\;\Varid{b}\mathrel{=}\Varid{Lens}_\Varid{1}\;(\Varid{s}\to \Varid{a})\;(\Varid{b}\to \Varid{s}\to \Varid{t}){}\<[E]%
\\[\blanklineskip]%
\>[B]{}(\mathbin{\circ})\mathbin{::}\Varid{Lens}_\Varid{1}\;\Varid{s}\;\Varid{t}\;\Varid{c}\;\Varid{d}\to \Varid{Lens}_\Varid{1}\;\Varid{c}\;\Varid{d}\;\Varid{a}\;\Varid{b}\to \Varid{Lens}_\Varid{1}\;\Varid{s}\;\Varid{t}\;\Varid{a}\;\Varid{b}{}\<[E]%
\\
\>[B]{}(\Varid{Lens}_\Varid{1}\;\Varid{get}_{1}\;\Varid{set}_{1})\mathbin{\circ}(\Varid{Lens}_\Varid{1}\;\Varid{get}_{2}\;\Varid{set}_{2})\mathrel{=}\Varid{Lens}_\Varid{1}\;{}\<[47]%
\>[47]{}(\Varid{get}_{2}\hsdot{\mathbin{\cdot}}{.}\Varid{get}_{1})\;{}\<[E]%
\\
\>[47]{}(\lambda \Varid{b}\;\Varid{s}\to \Varid{set}_{1}\;(\Varid{set}_{2}\;\Varid{b}\;(\Varid{get}_{1}\;\Varid{s}))\;\Varid{s}){}\<[E]%
\ColumnHook
\end{hscode}\resethooks
In fact, the naive encoding given above for lenses does not produce the best results.
Whilst it does collapse a sequence of compositions appropiately, the type ensures
that in order to implement a modification operation, we must perform a \ensuremath{\Varid{get}} followed
by a \ensuremath{\Varid{set}} and hence deconstruct \ensuremath{\Varid{s}} twice.
We can get around this problem by using the existential encoding which means
that we can directly implement an updating function by only deconstructing the
source once.
\begin{hscode}\SaveRestoreHook
\column{B}{@{}>{\hspre}l<{\hspost}@{}}%
\column{E}{@{}>{\hspre}l<{\hspost}@{}}%
\>[B]{}\keyword{data}\;\Varid{Lens}_\Varid{2}\;\Varid{s}\;\Varid{t}\;\Varid{a}\;\Varid{b}\mathrel{=}\forall \Varid{c}\hsforall \hsdot{\mathbin{\cdot}}{.}\Varid{Lens}_\Varid{2}\;(\Varid{s}\to (\Varid{a},\Varid{c}))\;((\Varid{b},\Varid{c})\to \Varid{t}){}\<[E]%
\ColumnHook
\end{hscode}\resethooks
Intuitively, the \ensuremath{\Varid{get}} function separates \ensuremath{\Varid{s}} into the the part we are
focusing on of type \ensuremath{\Varid{a}} and its complement \ensuremath{\Varid{c}}.
In turn, the \ensuremath{\Varid{set}} function recombines a value of type \ensuremath{\Varid{b}} with the complement.
\begin{hscode}\SaveRestoreHook
\column{B}{@{}>{\hspre}l<{\hspost}@{}}%
\column{5}{@{}>{\hspre}l<{\hspost}@{}}%
\column{15}{@{}>{\hspre}l<{\hspost}@{}}%
\column{21}{@{}>{\hspre}l<{\hspost}@{}}%
\column{42}{@{}>{\hspre}l<{\hspost}@{}}%
\column{67}{@{}>{\hspre}l<{\hspost}@{}}%
\column{E}{@{}>{\hspre}l<{\hspost}@{}}%
\>[B]{}(\mathbin{\bullet})\mathbin{::}\Varid{Lens}_\Varid{2}\;\Varid{s}\;\Varid{t}\;\Varid{c}\;\Varid{d}\to \Varid{Lens}_\Varid{2}\;\Varid{c}\;\Varid{d}\;\Varid{a}\;\Varid{b}\to \Varid{Lens}_\Varid{2}\;\Varid{s}\;\Varid{t}\;\Varid{a}\;\Varid{b}{}\<[E]%
\\
\>[B]{}(\Varid{Lens}_\Varid{2}\;\Varid{get}_{1}\;\Varid{set}_{1})\mathbin{\bullet}(\Varid{Lens}_\Varid{2}\;\Varid{get}_{2}\;\Varid{set}_{2})\mathrel{=}\Varid{Lens}_\Varid{2}\;\Varid{get}\;\Varid{set}\;\keyword{where}{}\<[E]%
\\
\>[B]{}\hsindent{5}{}\<[5]%
\>[5]{}\Varid{get}\;\Varid{s}\mathrel{=}{}\<[15]%
\>[15]{}\keyword{let}\;{}\<[21]%
\>[21]{}(\Varid{c},\Varid{com}_{1})\mathrel{=}\Varid{get}_{1}\;\Varid{s};{}\<[42]%
\>[42]{}(\Varid{a},\Varid{com}_{2})\mathrel{=}\Varid{get}_{2}\;\Varid{c}\;\keyword{in}\;{}\<[67]%
\>[67]{}(\Varid{a},(\Varid{com}_{1},\Varid{com}_{2})){}\<[E]%
\\
\>[B]{}\hsindent{5}{}\<[5]%
\>[5]{}\Varid{set}\;(\Varid{b},(\Varid{com}_{1},\Varid{com}_{2}))\mathrel{=}\Varid{set}_{1}\;((\Varid{set}_{2}\;(\Varid{b},\Varid{com}_{2})),\Varid{com}_{1}){}\<[E]%
\\[\blanklineskip]%
\>[B]{}\Varid{modify}\mathbin{::}\Varid{Lens}_\Varid{2}\;\Varid{s}\;\Varid{t}\;\Varid{a}\;\Varid{b}\to (\Varid{a}\to \Varid{b})\to (\Varid{s}\to \Varid{t}){}\<[E]%
\\
\>[B]{}\Varid{modify}\;(\Varid{Lens}_\Varid{2}\;\Varid{get}\;\Varid{set})\;\Varid{f}\;\Varid{s}\mathrel{=}\keyword{let}\;(\Varid{a},\Varid{c})\mathrel{=}\Varid{get}\;\Varid{s}\;\keyword{in}\;\Varid{set}\;((\Varid{f}\;\Varid{a}),\Varid{c}){}\<[E]%
\ColumnHook
\end{hscode}\resethooks

Using this definition, chained modifications can be fused into a single
function. Thus, in our implementation, the lenses we use are of this latter
encoding.  Once we have fused them together, we turn them into whichever
encoding that we want the library to produce. By default, it is the van
Laarhoven encoding as found in the \text{\ttfamily lens} library~\cite{kmett_lens}.

\subsubsection{Traversals}

Optimising traversals in the same manner is slightly trickier as we must find
an encoding of a traversal which does not require a recursive composition operator.
In order to do this, we use a van Laarhoven style
representation. The composition operator for these traversals is the function
composition operator. However, this is not sufficient, the downside of
using this composition operator is that it does not perform normalisation as
happened with lenses. It is necessary to appeal to the \ensuremath{\Conid{Applicative}} laws
in order to rearrange and normalise these compositions. The following
technique is due to Eric Mertens and can be found implemented in the \text{\ttfamily lens} library.

A van Laarhoven \ensuremath{\Conid{Traversal}} is a function with the following type.

\begin{hscode}\SaveRestoreHook
\column{B}{@{}>{\hspre}l<{\hspost}@{}}%
\column{E}{@{}>{\hspre}l<{\hspost}@{}}%
\>[B]{}\keyword{type}\;\Conid{Traversal}\;\Varid{s}\;\Varid{t}\;\Varid{a}\;\Varid{b}\mathrel{=}\forall \Varid{g}\hsforall \hsdot{\mathbin{\cdot}}{.}\Conid{Applicative}\;\Varid{g}\Rightarrow (\Varid{a}\to \Varid{g}\;\Varid{b})\to \Varid{s}\to \Varid{g}\;\Varid{t}{}\<[E]%
\ColumnHook
\end{hscode}\resethooks

The result type of these functions is a value constructed using \ensuremath{\Conid{Applicative}}
operators.
\ensuremath{\Conid{Applicative}} expressions have a normal form of a single \ensuremath{\Varid{pure}} followed by
a sequence of left-associated applications using the combinator \ensuremath{\mathbin{{\langle}{*}{\rangle}}}~\cite{Mcbride:2008:Applicative}.
In order to rewrite this normal form, we must re-associate all uses of \ensuremath{\mathbin{{\langle}{*}{\rangle}}}
and then fuse together all uses of \ensuremath{\Varid{pure}}.

This first step is achieved by instantiating \ensuremath{\Varid{g}} to be \ensuremath{\Conid{Curried}}.

\begin{hscode}\SaveRestoreHook
\column{B}{@{}>{\hspre}l<{\hspost}@{}}%
\column{3}{@{}>{\hspre}l<{\hspost}@{}}%
\column{E}{@{}>{\hspre}l<{\hspost}@{}}%
\>[B]{}\keyword{data}\;\Conid{Curried}\;\Varid{f}\;\Varid{a}\mathrel{=}\Conid{Curried}\;\{\mskip1.5mu \Varid{runCurried}\mathbin{::}\forall \Varid{r}\hsforall \hsdot{\mathbin{\cdot}}{.}\Varid{f}\;(\Varid{a}\to \Varid{r})\to \Varid{f}\;\Varid{r}\mskip1.5mu\}{}\<[E]%
\\[\blanklineskip]%
\>[B]{}\keyword{instance}\;\Conid{Functor}\;\Varid{f}\Rightarrow \Conid{Functor}\;(\Conid{Curried}\;\Varid{f})\;\keyword{where}{}\<[E]%
\\
\>[B]{}\hsindent{3}{}\<[3]%
\>[3]{}\Varid{fmap}\;\Varid{f}\;(\Conid{Curried}\;\Varid{v})\mathrel{=}\Conid{Curried}\;(\lambda \Varid{far}\to \Varid{v}\;(\Varid{fmap}\;(\hsdot{\mathbin{\cdot}}{.}\Varid{f})\;\Varid{far})){}\<[E]%
\\[\blanklineskip]%
\>[B]{}\keyword{instance}\;\Conid{Functor}\;\Varid{f}\Rightarrow \Conid{Applicative}\;(\Conid{Curried}\;\Varid{f})\;\keyword{where}{}\<[E]%
\\
\>[B]{}\hsindent{3}{}\<[3]%
\>[3]{}\Varid{pure}\;\Varid{a}\mathrel{=}\Conid{Curried}\;(\lambda \Varid{far}\to \Varid{fmap}\;(\mathbin{\$}\Varid{a})\;\Varid{far}){}\<[E]%
\\
\>[B]{}\hsindent{3}{}\<[3]%
\>[3]{}\Conid{Curried}\;\Varid{mf}\mathbin{{\langle}{*}{\rangle}}\Conid{Curried}\;\Varid{ma}\mathrel{=}\Conid{Curried}\;(\Varid{ma}\hsdot{\mathbin{\cdot}}{.}\Varid{mf}\hsdot{\mathbin{\cdot}}{.}\Varid{fmap}\;(\hsdot{\mathbin{\cdot}}{.})){}\<[E]%
\ColumnHook
\end{hscode}\resethooks

It is the definition of \ensuremath{\mathbin{{\langle}{*}{\rangle}}} which performs the reassociation.
Notice that the \ensuremath{\Conid{Applicative}} instance for \ensuremath{\Conid{Curried}} delegates all calls to \ensuremath{\Varid{pure}} to
the underlying functor. We will fuse those together with an additional layer
termed \ensuremath{\Conid{Yoneda}} which intercepts all the calls to \ensuremath{\Varid{fmap}} and fuses them together.

\begin{hscode}\SaveRestoreHook
\column{B}{@{}>{\hspre}l<{\hspost}@{}}%
\column{3}{@{}>{\hspre}l<{\hspost}@{}}%
\column{E}{@{}>{\hspre}l<{\hspost}@{}}%
\>[B]{}\keyword{data}\;\Conid{Yoneda}\;\Varid{f}\;\Varid{a}\mathrel{=}\Conid{Yoneda}\;\{\mskip1.5mu \Varid{runYoneda}\mathbin{::}\forall \Varid{r}\hsforall \hsdot{\mathbin{\cdot}}{.}(\Varid{a}\to \Varid{r})\to \Varid{f}\;\Varid{r}\mskip1.5mu\}{}\<[E]%
\\[\blanklineskip]%
\>[B]{}\keyword{instance}\;\Conid{Functor}\;(\Conid{Yoneda}\;\Varid{f})\;\keyword{where}{}\<[E]%
\\
\>[B]{}\hsindent{3}{}\<[3]%
\>[3]{}\Varid{fmap}\;\Varid{f}\;(\Conid{Yoneda}\;\Varid{v})\mathrel{=}\Conid{Yoneda}\;(\lambda \Varid{k}\to \Varid{v}\;(\Varid{k}\hsdot{\mathbin{\cdot}}{.}\Varid{f})){}\<[E]%
\\[\blanklineskip]%
\>[B]{}\keyword{instance}\;\Conid{Applicative}\;\Varid{f}\Rightarrow \Conid{Applicative}\;(\Conid{Yoneda}\;\Varid{f})\;\keyword{where}{}\<[E]%
\\
\>[B]{}\hsindent{3}{}\<[3]%
\>[3]{}\Varid{pure}\;\Varid{a}\mathrel{=}\Conid{Yoneda}\;(\lambda \Varid{f}\to \Varid{pure}\;(\Varid{f}\;\Varid{a})){}\<[E]%
\\
\>[B]{}\hsindent{3}{}\<[3]%
\>[3]{}\Conid{Yoneda}\;\Varid{m}\mathbin{{\langle}{*}{\rangle}}\Conid{Yoneda}\;\Varid{n}\mathrel{=}\Conid{Yoneda}\;(\lambda \Varid{f}\to \Varid{m}\;(\Varid{f}\hsdot{\mathbin{\cdot}}{.})\mathbin{{\langle}{*}{\rangle}}\Varid{n}\;\Varid{id}){}\<[E]%
\ColumnHook
\end{hscode}\resethooks

This time, we notice that \ensuremath{\Conid{Yoneda}} just delegates the definitions of the \ensuremath{\Conid{Applicative}}.
Putting this together, we instantiate \ensuremath{\Varid{g}} to be \ensuremath{\Conid{Curried}\;(\Conid{Yoneda}\;\Varid{g})} and then
use \ensuremath{\Varid{lowerCurriedYoneda}} in order to return to a simple type parameterised by
an \ensuremath{\Conid{Applicative}} constraint.

\begin{hscode}\SaveRestoreHook
\column{B}{@{}>{\hspre}l<{\hspost}@{}}%
\column{E}{@{}>{\hspre}l<{\hspost}@{}}%
\>[B]{}\Varid{liftCurriedYoneda}\mathbin{::}\Conid{Applicative}\;\Varid{g}\Rightarrow \Varid{g}\;\Varid{a}\to \Conid{Curried}\;(\Conid{Yoneda}\;\Varid{g})\;\Varid{a}{}\<[E]%
\\
\>[B]{}\Varid{lowerCurriedYoneda}\mathbin{::}\Conid{Applicative}\;\Varid{g}\Rightarrow \Conid{Curried}\;(\Conid{Yoneda}\;\Varid{g})\;\Varid{a}\to \Varid{g}\;\Varid{a}{}\<[E]%
\ColumnHook
\end{hscode}\resethooks

This process performs the reassociating and fusion that we desired.
However, in practice, it is difficult to be sure that the
compiler will remove this overhead. On the other hand, it does not affect performance
in common use cases such as modifying or summarising. This is because when \ensuremath{\Varid{g}}
is instantiated to a known \ensuremath{\Conid{Applicative}}, the \ensuremath{\Conid{Applicative}} methods can be
inlined as they are not recursive. We usually instantiate to \ensuremath{\Varid{g}} to either
\ensuremath{\Conid{Const}} or \ensuremath{\Conid{Identity}} which are completely eliminated.

Using similar techniques to traversals,
we could optimise the van Laarhoven or profunctor representation
of lenses and prisms but these simple minded techniques are the most reliable
and very effective in generating good programs without impacting compile times significantly.

\section{Benchmarks}
\label{sec:benchmarks}

We compare the performance of \text{\ttfamily generic\char45{}lens} with hand-written code
as well as five other generic programming libraries which derive
traversals for data types. There are no other libraries which derive
lenses or prisms in a similar way so we could not compare this
aspect of the library.
\begin{description}
\item[\text{\ttfamily hand}]
Hand-written definitions in an idiomatic direct style.

\item[\normalfont{\textbf{(gl)}} \text{\ttfamily generic\char45{}lens}]
The library which we describe in this paper.

\item[\normalfont{(gp)} \text{\ttfamily geniplate\char45{}0\char46{}7\char46{}6}~\cite{Augustsson:2018:Geniplate}]
A library which provides a similar interface to the \text{\ttfamily uniplate} library
below but uses
Template Haskell in order to generate traversals.

\item[\normalfont{(up)} \text{\ttfamily uniplate\char45{}1\char46{}6\char46{}12}~\cite{Mitchell:2007:Uniform}]
A library which provides an interface for traversing data.
A traversal for a data type which has a \ensuremath{\Conid{Data}} instance is derived by
using \text{\ttfamily Data\char46{}Generics\char46{}Uniplate\char46{}Data}.

\item[\normalfont{(lens)} \text{\ttfamily lens\char45{}4\char46{}16}~\cite{kmett_lens}]
This library provides a reimplementation of the \text{\ttfamily uniplate} interface to
generate van Laarhoven style traversals rather than uniplate traversals.

\item[\normalfont{(syb)} \text{\ttfamily syb\char45{}0\char46{}7}~\cite{Lammel:2003:SYB}]
One of the first generics libraries using the \ensuremath{\Conid{Data}} type class to dynamically decide
which nodes to traverse.

\item[\normalfont{(ol)} \text{\ttfamily one\char45{}liner\char45{}1\char46{}0}~\cite{Visscher:2018:OL}]
A library implementing profunctor style generic traversals using generics
in a similar style to \text{\ttfamily generic\char45{}lens}. It generates the most general
constrained traversal which we instantiate suitably to turn it into an
ordinary traversal.
\end{description}
We implement a collection of benchmarks which modify, update, and summarise
data types of three different sizes. \ensuremath{\Conid{Tree}} is a simple data type representing binary trees with
\ensuremath{\mathrm{2}} constructors. \ensuremath{\Conid{Logic}} is a deep embedding of propositional logic with \ensuremath{\mathrm{6}} constructors.
\ensuremath{\Conid{HsModule}} is a large data type representing a Haskell syntax tree with many constructors.

Our benchmarks have been compiled with \text{\ttfamily \char45{}fexpose\char45{}all\char45{}unfoldings} and
enable a later specialisation pass. The former ensures more
predictable cross-module inlining. The latter is a more aggressive
change which in particular helps the HsMod benchmark by performing a specialisation pass
towards the end of the compilation.

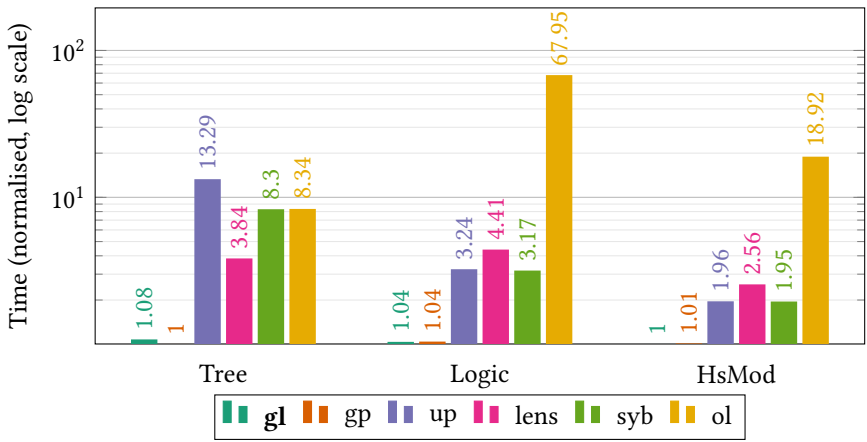
\begin{figure}
\begin{tikzpicture}
  \begin{axis}[
    xtick = data,
    xticklabels = { Tree, Logic, HsMod },
    major x tick style = transparent,
    width  = 0.85*\textwidth,
    height = 6cm,
    enlarge x limits=0.25,
    enlarge y limits={upper, value=0.25},
    ylabel = {Time (normalised, log scale)},
    ybar,
    cycle list name = Dark2-8,
    every axis plot/.append style={fill,draw=none,no markers},
    ymode = log,
    ymajorgrids = true,
    yminorgrids = true,
    minor y grid style = {line width=0.05pt, gray!20},
    legend cell align=left,
    legend columns=-1,
    legend style={
      at={(0.5,-0.15)},
      anchor= north,
      column sep=1ex,
    },
    point meta=rawy,
    every node near coord/.append style={rotate=90, anchor=west},
    nodes near coords,
    nodes near coords align = {vertical}
  ]

\pgfplotstableread[col sep=comma]{bench-results/hand.csv}{\hand}
\foreach\filename in {
bench-results/gl.csv,
bench-results/geniplate.csv,
bench-results/uniplate.csv,
bench-results/lens.csv,
bench-results/syb.csv,
bench-results/one-liner.csv}{%
\pgfplotstableread[col sep=comma]{\filename}\thistable
\pgfplotstablecreatecol[create col/copy column from table=\hand{Mean}]{y scale}\thistable
\addplot table [x expr= \coordindex, y expr={\thisrow{y} / \thisrow{y scale}}] \thistable;
};
\legend{\textbf{gl}, gp, up, lens, syb, ol}

	\end{axis}
\end{tikzpicture}
\caption{Benchmarks showing execution time normalised as a factor of
hand-written code}
\label{fig:benchmarks}
\end{figure}

We show three results in Figure~\ref{fig:benchmarks}. The y-axis is a
log scale where we normalise against the hand-written code. The
number above each column indicates the time relative to the
hand-written code. For example, a value of 2 indicates that
the benchmark took twice as much time as the hand-written definition.
We include one modification benchmark for each different data type to indicate the
relative performance of the libraries.

One should notice how the performance of our \text{\ttfamily generic\char45{}lens} library,
labelled \textbf{(gl)}, is comparable to the hand written examples and
\text{\ttfamily geniplate} (gp) which uses Template
Haskell to analyse the data definitions and to produce the optimal
code. On the other hand, SYB (syb) is consistently very slow. \text{\ttfamily one\char45{}liner} (ol)
performs an order of magnitude worse than SYB in these benchmarks.
Experiments at higher inlining thresholds indicate that the
performance can be comparable with our library%
.

We also have a large suite of other benchmarks which we have used to validate our
approach. These include effectful traversals, summarising and traversing dense and
sparse structures. We observe that in all these cases, the performance is very
close to the hand-written definitions.

\mpi{the renumberInt benchmark is one using a state monad}

\mpi{I think Logic is as fast as hand-written with unfolding-threshold-110}

% include Considerations.lhs
\section{Related Work}
\label{sec:related}

The scrap your boilerplate work pioneered the generation of code
suitable for generic traversals~\cite{Lammel:2003:SYB,Lammel:2004:SMB}. Our
work can be seen as the latest refinement in a long line of techniques.
We believe that we are the first to provide an efficient embeddeding in Haskell using
existing language features which does not rely on Template Haskell~\cite{Adams:2012:Template}.
Furthermore, the type-changing variants which are
enabled by our use of generics is novel.
\citet{Magalhaes:2014:multiple} considered multiple parameters, but has a
different solution to ours and is less flexible.

Our method is to use the generic structure of data, which was prominently
investigated by~\citet{Hinze:2006:generics}, and has led to an implementation
in GHC that allows generic type classes to be
derived~\cite{Magalhaes:2010:generic}. \ensuremath{\Conid{\Conid{GHC}.Generics}} was intended to be a low-level
way for library authors to implement generic programming libraries.
Later \citet{Edsko:2014:SOP} implemented \text{\ttfamily generics\char45{}sop},
a high-level interface. However, it is not efficient as library
authors have to use recursive functions to convert from their representation.
\citet{Visscher:2018:OL} have implemented the low-level \text{\ttfamily one\char45{}liner} library,
which provides constrained profunctor traversals. As we saw in the benchmarks, we predict their performance
could be comparable to ours with suitable compiler hints. In our interface we already
provide several different traversal schemes rather than just the constrained variant.
We also consider problems such as removing redundant traversals and
type inference.

Lenses have quickly been adopted into the Haskell ecosystem since~\cite{2009:vanLaarhoven:cps}
and have
been described from first principles in~\cite{Pickering:2017:optics}.
There are many different libraries implementing lenses in the Haskell
ecosystem but by far the most prominent and well used is \text{\ttfamily lens}~\cite{kmett_lens}.
The interface our library provides is compatible with their library and heavily
inspired by it. The \text{\ttfamily lens} library can also derive lenses using Template Haskell.
We find this unsatisfactory as
users must decide up-front which lenses they want for their data types. As a result,
users usually invent named fields with a specific naming strategy and derive
lenses with names based from them. They can't derive the other queries we specify,
in part because defining all such lenses up front would incur namespace pollution.
It is more desirable to have the flexibility we provide in order to specify
precisely at use-sites the mode of inspection.

\citet{Yallop:2017:staged} discusses how structured multi-staged
programming techniques can be used to improve SYB. The work observes the
benefits which can be gained from a simple binding time analysis.
The insight is that when applied, the types a traversal must deal with
are already known.
Once the target of the traversal is known, a specialised version for
that specific structure is created. Further, if the traversal function
is also statically known, it can be partially evaluated at each node to
eliminate the dynamic type checks.
The constrained traversal can be seen as an implementation of
the same idea in the current work.
By knowing the type, we generate a specialised traversal which
targets all fields but has the right unrolled structure.
By instantiating the constraint (which we do statically) we can then select
which parts of the data type to target.

They further develop this approach in order to analyse additional
local transformations which can be applied in order to improve the
generated code.
In section~\ref{sec:performance} we explained how compiler optimisations
performed by GHC amount to achieving the same thing.
As one particular example, they observe the need for let-insertion to
deal with recursive definitions. Their solution is to use a fixpoint combinator
which supports memoisation. They then insert let bindings for each recursive call.
For us, this memoisation is performed in the same manner by the
constraint solver.

Their work is valuable as it carefully analyses the precise optimisations needed
to create optimal code. Our work is complementary as we observe that with an
automatic partial evaluator (GHC's optimiser), it is sufficient to instruct
a simple unrolling before passing the code to be automatically optimised
once it is no longer recursive. Like ours, Yallop's implementation also hinges on
an observation that implicit information can be treated as static. It is this
fact which allows much of the static overhead to be eliminated automatically.
\cite{Magalhaes:2012:Optimisation} also considered the behaviour of the optimiser
on similar generic programs.

\citet{Adams:2014:optimizing} analysed the poor
performance of SYB and performed code optimizations
expressed in HERMIT~\cite{Farmer:2015:Hermit} to get better performance, in their extended work
they show how this can be done with an adapted version of
GHC~\cite{Adams:2015:optimizing}. A large problem with their approach is that
they rely on being able to symbolically evaluate type casts. The implementation
of \ensuremath{\Conid{Typeable}}~\cite{Jones:2016:Reflection} implements type equality by comparing
fingerprints using primitive operations. There is no hope for a compiler to evaluate
these operations without additional guidance. Our solution does not generate any
domain specific idiosyncracies which we need to modify the optimiser in order to
eliminate.

\bibliography{bibliography}


\begin{thebibliography}{36}

%%% the Web address.

\ifx \showCODEN    \undefined \def \showCODEN     #1{\unskip}     \fi
\ifx \showDOI      \undefined \def \showDOI       #1{#1}\fi
\ifx \showISBNx    \undefined \def \showISBNx     #1{\unskip}     \fi
\ifx \showISBNxiii \undefined \def \showISBNxiii  #1{\unskip}     \fi
\ifx \showISSN     \undefined \def \showISSN      #1{\unskip}     \fi
\ifx \showLCCN     \undefined \def \showLCCN      #1{\unskip}     \fi
\ifx \shownote     \undefined \def \shownote      #1{#1}          \fi
\ifx \showarticletitle \undefined \def \showarticletitle #1{#1}   \fi
\ifx \showURL      \undefined \def \showURL       {\relax}        \fi
% invisible to TeX
\providecommand\bibfield[2]{#2}
\providecommand\bibinfo[2]{#2}
\providecommand\natexlab[1]{#1}
\providecommand\showeprint[2][]{arXiv:#2}

\bibitem[\protect\citeauthoryear{Adams and DuBuisson}{Adams and
  DuBuisson}{2012}]%
        {Adams:2012:Template}
\bibfield{author}{\bibinfo{person}{Michael~D Adams} {and}
  \bibinfo{person}{Thomas~M DuBuisson}.} \bibinfo{year}{2012}\natexlab{}.
\newblock \showarticletitle{Template your boilerplate: Using Template Haskell
  for efficient generic programming}. In \bibinfo{booktitle}{\emph{ACM SIGPLAN
  Notices}}, Vol.~\bibinfo{volume}{47}. ACM, \bibinfo{pages}{13--24}.
\newblock


\bibitem[\protect\citeauthoryear{Adams, Farmer, and Magalh\~{a}es}{Adams
  et~al\mbox{.}}{2014}]%
        {Adams:2014:optimizing}
\bibfield{author}{\bibinfo{person}{Michael~D. Adams}, \bibinfo{person}{Andrew
  Farmer}, {and} \bibinfo{person}{Jos{\'e}~Pedro Magalh\~{a}es}.}
  \bibinfo{year}{2014}\natexlab{}.
\newblock \showarticletitle{Optimizing SYB is Easy!}. In
  \bibinfo{booktitle}{\emph{Proceedings of the ACM SIGPLAN 2014 Workshop on
  Partial Evaluation and Program Manipulation}} \emph{(\bibinfo{series}{PEPM
  '14})}. \bibinfo{publisher}{ACM}, \bibinfo{address}{New York, NY, USA},
  \bibinfo{pages}{71--82}.
\newblock
\showISBNx{978-1-4503-2619-3}
\urldef\tempurl%
\url{https://doi.org/10.1145/2543728.2543730}
\showDOI{\tempurl}


\bibitem[\protect\citeauthoryear{Adams, Farmer, and Magalh\~{a}es}{Adams
  et~al\mbox{.}}{2015}]%
        {Adams:2015:optimizing}
\bibfield{author}{\bibinfo{person}{Michael~D. Adams}, \bibinfo{person}{Andrew
  Farmer}, {and} \bibinfo{person}{Jos{\'e}~Pedro Magalh\~{a}es}.}
  \bibinfo{year}{2015}\natexlab{}.
\newblock \showarticletitle{Optimizing SYB Traversals is Easy!}
\newblock \bibinfo{journal}{\emph{Sci. Comput. Program.}}
  \bibinfo{volume}{112}, \bibinfo{number}{P2} (\bibinfo{date}{Nov.}
  \bibinfo{year}{2015}), \bibinfo{pages}{170--193}.
\newblock
\showISSN{0167-6423}
\urldef\tempurl%
\url{https://doi.org/10.1016/j.scico.2015.09.003}
\showDOI{\tempurl}


\bibitem[\protect\citeauthoryear{Augustsson}{Augustsson}{2018}]%
        {Augustsson:2018:Geniplate}
\bibfield{author}{\bibinfo{person}{Lennart Augustsson}.}
  \bibinfo{year}{2018}\natexlab{}.
\newblock \bibinfo{title}{\texttt{geniplate-mirror-0.7.6} library}.
\newblock   (\bibinfo{year}{2018}).
\newblock
\urldef\tempurl%
\url{http://hackage.haskell.org/package/geniplate-mirror-0.7.6}
\showURL{%
\tempurl}


\bibitem[\protect\citeauthoryear{Bird, Gibbons, Mehner, Voigtlaender, and
  Schrijvers}{Bird et~al\mbox{.}}{2013}]%
        {Bird:2013:Traversals}
\bibfield{author}{\bibinfo{person}{Richard Bird}, \bibinfo{person}{Jeremy
  Gibbons}, \bibinfo{person}{Stefan Mehner}, \bibinfo{person}{Janis
  Voigtlaender}, {and} \bibinfo{person}{Tom Schrijvers}.}
  \bibinfo{year}{2013}\natexlab{}.
\newblock \showarticletitle{Understanding Idiomatic Traversals Backwards and
  Forwards}. In \bibinfo{booktitle}{\emph{Haskell Symposium}}.
\newblock
\urldef\tempurl%
\url{http://www.comlab.ox.ac.uk/jeremy.gibbons/publications/uitbaf.pdf}
\showURL{%
\tempurl}


\bibitem[\protect\citeauthoryear{Bolingbroke}{Bolingbroke}{2011}]%
        {Bolingbroke_CK}
\bibfield{author}{\bibinfo{person}{Maximilian~C. Bolingbroke}.}
  \bibinfo{year}{2011}\natexlab{}.
\newblock \bibinfo{title}{Constraint Kinds for GHC}.
\newblock   (\bibinfo{year}{2011}).
\newblock
\urldef\tempurl%
\url{http://blog.omega-prime.co.uk/2011/09/10/constraint-kinds-for-ghc/}
\showURL{%
\tempurl}


\bibitem[\protect\citeauthoryear{Breitner, Eisenberg, Peyton~Jones, and
  Weirich}{Breitner et~al\mbox{.}}{2014}]%
        {Breitner:2014:safe-coercions}
\bibfield{author}{\bibinfo{person}{Joachim Breitner},
  \bibinfo{person}{Richard~A. Eisenberg}, \bibinfo{person}{Simon Peyton~Jones},
  {and} \bibinfo{person}{Stephanie Weirich}.} \bibinfo{year}{2014}\natexlab{}.
\newblock \showarticletitle{Safe Zero-cost Coercions for Haskell}.
\newblock \bibinfo{journal}{\emph{SIGPLAN Not.}} \bibinfo{volume}{49},
  \bibinfo{number}{9} (\bibinfo{date}{Aug.} \bibinfo{year}{2014}),
  \bibinfo{pages}{189--202}.
\newblock
\showISSN{0362-1340}
\urldef\tempurl%
\url{https://doi.org/10.1145/2692915.2628141}
\showDOI{\tempurl}


\bibitem[\protect\citeauthoryear{Chakravarty, Keller, {Peyton Jones}, and
  Marlow}{Chakravarty et~al\mbox{.}}{2005}]%
        {assoc05}
\bibfield{author}{\bibinfo{person}{Manuel M.~T. Chakravarty},
  \bibinfo{person}{Gabriele Keller}, \bibinfo{person}{Simon {Peyton Jones}},
  {and} \bibinfo{person}{Simon Marlow}.} \bibinfo{year}{2005}\natexlab{}.
\newblock \showarticletitle{Associated types with class}. In
  \bibinfo{booktitle}{\emph{POPL '05: Proceedings of the 32nd ACM
  SIGPLAN-SIGACT sysposium on Principles of programming languages}}.
  \bibinfo{publisher}{ACM Press}, \bibinfo{pages}{1--13}.
\newblock
\showISBNx{1-58113-830-X}
\urldef\tempurl%
\url{https://doi.org/10.1145/1040305.1040306}
\showDOI{\tempurl}


\bibitem[\protect\citeauthoryear{de~Vries and L{\"o}h}{de~Vries and
  L{\"o}h}{2014}]%
        {Edsko:2014:SOP}
\bibfield{author}{\bibinfo{person}{Edsko de Vries} {and}
  \bibinfo{person}{Andres L{\"o}h}.} \bibinfo{year}{2014}\natexlab{}.
\newblock \showarticletitle{True sums of products}. In
  \bibinfo{booktitle}{\emph{Proceedings of the 10th ACM SIGPLAN workshop on
  Generic programming}}. ACM, \bibinfo{pages}{83--94}.
\newblock


\bibitem[\protect\citeauthoryear{Eisenberg, Vytiniotis, Peyton~Jones, and
  Weirich}{Eisenberg et~al\mbox{.}}{2014}]%
        {Eisenberg:2014:closed}
\bibfield{author}{\bibinfo{person}{Richard~A. Eisenberg},
  \bibinfo{person}{Dimitrios Vytiniotis}, \bibinfo{person}{Simon Peyton~Jones},
  {and} \bibinfo{person}{Stephanie Weirich}.} \bibinfo{year}{2014}\natexlab{}.
\newblock \showarticletitle{Closed Type Families with Overlapping Equations}.
  In \bibinfo{booktitle}{\emph{Proceedings of the 41st ACM SIGPLAN-SIGACT
  Symposium on Principles of Programming Languages}}
  \emph{(\bibinfo{series}{POPL '14})}. \bibinfo{publisher}{ACM},
  \bibinfo{address}{New York, NY, USA}, \bibinfo{pages}{671--683}.
\newblock
\showISBNx{978-1-4503-2544-8}
\urldef\tempurl%
\url{https://doi.org/10.1145/2535838.2535856}
\showDOI{\tempurl}


\bibitem[\protect\citeauthoryear{Eisenberg, Weirich, and Ahmed}{Eisenberg
  et~al\mbox{.}}{2016}]%
        {Eisenberg:2016:visible}
\bibfield{author}{\bibinfo{person}{Richard~A. Eisenberg},
  \bibinfo{person}{Stephanie Weirich}, {and} \bibinfo{person}{Hamidhasan~G.
  Ahmed}.} \bibinfo{year}{2016}\natexlab{}.
\newblock \showarticletitle{Visible Type Application}. In
  \bibinfo{booktitle}{\emph{Proceedings of the 25th European Symposium on
  Programming Languages and Systems - Volume 9632}}.
  \bibinfo{publisher}{Springer-Verlag New York, Inc.}, \bibinfo{address}{New
  York, NY, USA}, \bibinfo{pages}{229--254}.
\newblock
\showISBNx{978-3-662-49497-4}
\urldef\tempurl%
\url{https://doi.org/10.1007/978-3-662-49498-1_10}
\showDOI{\tempurl}


\bibitem[\protect\citeauthoryear{Farmer}{Farmer}{2015}]%
        {Farmer:2015:Hermit}
\bibfield{author}{\bibinfo{person}{Andrew Farmer}.}
  \bibinfo{year}{2015}\natexlab{}.
\newblock \emph{\bibinfo{title}{{HERMIT:} Mechanized Reasoning during
  Compilation in the Glasgow Haskell Compiler}}.
\newblock \bibinfo{thesistype}{Ph.D. Dissertation}. \bibinfo{school}{University
  of Kansas, {USA}}.
\newblock
\urldef\tempurl%
\url{http://hdl.handle.net/1808/19416}
\showURL{%
\tempurl}


\bibitem[\protect\citeauthoryear{Hinze}{Hinze}{2006}]%
        {Hinze:2006:generics}
\bibfield{author}{\bibinfo{person}{Ralf Hinze}.}
  \bibinfo{year}{2006}\natexlab{}.
\newblock \showarticletitle{Generics for the Masses}.
\newblock \bibinfo{journal}{\emph{J. Funct. Program.}} \bibinfo{volume}{16},
  \bibinfo{number}{4-5} (\bibinfo{date}{July} \bibinfo{year}{2006}),
  \bibinfo{pages}{451--483}.
\newblock
\showISSN{0956-7968}
\urldef\tempurl%
\url{https://doi.org/10.1017/S0956796806006022}
\showDOI{\tempurl}


\bibitem[\protect\citeauthoryear{Jaskelioff and O'Connor}{Jaskelioff and
  O'Connor}{2015}]%
        {Jaskelioff:Representation:2014}
\bibfield{author}{\bibinfo{person}{Mauro Jaskelioff} {and}
  \bibinfo{person}{Russell O'Connor}.} \bibinfo{year}{2015}\natexlab{}.
\newblock \showarticletitle{A {Representation} {Theorem} for {Second}-{Order}
  {Functionals}}.
\newblock \bibinfo{journal}{\emph{Journal of Functional Programming}}
  \bibinfo{volume}{25}, \bibinfo{number}{e13} (\bibinfo{year}{2015}).
\newblock
\urldef\tempurl%
\url{https://doi.org/10.1017/S0956796815000088}
\showDOI{\tempurl}


\bibitem[\protect\citeauthoryear{Kmett}{Kmett}{2018}]%
        {kmett_lens}
\bibfield{author}{\bibinfo{person}{Edward Kmett}.}
  \bibinfo{year}{2018}\natexlab{}.
\newblock \bibinfo{title}{\texttt{lens-4.16} library}.
\newblock   (\bibinfo{year}{2018}).
\newblock
\urldef\tempurl%
\url{https://hackage.haskell.org/package/lens-4.16}
\showURL{%
\tempurl}


\bibitem[\protect\citeauthoryear{L{\"a}mmel and Peyton~Jones}{L{\"a}mmel and
  Peyton~Jones}{2003}]%
        {Lammel:2003:SYB}
\bibfield{author}{\bibinfo{person}{Ralf L{\"a}mmel} {and}
  \bibinfo{person}{Simon Peyton~Jones}.} \bibinfo{year}{2003}\natexlab{}.
\newblock \showarticletitle{Scrap Your Boilerplate: A Practical Design Pattern
  for Generic Programming}. In \bibinfo{booktitle}{\emph{Types in Languages
  Design and Implementation}}. \bibinfo{publisher}{ACM Press},
  \bibinfo{address}{New York, NY, USA}, \bibinfo{pages}{26--37}.
\newblock
\urldef\tempurl%
\url{https://doi.org/10.1145/640136.604179}
\showDOI{\tempurl}


\bibitem[\protect\citeauthoryear{L\"{a}mmel and Peyton~Jones}{L\"{a}mmel and
  Peyton~Jones}{2004}]%
        {Lammel:2004:SMB}
\bibfield{author}{\bibinfo{person}{Ralf L\"{a}mmel} {and}
  \bibinfo{person}{Simon Peyton~Jones}.} \bibinfo{year}{2004}\natexlab{}.
\newblock \showarticletitle{Scrap More Boilerplate: Reflection, Zips, and
  Generalised Casts}. In \bibinfo{booktitle}{\emph{Proceedings of the Ninth ACM
  SIGPLAN International Conference on Functional Programming}}
  \emph{(\bibinfo{series}{ICFP '04})}. \bibinfo{publisher}{ACM},
  \bibinfo{address}{New York, NY, USA}, \bibinfo{pages}{244--255}.
\newblock
\showISBNx{1-58113-905-5}
\urldef\tempurl%
\url{https://doi.org/10.1145/1016850.1016883}
\showDOI{\tempurl}


\bibitem[\protect\citeauthoryear{Magalh\~{a}es, Dijkstra, Jeuring, and
  L\"{o}h}{Magalh\~{a}es et~al\mbox{.}}{2010}]%
        {Magalhaes:2010:generic}
\bibfield{author}{\bibinfo{person}{Jos{\'e}~Pedro Magalh\~{a}es},
  \bibinfo{person}{Atze Dijkstra}, \bibinfo{person}{Johan Jeuring}, {and}
  \bibinfo{person}{Andres L\"{o}h}.} \bibinfo{year}{2010}\natexlab{}.
\newblock \showarticletitle{A Generic Deriving Mechanism for Haskell}. In
  \bibinfo{booktitle}{\emph{Proceedings of the Third ACM Haskell Symposium on
  Haskell}} \emph{(\bibinfo{series}{Haskell '10})}. \bibinfo{publisher}{ACM},
  \bibinfo{address}{New York, NY, USA}, \bibinfo{pages}{37--48}.
\newblock
\showISBNx{978-1-4503-0252-4}
\urldef\tempurl%
\url{https://doi.org/10.1145/1863523.1863529}
\showDOI{\tempurl}


\bibitem[\protect\citeauthoryear{Magalh{\~a}es}{Magalh{\~a}es}{2012}]%
        {Magalhaes:2012:Optimisation}
\bibfield{author}{\bibinfo{person}{Jos{\'e}~Pedro Magalh{\~a}es}.}
  \bibinfo{year}{2012}\natexlab{}.
\newblock \showarticletitle{Optimisation of generic programs through inlining}.
  In \bibinfo{booktitle}{\emph{Symposium on Implementation and Application of
  Functional Languages}}. Springer, \bibinfo{pages}{104--121}.
\newblock


\bibitem[\protect\citeauthoryear{Magalh{\~a}es}{Magalh{\~a}es}{2014}]%
        {Magalhaes:2014:multiple}
\bibfield{author}{\bibinfo{person}{Jos{\'e}~Pedro Magalh{\~a}es}.}
  \bibinfo{year}{2014}\natexlab{}.
\newblock \showarticletitle{Generic Programming with Multiple Parameters}. In
  \bibinfo{booktitle}{\emph{Functional and Logic Programming}},
  \bibfield{editor}{\bibinfo{person}{Michael Codish} {and}
  \bibinfo{person}{Eijiro Sumii}} (Eds.). \bibinfo{publisher}{Springer
  International Publishing}, \bibinfo{address}{Cham},
  \bibinfo{pages}{136--151}.
\newblock
\showISBNx{978-3-319-07151-0}


\bibitem[\protect\citeauthoryear{McBride and Paterson}{McBride and
  Paterson}{2008}]%
        {Mcbride:2008:Applicative}
\bibfield{author}{\bibinfo{person}{Conor McBride} {and} \bibinfo{person}{Ross
  Paterson}.} \bibinfo{year}{2008}\natexlab{}.
\newblock \showarticletitle{Applicative programming with effects}.
\newblock \bibinfo{journal}{\emph{Journal of functional programming}}
  \bibinfo{volume}{18}, \bibinfo{number}{1} (\bibinfo{year}{2008}),
  \bibinfo{pages}{1--13}.
\newblock


\bibitem[\protect\citeauthoryear{Mitchell and Runciman}{Mitchell and
  Runciman}{2007}]%
        {Mitchell:2007:Uniform}
\bibfield{author}{\bibinfo{person}{Neil Mitchell} {and} \bibinfo{person}{Colin
  Runciman}.} \bibinfo{year}{2007}\natexlab{}.
\newblock \showarticletitle{Uniform boilerplate and list processing}. In
  \bibinfo{booktitle}{\emph{Proceedings of the ACM SIGPLAN workshop on Haskell
  workshop}}. ACM, \bibinfo{pages}{49--60}.
\newblock


\bibitem[\protect\citeauthoryear{O'Connor}{O'Connor}{2011}]%
        {Oconnor:Functor:2011}
\bibfield{author}{\bibinfo{person}{Russell O'Connor}.}
  \bibinfo{year}{2011}\natexlab{}.
\newblock \showarticletitle{Functor is to Lens as Applicative is to Biplate:
  Introducing Multiplate}.
\newblock \bibinfo{journal}{\emph{CoRR}}  \bibinfo{volume}{abs/1103.2841}
  (\bibinfo{year}{2011}).
\newblock
\urldef\tempurl%
\url{https://doi.org/arXiv:1103.2841}
\showDOI{\tempurl}
\newblock
\shownote{Presented at WGP 2011.}


\bibitem[\protect\citeauthoryear{Peyton~Jones and Marlow}{Peyton~Jones and
  Marlow}{2002}]%
        {Jones:2002:Secrets}
\bibfield{author}{\bibinfo{person}{Simon Peyton~Jones} {and}
  \bibinfo{person}{Simon Marlow}.} \bibinfo{year}{2002}\natexlab{}.
\newblock \showarticletitle{Secrets of the glasgow haskell compiler inliner}.
\newblock \bibinfo{journal}{\emph{Journal of Functional Programming}}
  \bibinfo{volume}{12}, \bibinfo{number}{4-5} (\bibinfo{year}{2002}),
  \bibinfo{pages}{393--434}.
\newblock


\bibitem[\protect\citeauthoryear{Peyton~Jones, Vytiniotis, Weirich, and
  Shields}{Peyton~Jones et~al\mbox{.}}{2007}]%
        {PeytonJones:2007:PTI}
\bibfield{author}{\bibinfo{person}{Simon Peyton~Jones},
  \bibinfo{person}{Dimitrios Vytiniotis}, \bibinfo{person}{Stephanie Weirich},
  {and} \bibinfo{person}{Mark Shields}.} \bibinfo{year}{2007}\natexlab{}.
\newblock \showarticletitle{Practical Type Inference for Arbitrary-rank Types}.
\newblock \bibinfo{journal}{\emph{J. Funct. Program.}} \bibinfo{volume}{17},
  \bibinfo{number}{1} (\bibinfo{date}{Jan.} \bibinfo{year}{2007}),
  \bibinfo{pages}{1--82}.
\newblock
\showISSN{0956-7968}
\urldef\tempurl%
\url{https://doi.org/10.1017/S0956796806006034}
\showDOI{\tempurl}


\bibitem[\protect\citeauthoryear{Peyton~Jones, Weirich, Eisenberg, and
  Vytiniotis}{Peyton~Jones et~al\mbox{.}}{2016}]%
        {Jones:2016:Reflection}
\bibfield{author}{\bibinfo{person}{Simon Peyton~Jones},
  \bibinfo{person}{Stephanie Weirich}, \bibinfo{person}{Richard~A Eisenberg},
  {and} \bibinfo{person}{Dimitrios Vytiniotis}.}
  \bibinfo{year}{2016}\natexlab{}.
\newblock \showarticletitle{A reflection on types}.
\newblock In \bibinfo{booktitle}{\emph{A List of Successes That Can Change the
  World}}. \bibinfo{publisher}{Springer}, \bibinfo{pages}{292--317}.
\newblock


\bibitem[\protect\citeauthoryear{Peyton~Jones and Santos}{Peyton~Jones and
  Santos}{1998}]%
        {Jones:1998:Transformation}
\bibfield{author}{\bibinfo{person}{Simon~L Peyton~Jones} {and}
  \bibinfo{person}{Andr{\'e}L~M Santos}.} \bibinfo{year}{1998}\natexlab{}.
\newblock \showarticletitle{A transformation-based optimiser for Haskell}.
\newblock \bibinfo{journal}{\emph{Science of computer programming}}
  \bibinfo{volume}{32}, \bibinfo{number}{1-3} (\bibinfo{year}{1998}),
  \bibinfo{pages}{3--47}.
\newblock


\bibitem[\protect\citeauthoryear{Pickering, Gibbons, and Wu}{Pickering
  et~al\mbox{.}}{2017}]%
        {Pickering:2017:optics}
\bibfield{author}{\bibinfo{person}{Matthew Pickering}, \bibinfo{person}{Jeremy
  Gibbons}, {and} \bibinfo{person}{Nicolas Wu}.}
  \bibinfo{year}{2017}\natexlab{}.
\newblock \showarticletitle{Profunctor Optics: Modular Data Accessors}.
\newblock \bibinfo{journal}{\emph{Programming Journal}} \bibinfo{volume}{1},
  \bibinfo{number}{2} (\bibinfo{year}{2017}), \bibinfo{pages}{7}.
\newblock
\urldef\tempurl%
\url{https://doi.org/10.22152/programming-journal.org/2017/1/7}
\showDOI{\tempurl}


\bibitem[\protect\citeauthoryear{Sulzmann, Duck, Peyton~Jones, and
  Stuckey}{Sulzmann et~al\mbox{.}}{2007}]%
        {Sulzmann:2007:fundeps-chr}
\bibfield{author}{\bibinfo{person}{Martin Sulzmann},
  \bibinfo{person}{Gregory~J. Duck}, \bibinfo{person}{Simon Peyton~Jones},
  {and} \bibinfo{person}{Peter~J. Stuckey}.} \bibinfo{year}{2007}\natexlab{}.
\newblock \showarticletitle{Understanding Functional Dependencies via
  Constraint Handling Rules}.
\newblock \bibinfo{journal}{\emph{J. Funct. Program.}} \bibinfo{volume}{17},
  \bibinfo{number}{1} (\bibinfo{date}{Jan.} \bibinfo{year}{2007}),
  \bibinfo{pages}{83--129}.
\newblock
\showISSN{0956-7968}
\urldef\tempurl%
\url{https://doi.org/10.1017/S0956796806006137}
\showDOI{\tempurl}


\bibitem[\protect\citeauthoryear{van Laarhoven}{van Laarhoven}{2009}]%
        {2009:vanLaarhoven:cps}
\bibfield{author}{\bibinfo{person}{Twan van Laarhoven}.}
  \bibinfo{year}{2009}\natexlab{}.
\newblock \bibinfo{title}{{CPS}-Based Functional References}.
  (\bibinfo{date}{July} \bibinfo{year}{2009}).
\newblock
\urldef\tempurl%
\url{http://www.twanvl.nl/blog/haskell/cps-functional-references}
\showURL{%
\tempurl}


\bibitem[\protect\citeauthoryear{Visscher and Xia}{Visscher and Xia}{2018}]%
        {Visscher:2018:OL}
\bibfield{author}{\bibinfo{person}{Sjoerd Visscher} {and}
  \bibinfo{person}{Li-yao Xia}.} \bibinfo{year}{2018}\natexlab{}.
\newblock \bibinfo{title}{\texttt{one-liner-1.0} library}.
\newblock   (\bibinfo{year}{2018}).
\newblock
\urldef\tempurl%
\url{http://hackage.haskell.org/package/one-liner-1.0}
\showURL{%
\tempurl}


\bibitem[\protect\citeauthoryear{Vytiniotis, Peyton~Jones, Schrijvers, and
  Sulzmann}{Vytiniotis et~al\mbox{.}}{2011}]%
        {Vytiniotis:2011:outsidein}
\bibfield{author}{\bibinfo{person}{Dimitrios Vytiniotis},
  \bibinfo{person}{Simon Peyton~Jones}, \bibinfo{person}{Tom Schrijvers}, {and}
  \bibinfo{person}{Martin Sulzmann}.} \bibinfo{year}{2011}\natexlab{}.
\newblock \showarticletitle{Outsidein(x) Modular Type Inference with Local
  Assumptions}.
\newblock \bibinfo{journal}{\emph{J. Funct. Program.}} \bibinfo{volume}{21},
  \bibinfo{number}{4-5} (\bibinfo{date}{Sept.} \bibinfo{year}{2011}),
  \bibinfo{pages}{333--412}.
\newblock
\showISSN{0956-7968}
\urldef\tempurl%
\url{https://doi.org/10.1017/S0956796811000098}
\showDOI{\tempurl}


\bibitem[\protect\citeauthoryear{Wadler and Blott}{Wadler and Blott}{1989}]%
        {Wadler:1989:MAP}
\bibfield{author}{\bibinfo{person}{Phillip Wadler} {and}
  \bibinfo{person}{Stephen Blott}.} \bibinfo{year}{1989}\natexlab{}.
\newblock \showarticletitle{How to Make Ad-hoc Polymorphism Less Ad Hoc}. In
  \bibinfo{booktitle}{\emph{Proceedings of the 16th ACM SIGPLAN-SIGACT
  Symposium on Principles of Programming Languages}}
  \emph{(\bibinfo{series}{POPL '89})}. \bibinfo{publisher}{ACM},
  \bibinfo{address}{New York, NY, USA}, \bibinfo{pages}{60--76}.
\newblock
\showISBNx{0-89791-294-2}
\urldef\tempurl%
\url{https://doi.org/10.1145/75277.75283}
\showDOI{\tempurl}


\bibitem[\protect\citeauthoryear{Weirich, Vytiniotis, Peyton~Jones, and
  Zdancewic}{Weirich et~al\mbox{.}}{2011}]%
        {Weirich:2011:GTA}
\bibfield{author}{\bibinfo{person}{Stephanie Weirich},
  \bibinfo{person}{Dimitrios Vytiniotis}, \bibinfo{person}{Simon Peyton~Jones},
  {and} \bibinfo{person}{Steve Zdancewic}.} \bibinfo{year}{2011}\natexlab{}.
\newblock \showarticletitle{Generative Type Abstraction and Type-level
  Computation}. In \bibinfo{booktitle}{\emph{Proceedings of the 38th Annual ACM
  SIGPLAN-SIGACT Symposium on Principles of Programming Languages}}
  \emph{(\bibinfo{series}{POPL '11})}. \bibinfo{publisher}{ACM},
  \bibinfo{address}{New York, NY, USA}, \bibinfo{pages}{227--240}.
\newblock
\showISBNx{978-1-4503-0490-0}
\urldef\tempurl%
\url{https://doi.org/10.1145/1926385.1926411}
\showDOI{\tempurl}


\bibitem[\protect\citeauthoryear{Yallop}{Yallop}{2017}]%
        {Yallop:2017:staged}
\bibfield{author}{\bibinfo{person}{Jeremy Yallop}.}
  \bibinfo{year}{2017}\natexlab{}.
\newblock \showarticletitle{Staged Generic Programming}.
\newblock \bibinfo{journal}{\emph{Proc. ACM Program. Lang.}}
  \bibinfo{volume}{1}, \bibinfo{number}{ICFP}, Article \bibinfo{articleno}{29}
  (\bibinfo{date}{Aug.} \bibinfo{year}{2017}), \bibinfo{numpages}{29}~pages.
\newblock
\showISSN{2475-1421}
\urldef\tempurl%
\url{https://doi.org/10.1145/3110273}
\showDOI{\tempurl}


\bibitem[\protect\citeauthoryear{Yorgey, Weirich, Cretin, Peyton~Jones,
  Vytiniotis, and Magalh\~{a}es}{Yorgey et~al\mbox{.}}{2012}]%
        {Yorgey:2012:promotion}
\bibfield{author}{\bibinfo{person}{Brent~A. Yorgey}, \bibinfo{person}{Stephanie
  Weirich}, \bibinfo{person}{Julien Cretin}, \bibinfo{person}{Simon
  Peyton~Jones}, \bibinfo{person}{Dimitrios Vytiniotis}, {and}
  \bibinfo{person}{Jos{\'e}~Pedro Magalh\~{a}es}.}
  \bibinfo{year}{2012}\natexlab{}.
\newblock \showarticletitle{Giving Haskell a Promotion}. In
  \bibinfo{booktitle}{\emph{Proceedings of the 8th ACM SIGPLAN Workshop on
  Types in Language Design and Implementation}} \emph{(\bibinfo{series}{TLDI
  '12})}. \bibinfo{publisher}{ACM}, \bibinfo{address}{New York, NY, USA},
  \bibinfo{pages}{53--66}.
\newblock
\showISBNx{978-1-4503-1120-5}
\urldef\tempurl%
\url{https://doi.org/10.1145/2103786.2103795}
\showDOI{\tempurl}


\end{thebibliography}
\clearpage

\end{document}